\documentclass[fleqn,10pt]{article}
\usepackage{latexsym, graphicx, epsfig, amsmath, amssymb,amsfonts}
\usepackage{natbib,amsthm,version}
\usepackage{amsbsy,bm,multirow,enumerate}
\usepackage[titletoc,page]{appendix}
\usepackage[mathscr]{eucal}
\usepackage{mathtools}
\usepackage{xcolor}
\usepackage{subfigure}
\usepackage[utf8]{inputenc}
\usepackage[english]{babel}
\usepackage{float}
\usepackage{caption}
\usepackage{systeme}
\usepackage[affil-it]{authblk}
\usepackage{hyperref}

\DeclareCaptionFormat{cont}{#1 (continued)#2#3\par}

\begin{document}

\title{ 
	Implementing contact angle boundary conditions for second-order Phase-Field models of wall-bounded multiphase flows
	\footnote{\copyright $<$2022$>$. This manuscript version is made available under the CC-BY-NC-ND 4.0 license \url{http://creativecommons.org/licenses/by-nc-nd/4.0/}.}
	\footnote{This manuscript was accepted for publication in Journal of Computational Physics, Vol 471, Ziyang Huang, Guang Lin, Arezoo M. Ardekani, Implementing contact angle boundary conditions for second-order Phase-Field models of wall-bounded multiphase flows, Page 111619, Copyright Elsevier (2022).}
} 

\author[1]{
	Ziyang Huang%
	\thanks{Email: \texttt{ziyangh@umich.edu}.}}

\author[2,3]{
	Guang Lin%
	\thanks{Email: \texttt{guanglin@purdue.edu}; Corresponding author at Department of Mathematics, Purdue University, West Lafayette, IN 47907, USA.}}

\author[2]{
	Arezoo M. Ardekani%
	\thanks{Email: \texttt{ardekani@purdue.edu}; Corresponding author at School of Mechanical Engineering, Purdue University, West Lafayette, IN 47907, USA.}}

\affil[1]{
	Mechanical Engineering Department, University of Michigan, Ann Arbor, MI 48109, USA}
\affil[2]{
	School of Mechanical Engineering, Purdue University, West Lafayette, IN 47907, USA}
\affil[3]{
	Department of Mathematics, Purdue University, West Lafayette, IN 47907, USA}

\date{}

\maketitle


\begin{abstract}
In the present work, a general formulation is proposed to implement the contact angle boundary conditions for the second-order Phase-Field models, which is applicable to $N$-phase $(N \geqslant 2)$ moving contact line problems. 
To remedy the issue of mass change due to the contact angle boundary condition, a source term or Lagrange multiplier is added to the original second-order Phase-Field models, which is determined by the consistent and conservative volume distribution algorithm so that the summation of the order parameters and the \textit{consistency of reduction} are not influenced. 
To physically couple the proposed formulation to the hydrodynamics, especially for large-density-ratio problems, the consistent formulation is employed. 
The reduction-consistent conservative Allen-Cahn models are chosen as examples to illustrate the application of the proposed formulation.
The numerical scheme that preserves the consistency and conservation of the proposed formulation is employed to demonstrate its effectiveness.
Results produced by the proposed formulation are in good agreement with the exact and/or asymptotic solutions. The proposed method captures complex dynamics of moving contact line problems having large density ratios.
\end{abstract}

\vspace{0.05cm}

\section{Introduction}\label{Sec Introduction}
Moving contact line problems are ubiquitous in both natural phenomena and industrial applications. Various numerical models have been developed for this kind of problems, such as 
the front tracking method \citep{UnverdiTryggvason1992,Tryggvasonetal2001,ManservisiScardovelli2009,MuradogluTasoglu2010}, 
the level-set method \citep{OsherSethian1988,SethianSmereka2003,Spelt2005,ZhangYue2020}, 
the conservative level-set method \citep{OlssonKreiss2005,Olssonetal2007,Zahedietal2009,Satoetal2012}, and 
the volume-of-fluid (VOF) method \citep{HirtNichols1981,ScardovelliZaleski1999,Renardyetal2001,Afkhamietal2009,Yokoi2011}, and the contact angle boundary conditions therein. We refer interested readers to the comprehensive review \citep{Suietal2014}. 

In the present study, we focus on the Phase-Field (or Diffuse-Interface) models \citep{Andersonetal1998}, where the interface is represented as a transient layer with a small but finite thickness. Different from the sharp-interface models, which only include advection, diffusion in the Phase-Field models regularizes the singularity at the contact line. Such an additional effect can drive the contact line to move even though the no-slip boundary condition is assigned \citep{Seppecher1996,Jacqmin2000}. 
One commonly used procedure to derive the contact angle boundary conditions for the Phase-Field models is in the context of wall energy relaxation \citep{Jacqmin2000,Qianetal2006,Dong2012,Baietal2017a,Shenetal2020}, where the wall energy is minimized by the $L^2$ gradient flow. Such a procedure has been extended to include surfactant \citep{Zhuetal2020}, contact angle hysteresis \citep{Yue2020}, three fluid phases \citep{ShiWang2014,Shenetal2015,ZhangWang2016}, and $N$ ($N \geqslant 2$) fluid phases \citep{Dong2017}.
Alternatively, the contact angle boundary conditions can also be geometry-based \citep{DingSpelt2007,LeeKim2011,Loudetetal2020}, where the orientation of the interface is explicitly enforced, 
and the one in \citep{DingSpelt2007} has been extended to model contact lines formed by three fluid phases \citep{Zhangetal2016}.
Most of these contact angle boundary conditions can be in general written as an inhomogeneous Neumann boundary condition.
Among various Phase-Field models, the Cahn-Hilliard Phase-Field model \citep{CahnHilliard1958} is most popularly used to model moving contact line problems, since the contact angle boundary conditions can be directly applied without influencing the mass conservation. The Cahn-Hilliard model is a 4th-order partial differential equation (PDE) and therefore we also call it a 4th-order Phase-Field model here. To uniquely solve it, each boundary requires two boundary conditions, one of which is determined by mass conservation. Flexibility is given to the remaining one to control the morphology of the interface, which is achieved by implementing the contact angle boundary conditions. The popularity of implementing the Cahn-Hilliard model has motivated several theoretical analyses, e.g., in \citep{Jacqmin2000,Qianetal2006,Yueetal2010,YueFeng2011,Xuetal2018}, and comparison studies, e.g., in \citep{DingSpelt2007,Lacisetal2020}.

More recently, the second-order Phase-Field models, such as the conservative Phase-Field models \citep{ChiuLin2011,Mirjalilietal2020} and the conservative Allen-Cahn models \citep{BrasselBretin2011,Huangetal2020B},
have attracted lots of attention and became popular in modeling both two-phase flows, e.g., in \citep{ChiuLin2011,MirjaliliMani2021,JeongKim2017,JoshiJaiman2018,JoshiJaiman2018adapt,Huangetal2020CAC}, and $N$-phase ($N \geqslant 2$) flows, e.g., in \citep{Aiharaetal2019,Huetal2020,Huangetal2020B}. They are modified from the Allen-Cahn model \citep{AllenCahn1979} and enjoy several desirable properties that the Cahn-Hilliard model does not have, but are important in multiphase flow modeling, such as conserving volume enclosed by the interface, preserving under-resolved structures, and the maximum principle \citep{BrasselBretin2011,Kimetal2014,LeeKim2016,KimLee2017,Chaietal2018,Mirjalilietal2020,Huangetal2020CAC,Huangetal2020B}. Moreover, it is easier and more efficient to solve the 2nd-order model than the 4th-order one. 
However, difficulty appears when these 2nd-order Phase-Field models are used to model problems including moving contact lines, because only a single boundary condition is needed. This boundary condition is always determined by the mass conservation and the homogeneous Neumann boundary condition is normally required. Consequently, only $90^0$ contact angle can be assigned at the wall boundary, which strongly restricts the application of the second-order Phase-Field models. So far, the second-order Phase-Field models have not been able to share the fruitful progress made in the implementation of the contact angle boundary conditions for moving contact line problems.

The present study attempts to address this issue and proposes a novel and general formulation which has the following desirable properties:
\begin{itemize}
    \item It is valid for both two-phase and $N$-phase ($N > 2$) cases.
    \item It does not rely on the specific forms of the 2nd-order Phase-Field models and the contact angle boundary conditions.
    \item It grantees the \textit{consistency of reduction}, the mass conservation of each phase, and the summation of the volume fractions to be unity.
    \item It incorporates the \textit{consistency of mass conservation} and the \textit{consistency of mass and momentum transport} for large-density-ratio problems.
\end{itemize}
The idea is to introduce a Lagrange multiplier to the original Phase-Field model, so that the mass change due to the contact angle boundary condition is compensated. The Lagrange multiplier needs to be carefully designed to avoid producing voids, overfilling, or fictitious phases, and therefore the consistent and conservative volume distribution algorithm is employed \citep{Huangetal2020B}. Finally, the coupling to the hydrodynamics is accomplished by using the consistent formulation \citep{Huangetal2020CAC}, which is essential for large-density-ratio problems.
This general formulation is applied to the reduction-consistent conservative Allen-Cahn models \citep{Brackbilletal1992,Huangetal2020B}, and various tests are performed to demonstrate its effectiveness.

The \textit{consistency of reduction}, \textit{consistency of mass conservation}, and \textit{consistency of mass and momentum transport} are modeling principles followed in the present study. The \textit{consistency of reduction} \citep{BoyerMinjeaud2014,Dong2017,Dong2018,Huangetal2020B} requires that a $N$-phase model should be able to reduce to the corresponding $M$-phase ($1 \leqslant M \leqslant N-1$) model when ($N-M$) phases are absent. Fictitious phases can be produced if this principle is violated, as demonstrated in the references mentioned. The \textit{consistency of mass conservation} and \textit{consistency of mass and momentum transport} \citep{Huangetal2020,Huangetal2020CAC,Huangetal2020N} illustrate the mass and momentum transport in the Phase-Field models, which couples the Phase-Field models to the hydrodynamics. Violating these principles can produce density-ratio-dependent velocity fluctuations, as demonstrated in the references mentioned. These consistency conditions have been successfully implemented to not only two/$N$-phase flows \citep{Huangetal2020,Huangetal2020CAC,Huangetal2020N,Huangetal2020B} but also multiphase flows with mass transfer \citep{Huangetal2020NPMC} and solidification/melting \citep{Huangetal2021Solid}. Various problems having density ratios beyond $1,000$ have been tested in those references. We refer interested readers to \citep{Dong2018,Huangetal2020,Huangetal2020N} where the definitions and analyses of the consistency conditions are detailed.

The rest of the paper is organized as follows.
In Section \ref{Sec Definitions and governing equations}, the general formulation to include the contact angle boundary condition in the second-order Phase-Field models and its coupling to the hydrodynamics are elaborated, followed by its application to the conservative Allen-Cahn models. 
In Section \ref{Sec Discretizations}, the numerical methods to solve the complete system is briefly summarized.
In Section \ref{Sec Results}, various numerical tests are performed to demonstrate the proposed formulation in moving contact line problems.
In Section \ref{Sec Conclusions}, the present study is concluded and some possible future directions are introduced.

\section{Definitions and governing equations}\label{Sec Definitions and governing equations}
We first define the problem in Section \ref{Sec Basic definitions}. Then, the general formulation of implementing the contact angle boundary condition for a second-order Phase-Field model is proposed and elaborated in Section \ref{Sec Phase-Field}, along with its coupling to the hydrodynamics in Sections \ref{Sec Mass} and \ref{Sec Momentum}. Finally, two specific examples, one for two-phase problems and the other for $N$-phase problems, are provided in Section \ref{Sec CAC}, which are the applications of the proposed general formulation described in Section \ref{Sec Phase-Field} to the conservative Allen-Cahn models.

\subsection{Basic definitions}\label{Sec Basic definitions}
There are $N$ ($N \geqslant 2$) different incompressible and immiscible fluid phases inside domain $\Omega$, and their locations are labeled by a set of order parameters $\{\phi_p\}_{p=1}^N$. The order parameters need to follow the summation constraint:
\begin{equation}\label{Eq Summation}
\sum_{p=1}^N C_p=\sum_{p=1}^N \frac{1+\phi_p}{2} =1
\quad \mathrm{or} \quad
\sum_{p=1}^N \phi_p=2-N,
\end{equation}
where $\{C_p\}_{p=1}^N$ are the volume fractions of the phases and therefore their summation is always unity. In other words, void or overfilling is not allowed to appear. The densities and viscosities of the phases are denoted by $\{\rho_p\}_{p=1}^N$ and $\{\mu_p\}_{p=1}^N$, respectively. As a result, the mixture density and viscosity are
\begin{equation}\label{Eq Density}
\rho=\sum_{p=1}^N \rho_p \frac{1+\phi_p}{2},
\quad
\mu=\sum_{p=1}^N \mu_p \frac{1+\phi_p}{2}.
\end{equation}
Each pair of phases has a surface tension, for example, $\sigma_{p,q}$ denotes the surface tension at the interface of Phases $p$ and $q$. $\theta_{p,q}$ is the contact angle in between Phase $p$ and a wall boundary and formed by Phases $p$ and $q$. Notice that $\sigma_{p,q}$ ($=\sigma_{q,p}$) is symmetry, while $\theta_{p,q}$ and $\theta_{q,p}$ are supplementary angles, i.e., $\theta_{p,q}+\theta_{q,p}=\pi$, $1\leqslant p,q \leqslant N$. 
Since each phase is incompressible, the flow velocity is divergence-free \citep{Abelsetal2012,Dong2018,Huangetal2020N}, i.e.,
\begin{equation}\label{Eq Divergence}
\nabla \cdot \mathbf{u}=0.
\end{equation}
If there are only two phases, we denote $\phi_1=\phi$, $\phi_2=-\phi$, $\sigma=\sigma_{1,2}$, and $\theta=\theta_{1,2}$ for convenience. Consequently, one only needs to solve $\phi_1$ (or $\phi$), and $\phi_2$ is obtained automatically from Eq.(\ref{Eq Summation}), or equivalently $\phi_2=-\phi_1=-\phi$. Unless otherwise specified, the domain boundary $\partial \Omega$ is composed of wall boundaries, although periodic, inflow, or outflow boundaries can be incorporated, depending on specific problems.

\subsection{Governing equations}\label{Sec General formulations}
\begin{figure}[!t]
	\centering
	\includegraphics[scale=.4]{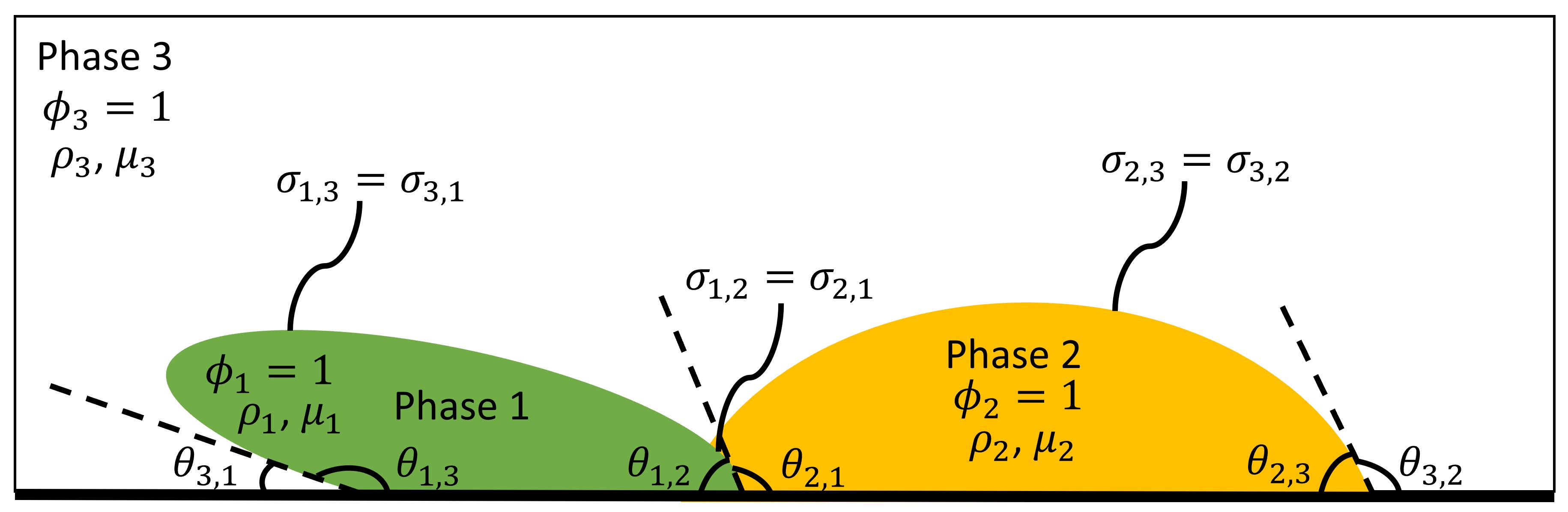}
	\caption{Sketch of the problem to be modeled by the second-order Phase-Field model and the contact angle boundary condition.
    \label{Fig Sketch}}
\end{figure}

The problem to be modeled by the second-order Phase-Field model and the contact angle boundary condition is sketched in Fig.\ref{Fig Sketch}. Here, the emphasis is on answering how to implement the contact angle boundary condition in the second-order Phase-Field model with the proposed general formulation. The hydrodynamics is included, following the \textit{consistency of mass conservation} and the \textit{consistency of mass and momentum transport} \citep{Huangetal2020,Huangetal2020CAC}.

\subsubsection{The proposed general formulation}\label{Sec Phase-Field}
The general form of the second-order Phase-Field model can be written as
\begin{equation}\label{Eq Phase-Field}
\frac{\partial \phi_p}{\partial t}
+
\nabla \cdot (\mathbf{u} \phi_p )
=
\mathcal{L}_p[\{\phi_q\}_{q=1}^N]
\quad \mathrm{in} \quad
\Omega,
\quad
1 \leqslant p \leqslant N,
\end{equation}
where $\mathcal{L}$ represents a functional of the order parameters, and the highest (spatial) derivatives included are the second-order derivatives. This is the reason that Eq.(\ref{Eq Phase-Field}) is called the second-order Phase-Field model. Because of the divergence-free velocity Eq.(\ref{Eq Divergence}), the convection term in Eq.(\ref{Eq Phase-Field}) has been written in a conservative form. To be physically admissible, $\mathcal{L}$ has the following properties:
\begin{eqnarray}\label{Eq Properties L}
\sum_{q=1}^N \mathcal{L}_q=0,
\quad
\int_\Omega \mathcal{L}_p d\Omega=0,
\quad
\mathcal{L}_p|_{\phi_p=-1}=0,
\quad
1 \leqslant p \leqslant N.
\end{eqnarray}
The first property in Eq.(\ref{Eq Properties L}) comes from the summation of the order parameters Eq.(\ref{Eq Summation}).
The second one implies the mass conservation of Phase $p$, which is shown more clearly after Eq.(\ref{Eq Phase-Field}) is integrated over $\Omega$:
\begin{equation}\label{Eq Integral phi_p}
\frac{d}{dt} \int_{\Omega} \phi_p d\Omega
+
\int_{\partial \Omega} \mathbf{n} \cdot \mathbf{u} \phi_p d\Gamma
=0,
\quad 
1 \leqslant p \leqslant N.
\end{equation}
Since $\mathcal{L}$ usually contains a diffusive term of type $\nabla^2 \phi$, to achieve the second property in Eq.(\ref{Eq Properties L}) and therefore the mass conservation of individual phases, i.e., Eq.(\ref{Eq Integral phi_p}), the homogeneous Neumann boundary condition is needed, i.e., $\int_\Omega \nabla^2 \phi d\Omega=\int_{\partial \Omega} \mathbf{n} \cdot \nabla \phi d\Gamma=0$. Such a boundary condition avoids the diffusive flux into wall boundaries.
Additionally with the impermeability condition, i.e., $\mathbf{n} \cdot \mathbf{u}=0$, at the domain boundary, we have $\frac{d}{dt}\int_\Omega \phi_p d\Omega=0$ from Eq.(\ref{Eq Integral phi_p}), which means the total mass of Phase $p$ in $\Omega$ will not change.
The last property in Eq.(\ref{Eq Properties L}) corresponds to the \textit{consistency of reduction} in the sense that Phase $p$ will not be produced if it is absent, i.e., $(\partial \phi_p/\partial t)|_{\phi_p=-1}=0$.
Notice that the convection term now becomes $\nabla \cdot (\mathbf{u} \phi_p)|_{\phi_p=-1}=-\nabla \cdot \mathbf{u}=0$.
However, in the present study, the contact angle boundary condition, i.e., 
\begin{equation}\label{Eq Contact angle boundary}
\mathbf{n} \cdot \nabla \phi_p = \mathcal{F}_p^w[\{\phi_q\}_{q=1}^N;\{\theta_{q,r}\}_{q,r=1}^N]
\quad \mathrm{at} \quad
\partial \Omega,
\quad
1 \leqslant p \leqslant N,
\end{equation}
needs to be implemented instead of the homogeneous Neumann boundary condition. As a result, the second property of $\mathcal{L}$ in Eq.(\ref{Eq Properties L}) is not guaranteed, and the mass conservation of each phase, i.e., Eq.(\ref{Eq Integral phi_p}), is probably violated. It should be noted that the notation in Eq.(\ref{Eq Contact angle boundary}) is simplified, and $\mathcal{F}^w$ and $\{\theta_{p,q}\}_{p,q=1}^N$ can be different at individual wall boundaries in practice. Similar to $\mathcal{L}$, physically admissible $\mathcal{F}^w$ has the following properties:
\begin{equation}\label{Eq Properties F}
\sum_{q=1}^N \mathcal{F}_q^w=0,
\quad
\mathcal{F}_p^w|_{\phi_p=-1}=0,
\quad
1 \leqslant p \leqslant N,
\end{equation}
compatible with the summation of the order parameters Eq.(\ref{Eq Summation}) and the \textit{consistency of reduction}, respectively.
This can be clearly seen after summing Eq.(\ref{Eq Contact angle boundary}) over the phases or setting $\phi_p=-1$ in Eq.(\ref{Eq Contact angle boundary}), and one obtains the respective two properties in Eq.(\ref{Eq Properties F}).

In order to implement the contact angle boundary condition Eq.(\ref{Eq Contact angle boundary}), while other physical principles are not violated, we propose to modify the second-order Phase-Field model Eq.(\ref{Eq Phase-Field}) to be
\begin{equation}\label{Eq Phase-Field contact}
\frac{\partial \phi_p}{\partial t}
+
\nabla \cdot (\mathbf{u} \phi_p )
=
\mathcal{L}_p[\{\phi_q\}_{q=1}^N]
+
L_p^w
\quad \mathrm{in} \quad
\Omega,
\quad
1 \leqslant p \leqslant N,
\end{equation}
where $L^w$ is the newly introduced Lagrange multiplier and has the following properties:
\begin{eqnarray}\label{Eq Properties L^w}
\sum_{q=1}^N L_q^w=0,
\quad
\int_{\Omega} L_p^w d\Omega = -\int_{\Omega} \mathcal{L}_p d\Omega=S_p
\quad \mathrm{with} \quad
\mathbf{n} \cdot \nabla \phi_p = \mathcal{F}_p^w
\quad \mathrm{at} \quad
\partial \Omega,
\quad
L_p^w|_{\phi_p=-1}=0,\\
\nonumber
1 \leqslant p \leqslant N,
\end{eqnarray}
to satisfy the summation of the order parameters Eq.(\ref{Eq Summation}), mass conservation of the phases Eq.(\ref{Eq Integral phi_p}), and \textit{consistency of reduction}, as explained below Eq.(\ref{Eq Properties L}). 
Here, we call $L^w$ a Lagrange multiplier, following studies like \citep{shen2011,BrasselBretin2011,Kimetal2014,LeeKim2016} which calls a source term added to the Phase-Field equation to enforce mass conservation a Lagrange multiplier.
Now, the question turns into determining $L^w$ that satisfies Eq.(\ref{Eq Properties L^w}). This question is successfully addressed by the consistent and conservative volume distribution algorithm in \citep{Huangetal2020B}. Specifically, $L^w$ is determined by
\begin{eqnarray}\label{Eq L^w}
L_p^w=\sum_{q=1}^N W_{p,q} B_q^w,
\quad
1 \leqslant p \leqslant N,\\
\nonumber
\sum_{q=1}^N \left(\int_{\Omega} W_{p,q} d\Omega \right) B_q^w=S_p,
\quad
W_{p,q}=\left\{
\begin{array}{cc}
     -(1+\phi_p)(1+\phi_q),& p \neq q,  \\
     (1+\phi_p)(1-\phi_q),&  p=q.
\end{array}
\right.
\end{eqnarray}
Notice that $B^w$ depends only on time and is solved from a $N$-by-$N$ symmetry and diagonally dominant linear system, thanks to the definition of $W_{p,q}$. 
Due to $\sum_{p=1}^N W_{p,q}=0$ from Eq.(\ref{Eq Summation}) and the definition of $W_{p,q}$ in Eq.(\ref{Eq L^w}), it is straightforward to show $\sum_{p=1}^N L_p^w=\sum_{q=1}^N B_q^w (\sum_{p=1}^N W_{p,q})=0$.
Satisfying the other two properties in Eq.(\ref{Eq Properties L^w}) by $L^w$ in Eq.(\ref{Eq L^w}) is obvious, and the related proofs are available in \citep{Huangetal2020B}. We refer interested readers to \citep{Huangetal2020B} for more details and analyses of the volume distribution algorithm which determines $L^w$ in Eq.(\ref{Eq L^w}).

When there are only two phases, as shown in \citep{Huangetal2020B}, one can obtain $L^w$ from Eq.(\ref{Eq L^w}) explicitly, i.e.,
\begin{eqnarray}\label{Eq L^w two-phase}
L_p^w=\frac{W_p}{\int_{\Omega} W_p d\Omega}S_p,
\quad 
W_p=1-\phi_p^2,
\quad
p=1,2.
\end{eqnarray}
As a result, we have the following two-phase second-order Phase-Field model with the contact angle boundary condition:
\begin{eqnarray}\label{Eq Phase-Field contact two-phase}
\frac{\partial \phi}{\partial t}
+
\nabla \cdot (\mathbf{u} \phi )
=
\mathcal{L}[\phi]+L^w
\quad \mathrm{in} \quad
\Omega,
\quad
\mathbf{n} \cdot \nabla \phi=\mathcal{F}^w[\phi;\theta]
\quad \mathrm{at} \quad
\partial \Omega,\\
\nonumber
L^w=\frac{W}{\int_{\Omega} W d\Omega} S,
\quad
S=-\int_{\Omega} \mathcal{L} d\Omega,
\quad
W=1-\phi^2.
\end{eqnarray}
Good performances of using $W_{p,q}$ in Eq.(\ref{Eq L^w}) and its two-phase reduction $W_p$ in Eq.(\ref{Eq L^w two-phase}) in Phase-Field models have been shown in previous studies, e.g., \citep{BrasselBretin2011,Kimetal2014,LeeKim2016,Huangetal2020B}. Their validity in two-/multi-phase flows has been evidenced, e.g., in \citep{Huangetal2020CAC,JeongKim2017,JoshiJaiman2018,JoshiJaiman2018adapt,Huangetal2020B} where physical results are reported.

The proposed general formulation is summarized as follows: given any physically admissible second-order Phase-Field model, i.e., Eq.(\ref{Eq Phase-Field}) satisfying Eq.(\ref{Eq Properties L}), and contact angle boundary condition, i.e., Eq.(\ref{Eq Contact angle boundary}) satisfying Eq.(\ref{Eq Properties F}), a new second-order Phase-Field model is developed, i.e., Eq.(\ref{Eq Phase-Field contact}) and Eq.(\ref{Eq L^w}), with the same contact angle boundary condition Eq.(\ref{Eq Contact angle boundary}), so that the summation of the order parameters Eq.(\ref{Eq Summation}), mass conservation of the phases Eq.(\ref{Eq Integral phi_p}), and \textit{consistency of reduction} are all satisfied. For two-phase problems, the proposed formulation becomes Eq.(\ref{Eq Phase-Field contact two-phase}).  

\subsubsection{Mass conservation and consistent formulation}\label{Sec Mass}
Before coupling to the hydrodynamics, we need to first determine the actual mass transport governed by the newly developed second-order Phase-Field model Eq.(\ref{Eq Phase-Field contact}) and the mixture density Eq.(\ref{Eq Density}). For a clear presentation, we combine $\mathcal{L}$ and $L^w$, i.e., $L=\mathcal{L}+L^w$, in Eq.(\ref{Eq Phase-Field contact}), and obtain 
\begin{equation}\label{Eq Phase-Field L}
\frac{\partial \phi_p}{\partial t}
+
\nabla \cdot (\mathbf{u} \phi_p)
=
L_p
\quad \mathrm{in} \quad
\Omega,
\quad
1 \leqslant p \leqslant N,
\end{equation}
\begin{equation}\label{Eq Integral L}
\int_{\Omega} L_p d\Omega =0,
\quad
1 \leqslant p \leqslant N,
\end{equation}
with the contact angle boundary condition Eq.(\ref{Eq Contact angle boundary}).
Next, we apply the consistent formulation \citep{Huangetal2020CAC}:
\begin{equation}\label{Eq Q}
\nabla \cdot (W_Q(\phi_p) \nabla Q_p )=L_p
\quad \mathrm{in} \quad
\Omega,
\quad
\mathbf{n} \cdot \nabla Q_p=0
\quad \mathrm{at} \quad \partial \Omega,
\quad
W_Q(\phi)=1-\phi^2,
\quad 
1 \leqslant p \leqslant N.
\end{equation}
Here, $Q$ is the auxiliary variable of the consistent formulation. The consistent formulation Eq.(\ref{Eq Q}) relates the non-local term $L$ to a local conservative form. More details about the consistent formulation are available in its original work \citep{Huangetal2020CAC}, and not repeated here. Notice that the homogeneous Neumann boundary condition of $Q$ is obtained from Eq.(\ref{Eq Integral L}).
After considering Eq.(\ref{Eq Phase-Field L}) and Eq.(\ref{Eq Q}), the newly proposed Phase-Field model Eq.(\ref{Eq Phase-Field contact}) is equivalent to
\begin{equation}\label{Eq Phase-Field conservative}
\frac{\partial \phi_p}{\partial t}+\nabla \cdot \mathbf{m}_{\phi_p}=0,
\end{equation}
where the Phase-Field flux $\mathbf{m}_\phi$ is
\begin{equation}\label{Eq Phase-Field flux}
\mathbf{m}_{\phi_p}=\mathbf{u}\phi_p-W_Q(\phi_p) \nabla Q_p,
\quad
1 \leqslant p \leqslant N.
\end{equation}
Following the formulation in \citep{Huangetal2020N}, we can immediately obtain the consistent mass flux:
\begin{equation}\label{Eq Mass flux}
\mathbf{m}=\sum_{p=1}^N \frac{\rho_p}{2} (\mathbf{u}+\mathbf{m}_{\phi_p}),
\end{equation}
which leads to the mass conservation equation:
\begin{equation}\label{Eq Mass}
\frac{\partial \rho}{\partial t}
+
\nabla \cdot \mathbf{m}
=0,
\end{equation}
after the mixture density Eq.(\ref{Eq Density}) is included.
The derivations in this section is based on the \textit{consistency of mass conservation} proposed and analyzed in \citep{Huangetal2020,Huangetal2020N,Huangetal2020CAC}.

\subsubsection{Momentum equation}\label{Sec Momentum}
The fluid motion is governed by the momentum equation:
\begin{equation}\label{Eq Momentum}
\frac{\partial (\rho \mathbf{u})}{\partial t}
+
\nabla \cdot (\mathbf{m} \otimes \mathbf{u})
=
-
\nabla P
+
\nabla \cdot \left[ \mu (\nabla \mathbf{u}+\nabla \mathbf{u}^T ) \right]
+
\rho \mathbf{g}
+
\mathbf{f}_s,
\end{equation}
where $P$ is the pressure, $\mathbf{g}$ is the gravity, and $\mathbf{f}_s$ is the surface tension force. Notice that the same mass flux $\mathbf{m}$, defined in Eq.(\ref{Eq Mass flux}), appears in both the mass conservation equation Eq.(\ref{Eq Mass}) and the inertial term of the momentum Eq.(\ref{Eq Momentum}), which is required by the \textit{consistency of mass and momentum transport} \citep{Huangetal2020,Huangetal2020CAC}. As a result, the momentum equation Eq.(\ref{Eq Momentum}) satisfies not only the momentum conservation but also kinetic energy conservation (neglecting the viscosity, gravity and surface tension) and Galilean invariance, see \citep{Huangetal2020N}. It should also be noted that simply using $\nabla \cdot (\rho \mathbf{u} \otimes \mathbf{u})$ as the nonlinear inertial term in the momentum equation cannot simultaneously achieve these physical properties .

In the present study, the surface tension force is
\begin{equation}\label{Eq Fs two-phase}
\mathbf{f}_s=\xi \nabla \phi,
\quad
\xi=\lambda \left( \frac{1}{\eta^2} g'(\phi)-\nabla^2 \phi \right),
\quad
\lambda=\frac{3}{2\sqrt{2}} \sigma \eta,
\quad
g(\phi)=\frac{1}{4}(1-\phi^2)^2,
\end{equation}
for two-phase problems, and
\begin{eqnarray}\label{Eq Fs}
\mathbf{f}_s=\frac{1}{2} \sum_{p=1}^N \xi_p \nabla \phi_p,
\quad
\xi_p=\sum_{q=1}^N \lambda_{p,q} \left[ \frac{1}{\eta^2} (g'_1(\phi_p)-g'_2(\phi_p+\phi_q) )+ \nabla^2 \phi_q \right],\\
\nonumber
\lambda_{p,q}=\frac{3}{2\sqrt{2}} \sigma_{p,q} \eta,
\quad
g_1(\phi)=\frac{1}{4} (1-\phi^2)^2,
\quad
g_2(\phi)=\frac{1}{4} \phi^2 (\phi+2)^2,
\end{eqnarray}
for multiphase problems. Here, $\lambda$ or $\lambda_{p,q}$ is the mixing energy density, $\eta$ is the interface thickness, $g(\phi)$, $g_1(\phi)$, and $g_2(\phi)$ are potential functions, and $g'(\phi)$, $g'_1(\phi)$, and $g'_2(\phi)$
are their derivatives with respect to $\phi$. Eq.(\ref{Eq Fs two-phase}) and Eq.(\ref{Eq Fs}) have been widely used in two-phase and $N$-phase flows, e.g., in \citep{Jacqmin1999,DongShen2012,Huangetal2020,Huangetal2020CAC,Dong2018,Huangetal2020N,HowardTartakovsky2020,Huetal2020}.

In summary, given any physically admissible 2nd-order Phase-Field model and contact angle boundary condition, i.e., $\mathcal{L}$ and $\mathcal{F}^w$, the governing equations include Eq.(\ref{Eq Phase-Field contact}) for the order parameters, Eq.(\ref{Eq Q}) from the consistent formulation, and Eq.(\ref{Eq Momentum}) and Eq.(\ref{Eq Divergence}) for the velocity and pressure. In the governing equations, the density (viscosity), consistent mass flux, and surface tension force are computed from Eq.(\ref{Eq Density}), Eq.(\ref{Eq Mass flux}), and Eq.(\ref{Eq Fs}) (or Eq.(\ref{Eq Fs two-phase})), respectively.

The proposed formulation has not considered the second law of thermodynamics, because many popularly used 2nd-order Phase-Field models, such as \citep{BrasselBretin2011,ChiuLin2011}, and the contact angle boundary conditions, particularly the geometry-based ones and the $N$-phase ones, e.g., \citep{DingSpelt2007,LeeKim2011,Loudetetal2020,Zhangetal2016,Dong2017}, are not explicitly shown to be consistent with the second law of thermodynamics. Moreover, it is still an open question to obtain a Lagrange multiplier that satisfies the constraints in Eq.(\ref{Eq Properties L^w}) and at the same time be consistent with the second law of thermodynamics. Actually, satisfying the constraints in Eq.(\ref{Eq Properties L^w}) alone is a very challenging task. We implement the algorithm in \citep{Huangetal2020B} to determine $L^w$ that satisfies all the constraints in Eq.(\ref{Eq Properties L^w}), but we are still unclear whether it is consistent with the second law of thermodynamics. Lastly, following the analyses in \citep{Huangetal2020N}, we would like to point out that, as long as the Phase-Field models with the contact angle boundary conditions are consistent with the second law of thermodynamics, such consistency will still be true after adding the momentum equation Eq.(\ref{Eq Momentum}).

\subsection{Application to the conservative Allen-Cahn models}\label{Sec CAC}
In the present study, the conservative Allen-Cahn models are considered as examples to demonstrate the effectiveness of the general formulation developed in Section \ref{Sec Phase-Field}. Specific formulations of $\mathcal{L}$ in the second-order Phase-Field model Eq.(\ref{Eq Phase-Field}) and $\mathcal{F}^w$ in the contact angle boundary condition Eq.(\ref{Eq Contact angle boundary}) are provided, and both two-phase and $N$-phase formulations are considered.

\subsubsection{Two-Phase model}\label{Sec Two-Phase}
The two-phase conservative Allen-Cahn model proposed in \citep{BrasselBretin2011} is considered, where $\mathcal{L}$ in the model is defined as
\begin{eqnarray}\label{Eq L CAC two-phase}
\mathcal{L}[\phi]
=M \lambda \left( \nabla^2 \phi - \frac{1}{\eta^2} g'(\phi) \right)+L^c,\\
\nonumber
L^c=W B^c,
\quad
W=1-\phi^2,
\quad
B^c=\frac{\int_\Omega \frac{M\lambda}{\eta^2} g'(\phi) d\Omega}{\int_{\Omega} W d\Omega}.
\end{eqnarray}
Here, $M$ is the mobility. One can easily show that $\mathcal{L}$ in Eq.(\ref{Eq L CAC two-phase}) satisfies all the conditions in Eq.(\ref{Eq Properties L}), and therefore it is physically admissible. The two-phase contact angle boundary condition considered is 
\begin{equation}\label{Eq F^w two-phase}
\mathcal{F}^w[\phi;\theta]=\frac{\sqrt{2}}{3\eta} \cos(\theta) g'_w(\phi),
\end{equation}
which is proposed in \citep{Jacqmin2000} from a wall functional. Here, $g_w(\phi)$ is an interpolation function satisfying $g_w(\pm1)=\pm1$ and $g'_w(\pm1)=0$, and we choose $g_w(\phi)=\sin \left(\frac{\pi}{2} \phi \right)$, like \citep{Shenetal2015,Baietal2017a,Huangetal2020,Shenetal2020}. Another choice of $g_w(\phi)$ is the Hermite polynomial, i.e., $g_w(\phi)=\frac{1}{2}\phi(3-\phi^2)$, used, e.g., in \citep{Jacqmin2000,Dong2012,ZhangWang2016,Yue2020}. Our tests do not find distinguishable difference of these two choices. Again, $\mathcal{F}^w$ in Eq.(\ref{Eq F^w two-phase}) is physically admissible since it satisfies Eq.(\ref{Eq Properties F}).  

Then, we apply the proposed formulations in Section \ref{Sec Phase-Field}, i.e., Eq.(\ref{Eq Phase-Field contact two-phase}), which introduces $L^w$ to the two-phase conservative Allen-Cahn model. Since both $L^c$ in Eq.(\ref{Eq L CAC two-phase}) and $L^w$ in Eq.(\ref{Eq Phase-Field contact two-phase}) share an identical weight function $W=1-\phi^2$, we can combine $L^c$ and $L^w$, i.e., $L^a=L^c+L^w$, as well as $B^c$ and $B^w$, i.e., $B=B^c+B^w$, for simplicity. As a result, we reach the following system:
\begin{eqnarray}\label{Eq CAC two-phase}
\frac{\partial \phi}{\partial t}
+
\nabla \cdot (\mathbf{u} \phi)
=
M \lambda \left( \nabla^2 \phi - \frac{1}{\eta^2} g'(\phi) \right)
+
L^a
\quad \mathrm{in} \quad
\Omega,
\quad
\mathbf{n} \cdot \nabla \phi=\frac{\sqrt{2}}{3\eta} \cos(\theta) g'_w(\phi)
\quad \mathrm{at} \quad
\partial \Omega,\\
\nonumber
L^a=W B,
\quad
W=1-\phi^2,
\quad
B=\frac{\int_\Omega M \lambda \left( \frac{1}{\eta^2} g'(\phi)-\nabla^2 \phi \right) d\Omega}{\int_{\Omega} W d\Omega},
\quad
g_w(\phi)=\sin \left(\frac{\pi}{2} \phi \right).
\end{eqnarray}
One can understand $L^a$ in Eq.(\ref{Eq CAC two-phase}) serving as a Lagrange multiplier to compensate the mass change inside $\Omega$ from $\frac{M\lambda}{\eta^2}g'(\phi)$ and at $\partial \Omega$ from the contact angle boundary condition.

\subsubsection{$N$-Phase model}\label{Sec N-Phase}
Here, we consider the reduction-consistent multiphase conservative Allen-Cahn model proposed in \citep{Huangetal2020B}, where $\mathcal{L}$ is defined as
\begin{eqnarray}\label{Eq L CAC N-phase}
\mathcal{L}_p[\{\phi_q\}_{q=1}^N]=
M \lambda_0 \left( \nabla^2 \phi_p - \frac{1}{\eta^2} \left(g'(\phi_p)-\frac{1+\phi_p}{2} L^s\right) \right)
+
L_p^c,
\quad
1 \leqslant p \leqslant N,\\
\nonumber
L^s=\sum_{q=1}^N g'(\phi_q),
\quad
L_p^c=\sum_{q=1}^N W_{p,q} B_q^c,
\quad
\int_{\Omega} L_p^c d\Omega
=\int_{\Omega} \frac{M \lambda_0}{\eta^2} \left(g'(\phi_p)-\frac{1+\phi_p}{2} L^s\right) d\Omega.
\end{eqnarray}
Here, $\lambda_0=\max(\lambda_{p,q})$, and $L^c$ is also determined from the consistent and conservative volume distribution algorithm \citep{Huangetal2020B}. Therefore $\mathcal{L}$ in Eq.(\ref{Eq L CAC N-phase}) satisfies all the conditions in Eq.(\ref{Eq Properties L}) and is physically admissible. We employ the reduction-consistent contact angle boundary condition proposed in \citep{Dong2017}, whose formulation is
\begin{eqnarray}\label{Eq F^w N-phase}
\mathcal{F}_p^w[\{\phi_q\}_{q=1}^N;\{\theta_{q,r}\}_{q,r=1}^N]= \sum_{q=1}^N \zeta_{p,q} \frac{1+\phi_p}{2} \frac{1+\phi_q}{2},
\quad
1 \leqslant p \leqslant N,\\
\nonumber
\zeta_{p,q}=\frac{2\sqrt{2}}{\eta} \cos(\theta_{p,q}).
\end{eqnarray}
Notice that $\zeta_{p,q}$ is antisymmetric, i.e., $\zeta_{p,q}=-\zeta_{q,p}$. Therefore, $\mathcal{F}^w$ in Eq.(\ref{Eq F^w N-phase}) also satisfies Eq.(\ref{Eq Properties F}) and is physically admissible.

Similar to the two-phase case in Section \ref{Sec Two-Phase}, we combine $L^c$ and $L^w$ after applying the general formulation proposed in Section \ref{Sec Phase-Field}, i.e., Eq.(\ref{Eq Phase-Field contact}). Therefore, we have the following system:
\begin{eqnarray}\label{Eq CAC N-phase}
\frac{\partial \phi_p}{\partial t}
+
\nabla \cdot (\mathbf{u} \phi_p)
=
M \lambda_0 \left( \nabla^2 \phi_p - \frac{1}{\eta^2} \left(g'(\phi_p)-\frac{1+\phi_p}{2} L^s\right) \right)
+
L_p^a
\quad \mathrm{in} \quad
\Omega,\\
\nonumber
\mathbf{n} \cdot \nabla \phi_p=\sum_{q=1}^N \zeta_{p,q} \frac{1+\phi_p}{2} \frac{1+\phi_q}{2}
\quad \mathrm{at} \quad
\partial \Omega,
\quad
1 \leqslant p \leqslant N,\\
\nonumber
L^s=\sum_{q=1}^N g'(\phi_q),
\quad
L_p^a=\sum_{q=1}^N W_{p,q} B_q,
\quad
\int_\Omega L_p^a d\Omega=\int_{\Omega} M \lambda_0 \left( \frac{1}{\eta^2} \left(g'(\phi_p)-\frac{1+\phi_p}{2} L^s\right)-\nabla^2 \phi_p \right) d\Omega,\\
\nonumber
W_{p,q}=\left\{
\begin{array}{cc}
     -(1+\phi_p)(1+\phi_q),& p \neq q,  \\
     (1+\phi_p)(1-\phi_q),&  p=q.
\end{array}
\right.,
\quad
\zeta_{p,q}=\frac{2\sqrt{2}}{\eta} \cos(\theta_{p,q}).
\end{eqnarray}
$L^a$ in Eq.(\ref{Eq CAC N-phase}) distributes the mass change due to both the Allen-Cahn model and the contact angle boundary condition consistently and conservatively, thanks to the volume distribution algorithm in \citep{Huangetal2020B}. Based on the analyses in \citep{Huangetal2020B,Dong2017}, Eq.(\ref{Eq CAC N-phase}) will exactly reduce to Eq.(\ref{Eq CAC two-phase}) with $g_w(\phi)$ the Hermite polynomial, i.e., $g_w(\phi)=\frac{1}{2}\phi(3-\phi^2)$, when there are only two phases. Then, Eq.(\ref{Eq CAC two-phase}) or Eq.(\ref{Eq CAC N-phase}) is coupled to the hydrodynamics following Section \ref{Sec Mass} and Section \ref{Sec Momentum}.

Here, Eq.(\ref{Eq CAC two-phase}) and Eq.(\ref{Eq CAC N-phase}) are the specific forms of Eq.(\ref{Eq Phase-Field contact}) of the proposed general formulation in Section~\ref{Sec Phase-Field}, based on the conservative Allen-Cahn models. The rest of the governing equations have already been summarized at the end of Section~\ref{Sec Momentum}.

\section{Discretizations}\label{Sec Discretizations}
Details of applying the consistent formulation discretely and solving the momentum equation consistently and conservatively are available in our previous works \citep{Huangetal2020,Huangetal2020CAC}. The balanced-force method \citep{Huangetal2020,Huangetal2020N} is used to compute the surface tension force in Eq.(\ref{Eq Fs two-phase}) and Eq.(\ref{Eq Fs}).
The (modified) conservative Allen-Cahn equations Eq.(\ref{Eq CAC two-phase}) and Eq.(\ref{Eq CAC N-phase}) are numerically solved from the 2nd-order schemes in \citep{Huangetal2020CAC,Huangetal2020B}, where the Allen-Cahn model, i.e., the one neglecting all the Lagrange multipliers, is first solved, and then the Lagrange multipliers are obtained from satisfying the summation of the order parameters Eq.(\ref{Eq Summation}), the mass conservation Eq.(\ref{Eq Integral L}), and the \textit{consistency of reduction}. All the integrals are computed using the mid-point rule. 
The schemes are semi-implicit based on the 2nd-order backward difference in time. The convection terms are treated explicitly with the 5th-order WENO scheme \citep{JiangShu1996}, and the diffusion terms are treated implicitly with the 2nd-order central difference \citep{FerzigerPeric2001}. The non-linear term $g'(\phi)$ in both Eq.(\ref{Eq CAC two-phase}) and Eq.(\ref{Eq CAC N-phase}) is first linearized and then treated implicitly. More details of the schemes can be found in \citep{Huangetal2020CAC,Huangetal2020B}.

The only difference in the present study appears at the boundary condition of the order parameters, where the homogeneous Neumann boundary condition is replaced by the contact angle boundary condition. This requires only minor changes, and the contact angle boundary condition is implemented explicitly following \citep{Huangetal2020,Huangetal2020N,Dong2012,Dong2017}, i.e.,
\begin{equation}\label{Eq contact angle discrete}
\mathbf{n} \cdot \nabla \phi_p^{n+1}=\mathcal{F}_p^w[\{\phi_q^{*,n+1}\}_{q=1}^N;\{\theta_{q,r}\}_{q,r=1}^N],
\end{equation}
where $\phi^{*,n+1}$ is an explicit evaluation of $\phi^{n+1}$ from $\phi^n$, $\phi^{n-1}$ etc. Specifically, $\phi^{*,n+1}=\phi^n$ is the first-order estimate, i.e., $\phi^{n+1}-\phi^{*,n+1}=\phi^{n+1}-\phi^{n} \sim O(\Delta t)$, and $\phi^{*,n+1}=2\phi^{n}-\phi^{n-1}$ is the second-order estimate, i.e., $\phi^{n+1}-\phi^{*,n+1}=\phi^{n+1}-2\phi^n+\phi^{n-1} \sim O(\Delta t^2)$. We use the second-order estimate in the present study.

In summary, the solution procedure is as follows:
\begin{enumerate}
    \item Solve Eq.(\ref{Eq CAC N-phase}) (or Eq.(\ref{Eq CAC two-phase})) with the scheme in \citep{Huangetal2020B} (or \citep{Huangetal2020CAC}) and the boundary condition Eq.(\ref{Eq contact angle discrete}) to update the order parameters.
    \item Solve Eq.(\ref{Eq Q}) with the scheme in \citep{Huangetal2020CAC} and then use Eq.(\ref{Eq Phase-Field flux}) to obtain the Phase-Field flux.
    \item Compute the density and viscosity with Eq.(\ref{Eq Density}), the consistent mass flux with Eq.(\ref{Eq Mass flux}), the surface tension force in Eq.(\ref{Eq Fs}) (or Eq.(\ref{Eq Fs two-phase})) with the balanced-force method \citep{Huangetal2020,Huangetal2020N}.
    \item Solve Eq.(\ref{Eq Momentum}) and Eq.(\ref{Eq Divergence}) to update the velocity and pressure with the scheme in \citep{Huangetal2020}.
\end{enumerate}
The chosen scheme has been carefully analyzed and verified, and we refer interested readers to \citep{Huangetal2020,Huangetal2020N,Huangetal2020CAC,Huangetal2020B} for more details.

\section{Results}\label{Sec Results}
Here, we mainly focus on demonstrating the effectiveness of the proposed general formulation in Section~\ref{Sec Phase-Field}, which applies to the conservative Allen-Cahn models in Section \ref{Sec CAC}, on modeling problems including moving contact lines. 
When setting up a case, it is sometimes more convenient to non-dimensionalize the governing equations in Section~\ref{Sec General formulations} and Section~\ref{Sec CAC}.
Given a density scale $\rho_{\mathrm{ref}}$, length scale $L_{\mathrm{ref}}$, and acceleration scale $a_{\mathrm{ref}}$, one can determine the scales of other variables in the governing equations, and they are listed in Table~\ref{Table Scales}, specifically to the conservative Allen-Cahn models.
Using those scales in Table~\ref{Table Scales}, one is able to obtain the dimensionless governing equations. 
The procedure is the same if $\rho_{\mathrm{ref}}$, $L_{\mathrm{ref}}$, and $u_{\mathrm{ref}}$ is given, and now $a_{\mathrm{ref}}$ becomes $u_{\mathrm{ref}}^2/L_{\mathrm{ref}}$ from Table~\ref{Table Scales}.
\begin{table*}[!t]
\caption{Scales of the variables in the conservative Allen-Cahn models given a density scale $\rho_{\mathrm{ref}}$, length scale $L_{\mathrm{ref}}$, and acceleration scale $a_{\mathrm{ref}}$}
    \centering
    \includegraphics[scale=.35]{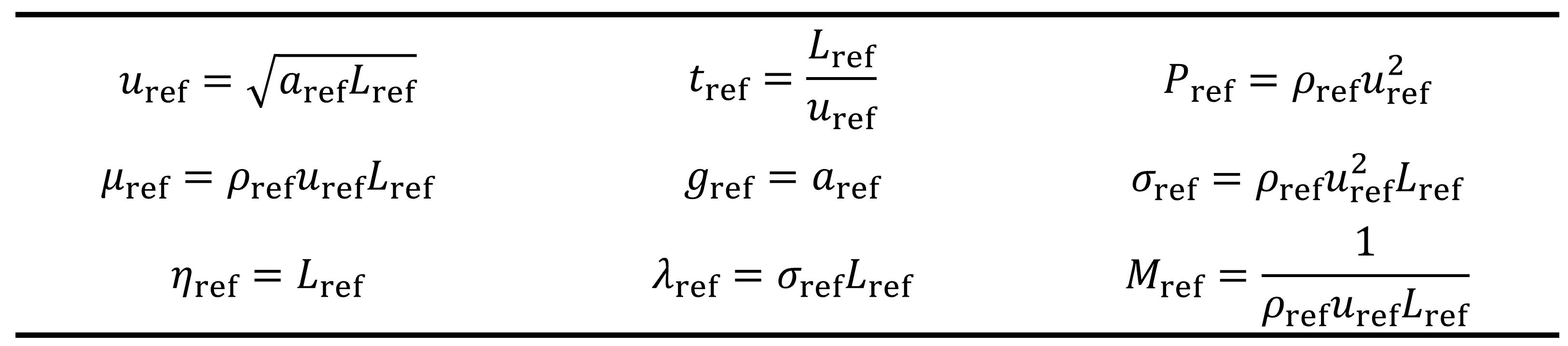}
    \label{Table Scales}
\end{table*}
The initial velocity is $\mathbf{u}=\mathbf{0} \mathrm{m/s}$ and we set $M\lambda=10^{-3} L_\mathrm{ref} u_\mathrm{ref}$ and $\eta=h$ unless otherwise specified, where $h$ denotes the grid size.

\subsection{Equilibrium drop}\label{Sec Equilibrium drop}
Here, we consider a semicircle liquid drop sliding on a horizontal solid wall using the two-phase model Eq.(\ref{Eq CAC two-phase}). 
The water drop initially has a radius of $R_0=8 \mathrm{mm}$, and is surrounded by the air. The material properties of the water and air considered are listed in Table \ref{Table ED}. The viscosities of the water and air are increased $20$ and $10$ times, respectively, in order to reach the equilibrium more quickly. Such a modification will not affect the conclusions drawn from the present section.
\begin{table*}[!t]
\caption{Material properties in the equilibrium drop}
    \centering
    \includegraphics[scale=.3]{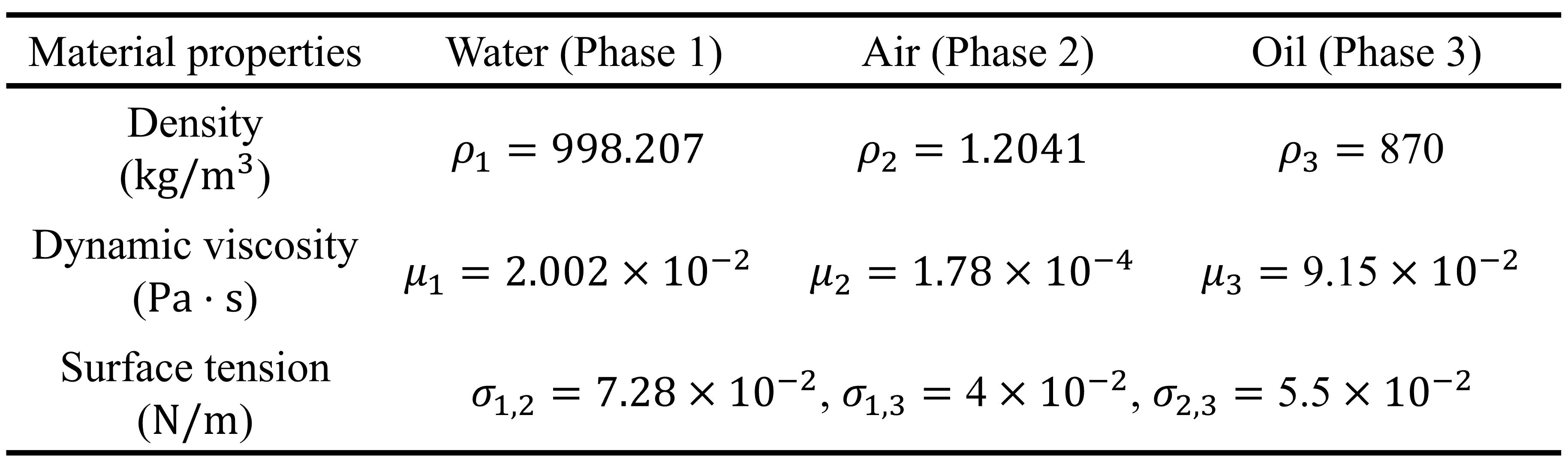}
    \label{Table ED}
\end{table*}
Considering the inertia-capillary velocity scale from the Weber number $We=\rho_1 U^2 R_0/\sigma_{1,2}=1$, we have the Reynolds number $Re=\rho_1 U R_0/\mu_1=38$, and capillary number $Ca=\mu_1 U/\sigma_{1,2}=0.0263$.
When the gravity is neglected, i.e., $|\mathbf{g}|=0$, one can obtain the final shape of the drop exactly using the mass conservation and the contact angle \citep{deGennesetal2003}. The exact solution is
\begin{equation}\label{Eq EDrop}
R_d=R_0 \sqrt{\frac{\pi/2}{\theta-\sin(\theta)\cos(\theta)}},
\quad
H_d=R_d (1-\cos(\theta) ),
\quad
L_d=2R_d \sin(\theta),
\end{equation}
where $R_d$, $H_d$, and $L_d$ are the final radius, height, and spreading length of the drop, respectively, and $\theta$ is the contact angle. 
Based on the asymptotic analysis for gravity-dominant cases \citep{deGennesetal2003}, the final height of the drop becomes
\begin{equation}\label{Eq EDrop g}
H_d
=
2 \sqrt{\frac{\sigma_{ds}}{\rho_d |\mathbf{g}|}} \sin(\theta/2),
\end{equation}
where $\rho_d$ is the density of the drop, $\sigma_{ds}$ is the surface tension between the drop and the surrounding phase, and $\mathbf{g}$ is the gravity pointing downward.

In the computation, we use $\rho_2$ (the air density), $L_\mathrm{ref}=5R_0=40\mathrm{mm}$, and $a_\mathrm{ref}=1\mathrm{m/s^2}$ as the density, length, and acceleration scales, respectively, to non-dimensionalize the governing equations. The length scale is chosen such that $\max(L_d)/L_\mathrm{ref} \sim O(1)$, where $\max(L_d)$ denotes the maximum final spreading length of the drop in all the considered cases. The acceleration scale is chosen for convenience when we investigate the effect of the gravity, as the dimensionless gravity will have the same value as the dimensional one. After the non-dimensionalization, the computational domain is $[-0.5,0.5]\times[0,0.3]$, and the drop is initially on the middle of the bottom wall. The lateral boundaries are periodic while they are no-slip walls at the top and bottom. Different contact angles are assigned at the bottom wall. We use $150\times 45$ grid cells and time step $\Delta t=1 \times 10^{-4}$ to discretize the space and time, respectively. All the results in this section are presented in their dimensionless forms.

\begin{figure}[!t]
	\centering
	\includegraphics[scale=.4]{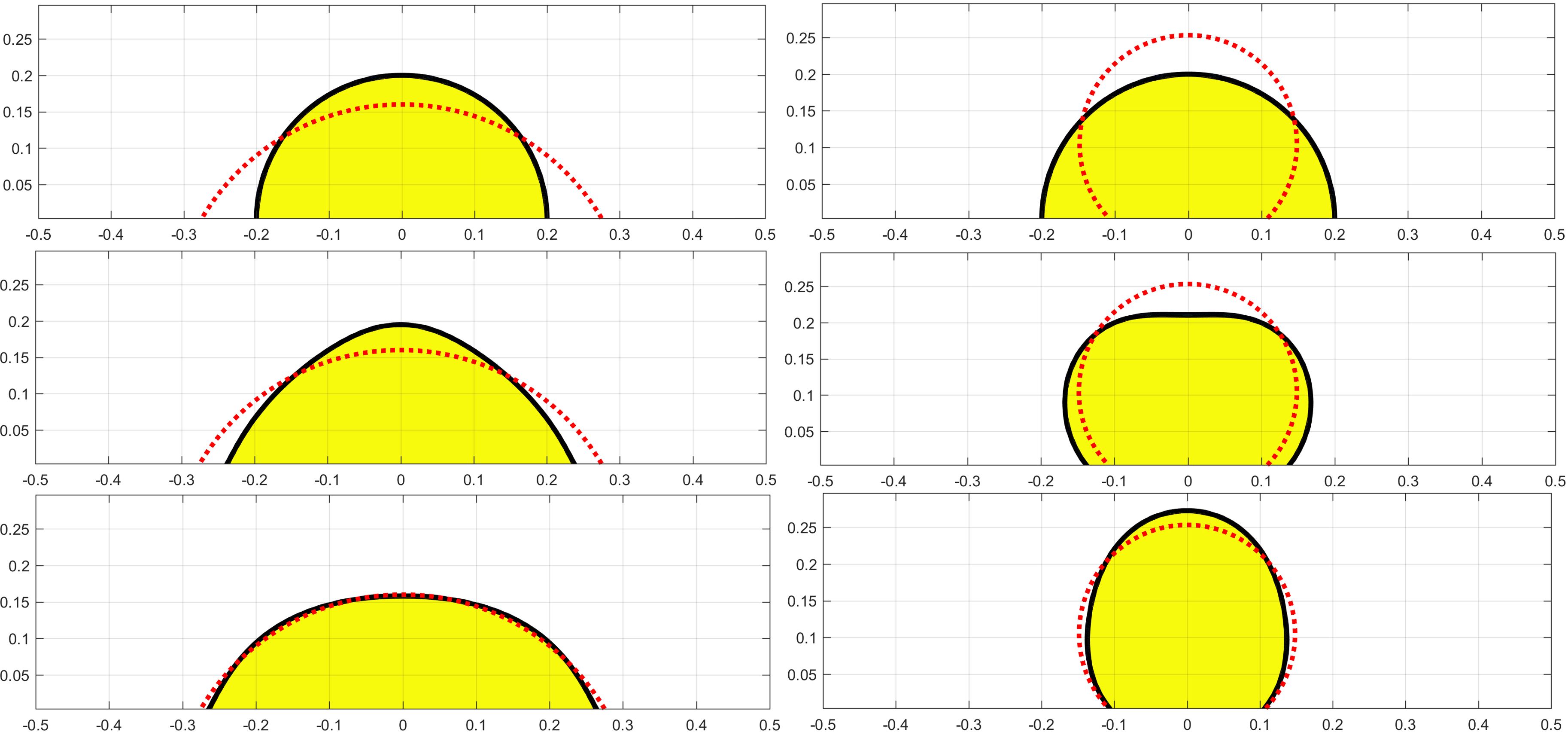}
	\includegraphics[scale=.4]{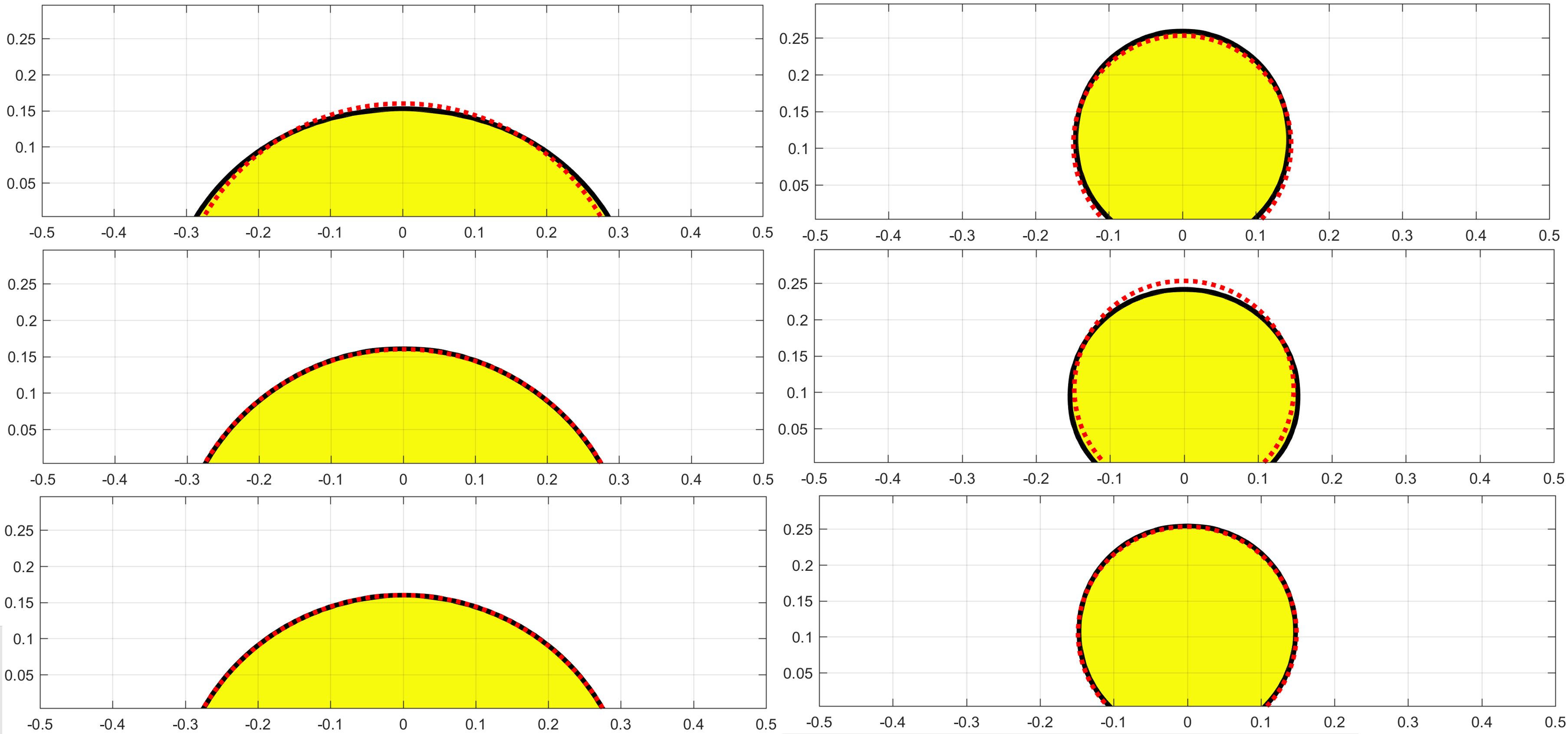}
	\caption{Evolution of the water drop using Eq.(\ref{Eq CAC two-phase}) with $|\mathbf{g}|=0$. Yellow: water ( Phase~1); White: air (Phase~2); Red dotted line: exact solution from Eq.(\ref{Eq EDrop}). Left column: $\theta=60^0$; Right column: $\theta=135^0$. From top to bottom, $t=0.0$, $t=0.2$, $t=0.4$, $t=1.0$, $t=1.4$, and $t=2.0$ (left) and $t=3.0$ (right).
    \label{Fig EDrop}}
\end{figure}

\begin{figure}[!t]
	\centering
	\includegraphics[scale=.45]{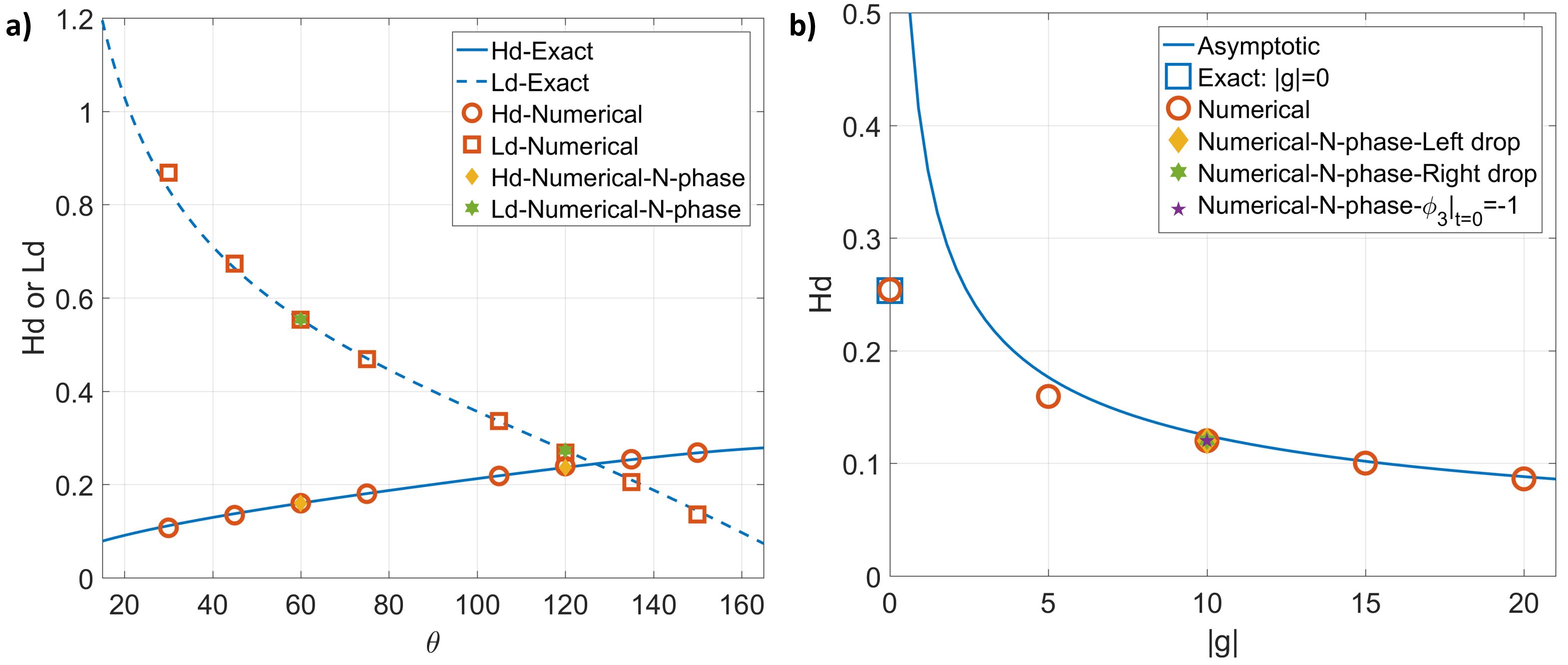}
	\caption{Height and spreading length of the drop. a) $H_d$ and $L_d$ versus $\theta$ with $|\mathbf{g}|=0$. b) $H_d$ versus $|\mathbf{g}|$ with $\theta=135^0$.
    \label{Fig EDrop HL}}
\end{figure}

\begin{figure}[!t]
	\centering
	\includegraphics[scale=.45]{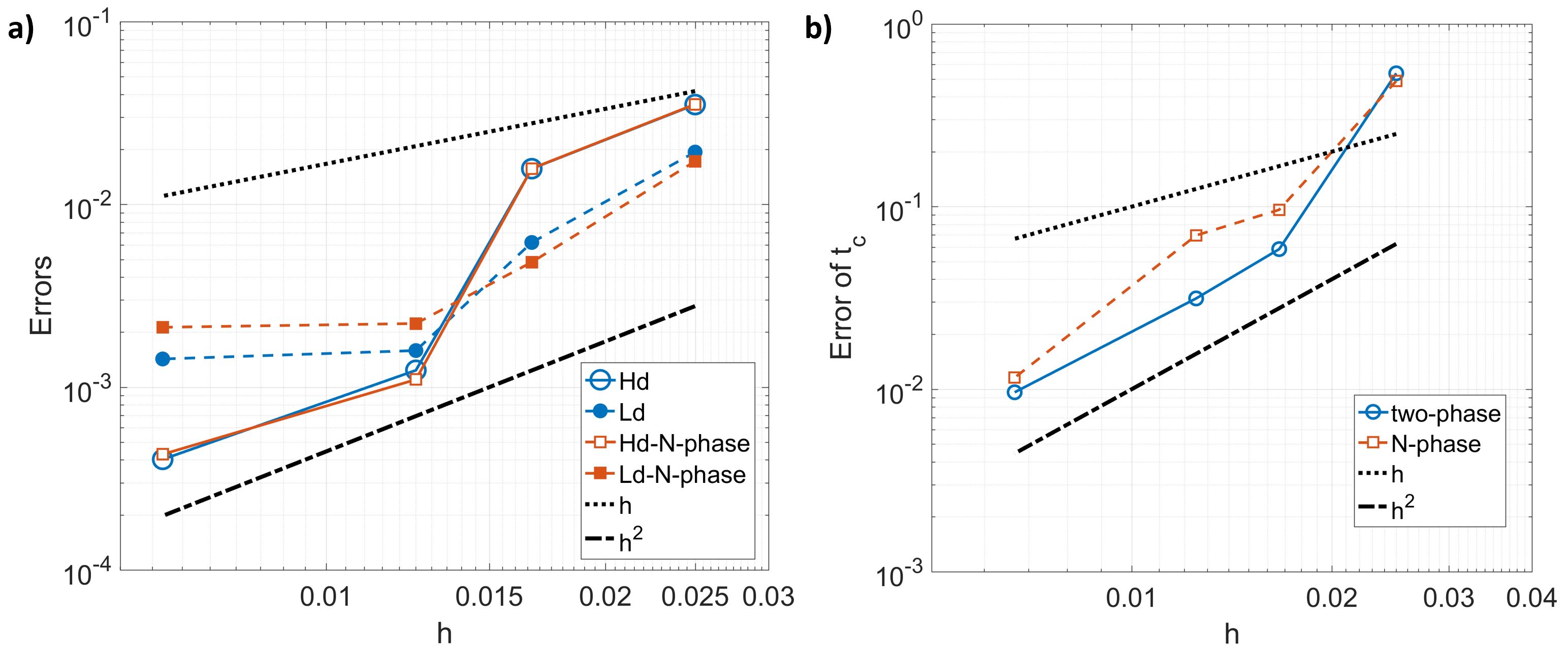}
	\caption{Convergence behaviors versus the grid size.
	a) Errors of the height and spreading length, and b) Errors of $t_c$ after which the kinetic energy ($E_K=\int_\Omega \frac{1}{2} \rho \mathbf{u}\cdot \mathbf{u} d\Omega$) is less than $10^{-5}$, from the cases of the water drop with $\theta=\theta_{1,2}=60^0$ and $|\mathbf{g}|=0$.
    \label{Fig EDrop Convergence}}
\end{figure}

We first neglect the gravity. Fig.\ref{Fig EDrop} shows the evolution of the drop with $\theta=60^0$ and $\theta=135^0$, along with the corresponding exact final solution from Eq.(\ref{Eq EDrop}). As expected, the drop starts with the semicircle shape, gradually approaches the final exact solution. The equilibrium shape agrees with the exact solution very well. 
We consider the zero contour of $\phi$ as the interface and measure the final height and spreading length of the drop for quantitative comparison. As shown in Fig.\ref{Fig EDrop HL} a), a good agreement with the exact solution from Eq.(\ref{Eq EDrop}) is obtained. Note that the height of the domain is changed to $0.5$ for $\theta=135^0$ and $\theta=150^0$, while the grid size remains unchanged.
To investigate the convergence with respect to grid refinement, the errors of the height and spreading length of the water drop versus the grid size is shown in Fig.\ref{Fig EDrop Convergence} a), using data from $\theta=60^0$. The observed convergence rate is between 1st- and 2nd-order. The saturated error of $L_d$ in Fig.\ref{Fig EDrop Convergence} a) can be caused by the evaluation of the spreading length. Since the distance from the bottom wall to the grid points nearest to it is a half of the grid size, the linear extrapolation is used to evaluate the interface location at the bottom wall, which introduces additional errors in the spreading length.
Moreover, interactions of the Phase-Field model and the contact angle boundary condition, both of which are non-linear, are also involved at the bottom wall. These complicated factors come into play, which makes the analyses of the saturation in Fig.4 a) very difficult.
Alternatively, we evaluate the time $t_c$ after which the kinetic energy ($E_K=\int_\Omega \frac{1}{2} \rho \mathbf{u}\cdot \mathbf{u} d\Omega$) is less than $10^{-5}$. Considering the finest-grid result as the reference value, a convergence rate near 2nd order is observed in Fig.\ref{Fig EDrop Convergence} b).
Additionally, we supplement, in Appendix~\ref{Appendix Manufactured}, a manufactured solution problem, which is commonly used to demonstrate the convergence. The convergence of the order parameter, velocity, and pressure with respect to the cell size is observed.

\begin{figure}[!t]
	\centering
	\includegraphics[scale=.4]{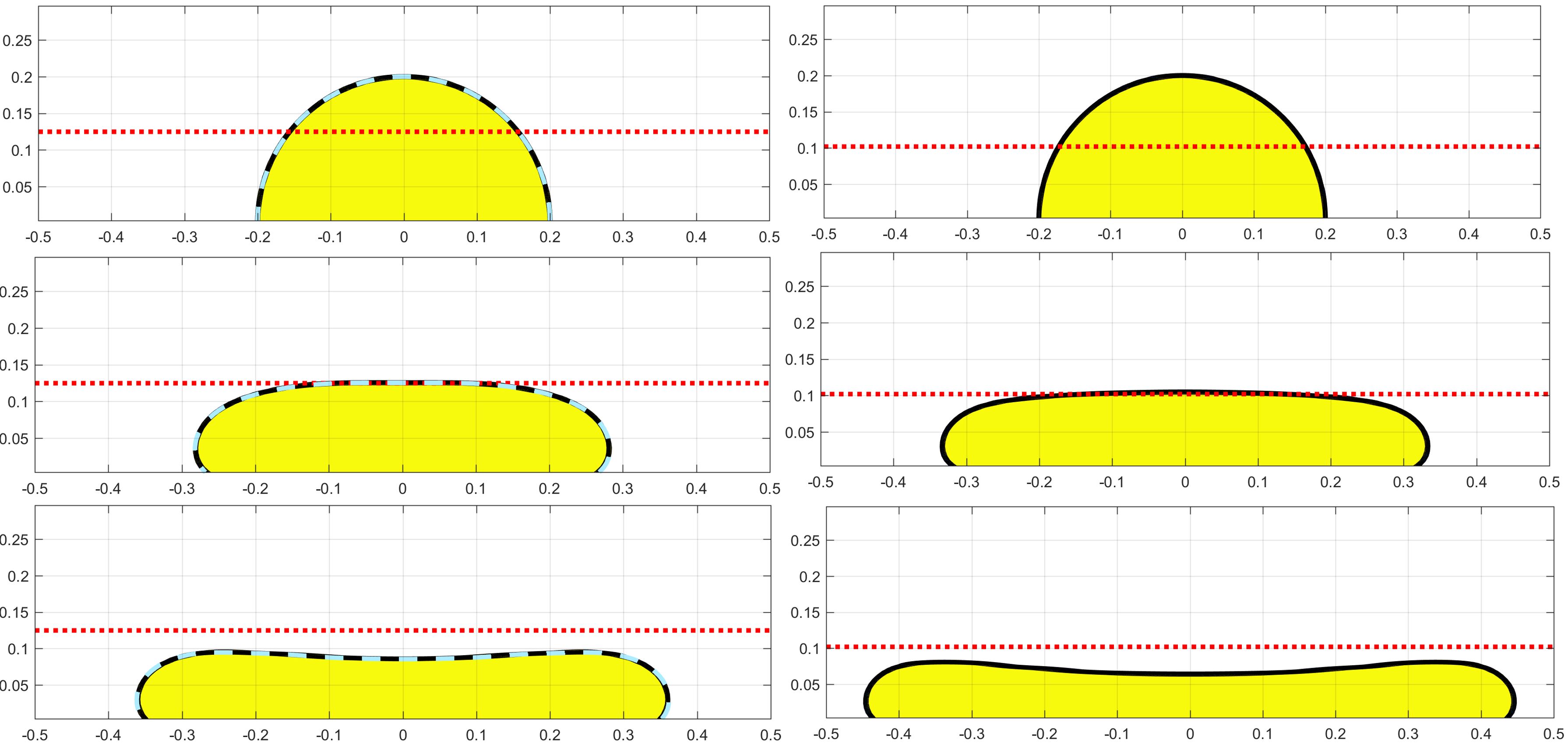}
	\includegraphics[scale=.4]{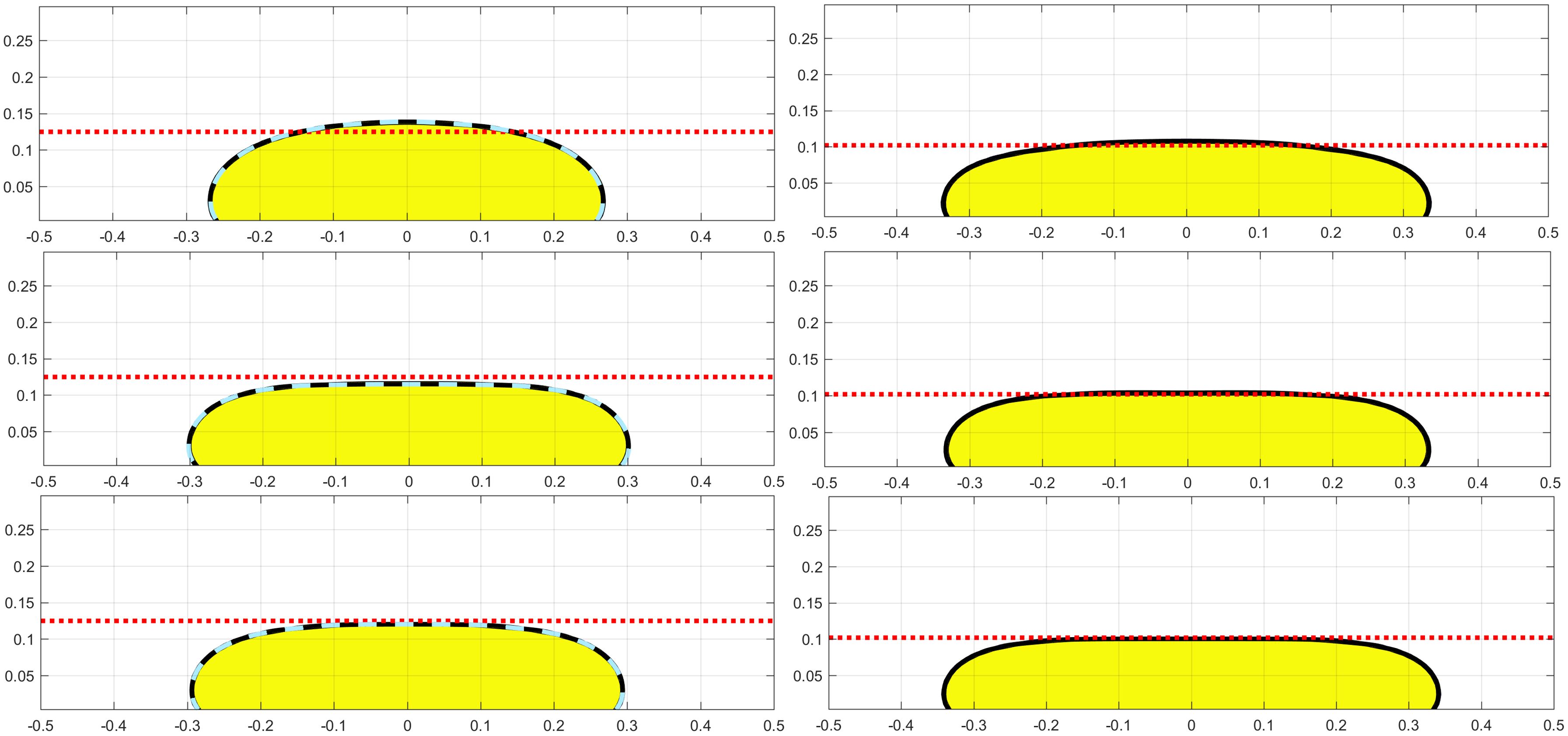}
	\caption{Evolution of the water drop using Eq.(\ref{Eq CAC two-phase}) with $\theta=135^0$. Yellow: water (Phase~1); White: air (Phase~2); Red dotted line: asymptotic solution from Eq.(\ref{Eq EDrop g}); Cyan dashed line: $N$-phase solution from Eq.(\ref{Eq CAC N-phase}) with $\phi_3|_{t=0} = -1$. Left column: $|\mathbf{g}|=10$; Right column: $|\mathbf{g}|=15$. From top to bottom, $t=0.0$, $t=0.2$, $t=0.4$, $t=1.0$, $t=1.4$, and $t=3.0$.
    \label{Fig EDrop g}}
\end{figure}

\begin{figure}[!t]
	\centering
	\includegraphics[scale=.29]{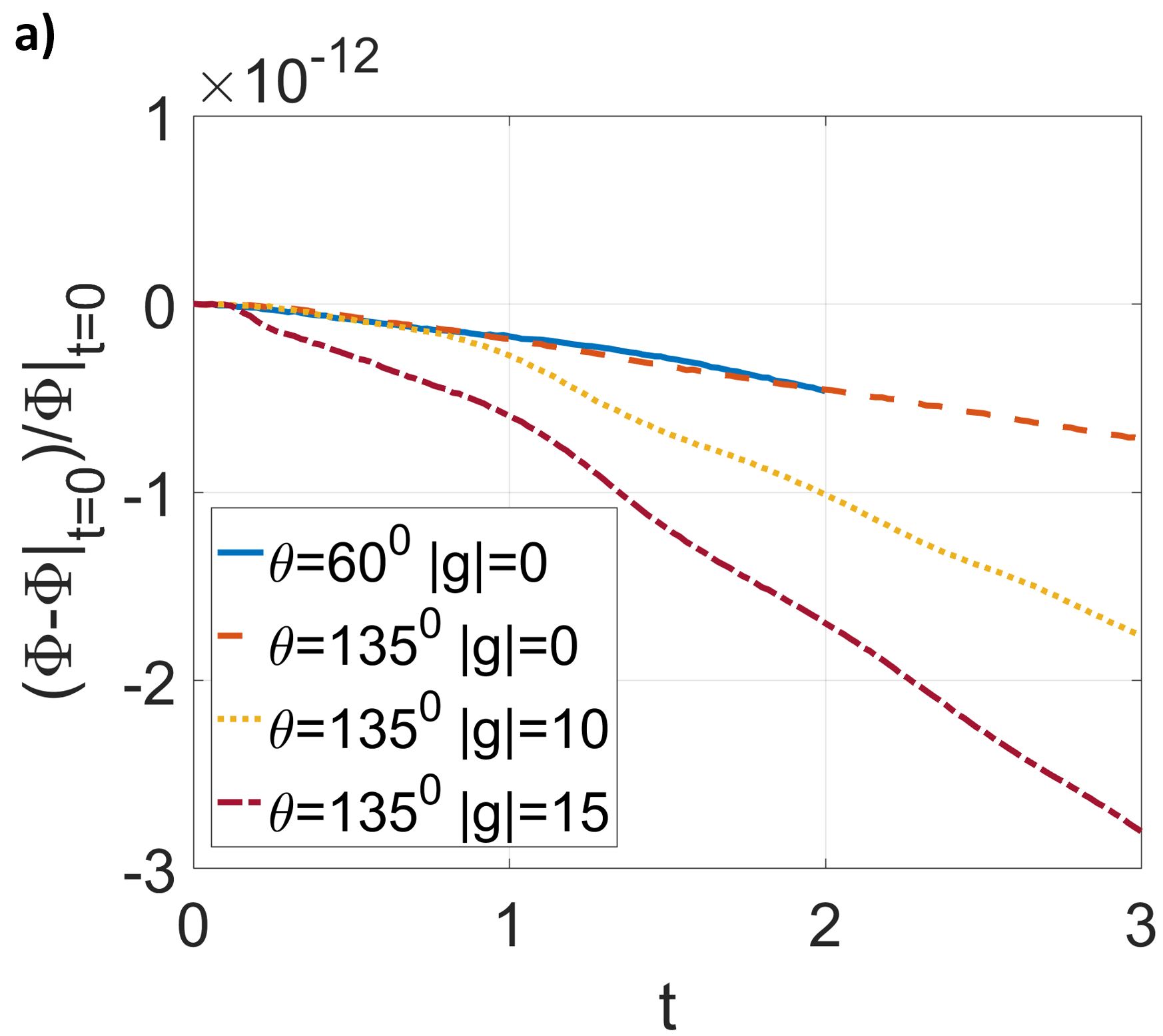}
	\includegraphics[scale=.29]{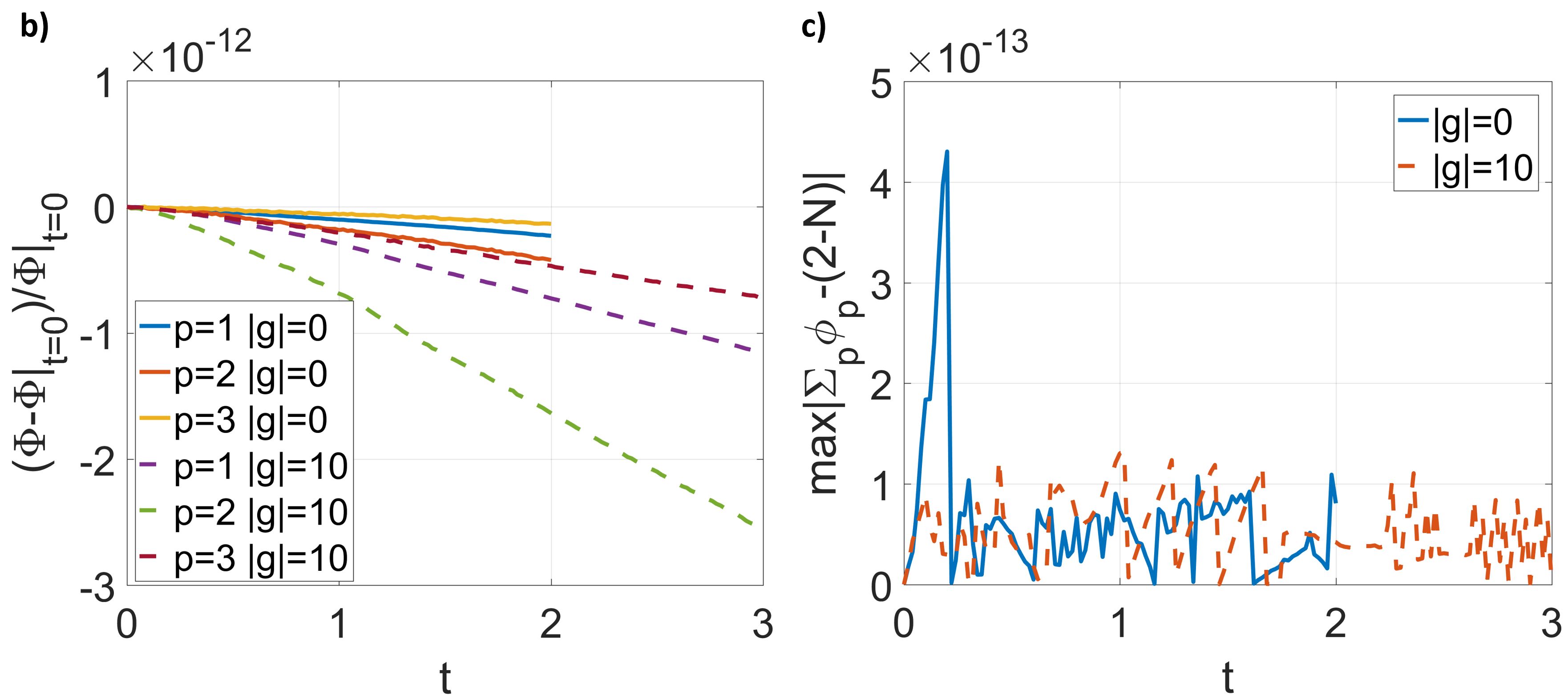}
	\caption{
	a) Relative changes of $\Phi$ ($=\int_{\Omega} \phi d\Omega$) from the two-phase solutions versus time. 
	b) Relative changes of $\Phi$ from the $N$-phase solutions versus time ($p$ is the phase index). 
	c) $\max|\sum_{p} \phi_p-(2-N)|$ from the $N$-phase solutions versus time.
    \label{Fig EDrop Mass}}
\end{figure}

Then, the effect of gravity is included, and the domain is changed to $[-0.6,0.6]\times[0,0.24]$ without changing the grid size.
Fig.\ref{Fig EDrop g} shows the evolution of the drop with $|\mathbf{g}|=10$ and $|\mathbf{g}|=15$, along with the prediction from Eq.(\ref{Eq EDrop g}). The contact angle is $\theta=135^0$. One can observe that the drop is flattened, having a pancake-like shape, when the gravity is added. The final height of the drop matches the asymptotic prediction. 
Fig.\ref{Fig EDrop HL} b) shows the final height of the drop versus the gravity, and our numerical prediction overall agrees well with both the exact solution Eq.(\ref{Eq EDrop}) without gravity and the asymptotic solution Eq.(\ref{Eq EDrop g}) with dominant gravity. 
Further, Fig.\ref{Fig EDrop Mass} a) demonstrates the mass conservation of the proposed formulation, where the relative changes of $\Phi$ ($\Phi=\int_\Omega\phi d\Omega$) of the four cases reported in Fig.\ref{Fig EDrop} and Fig.\ref{Fig EDrop g} are in the order of the round-off error.

\begin{figure}[!t]
	\centering
	\includegraphics[scale=.4]{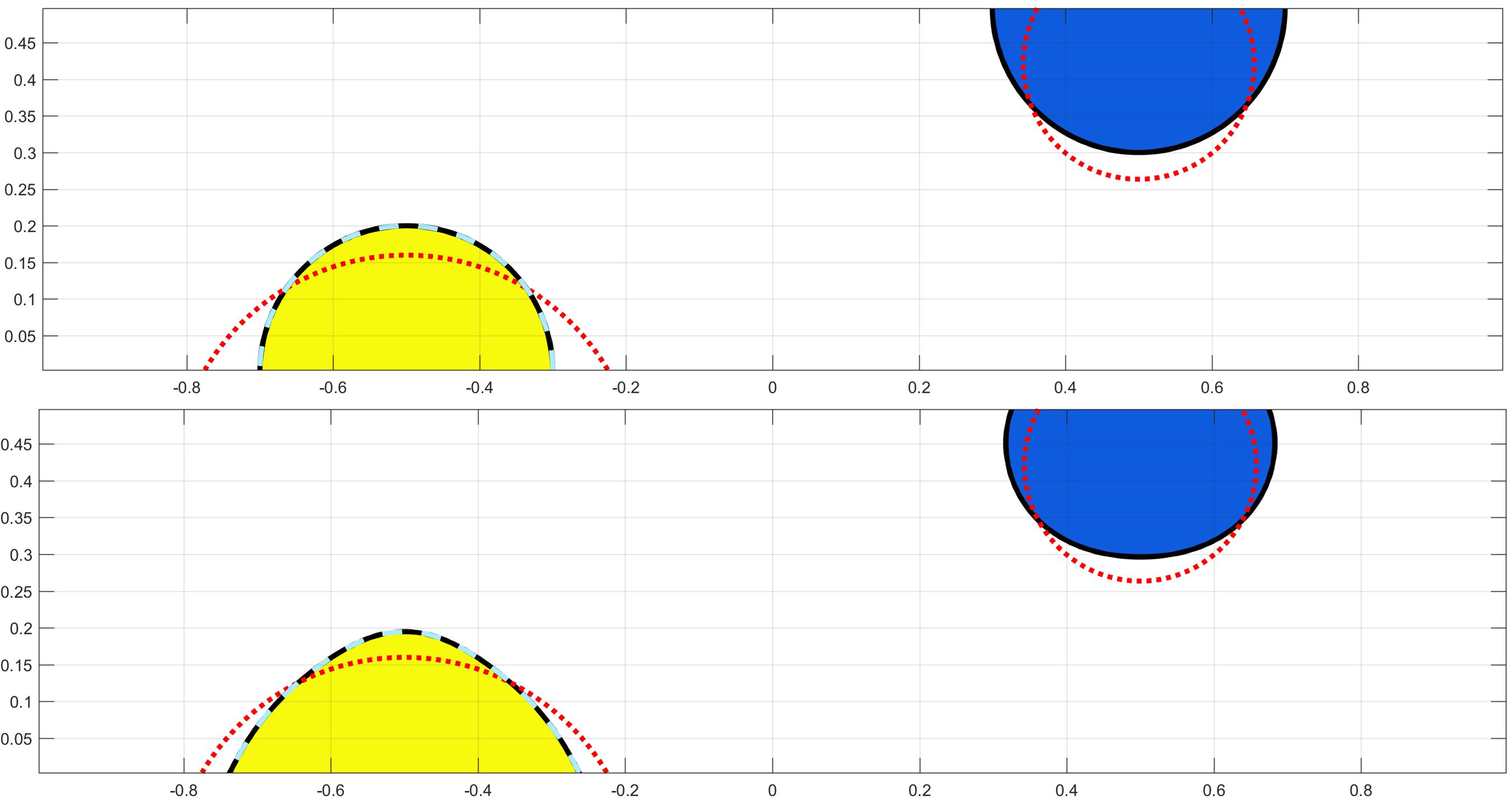}
	\includegraphics[scale=.4]{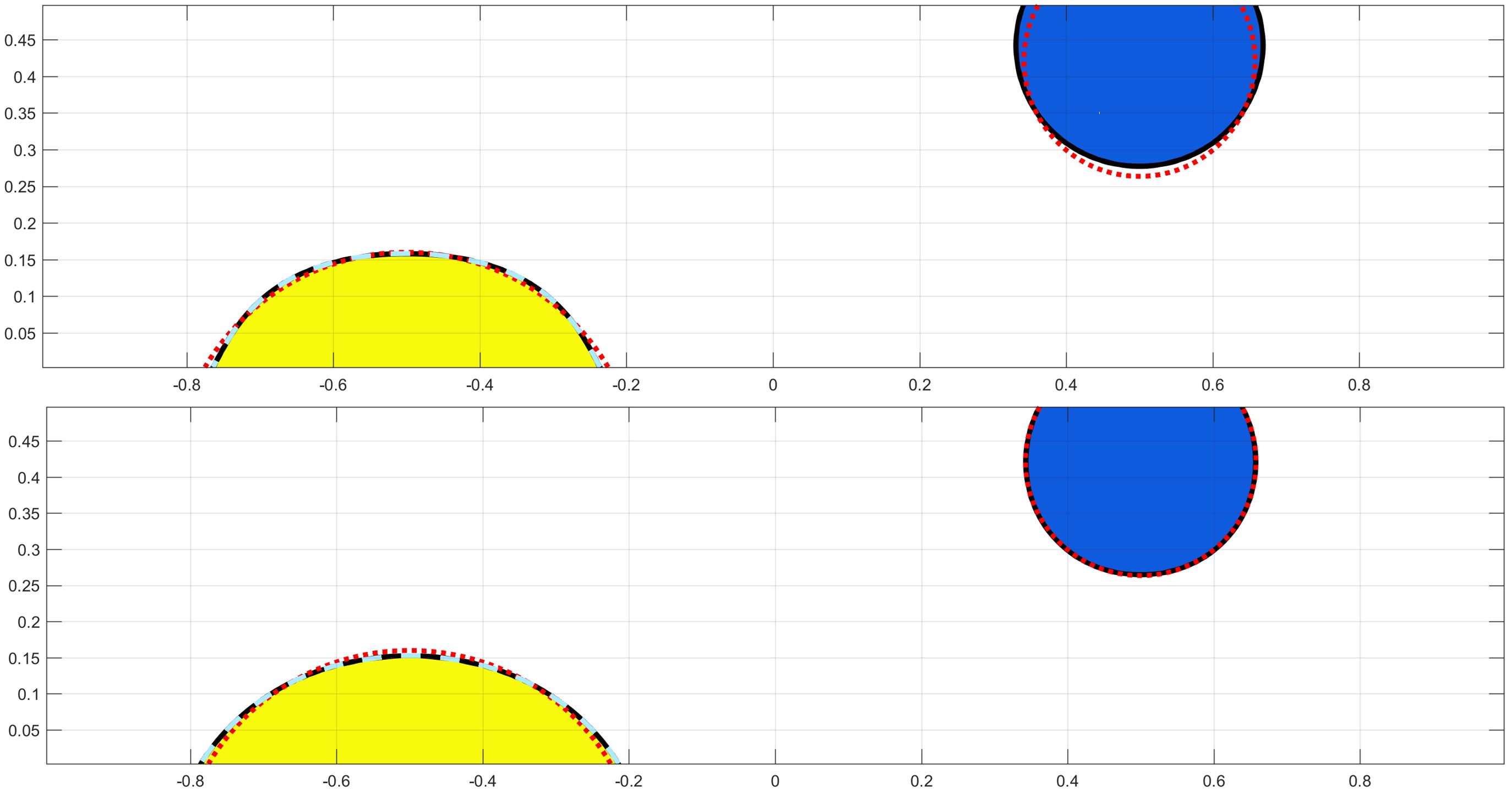}
	\includegraphics[scale=.4]{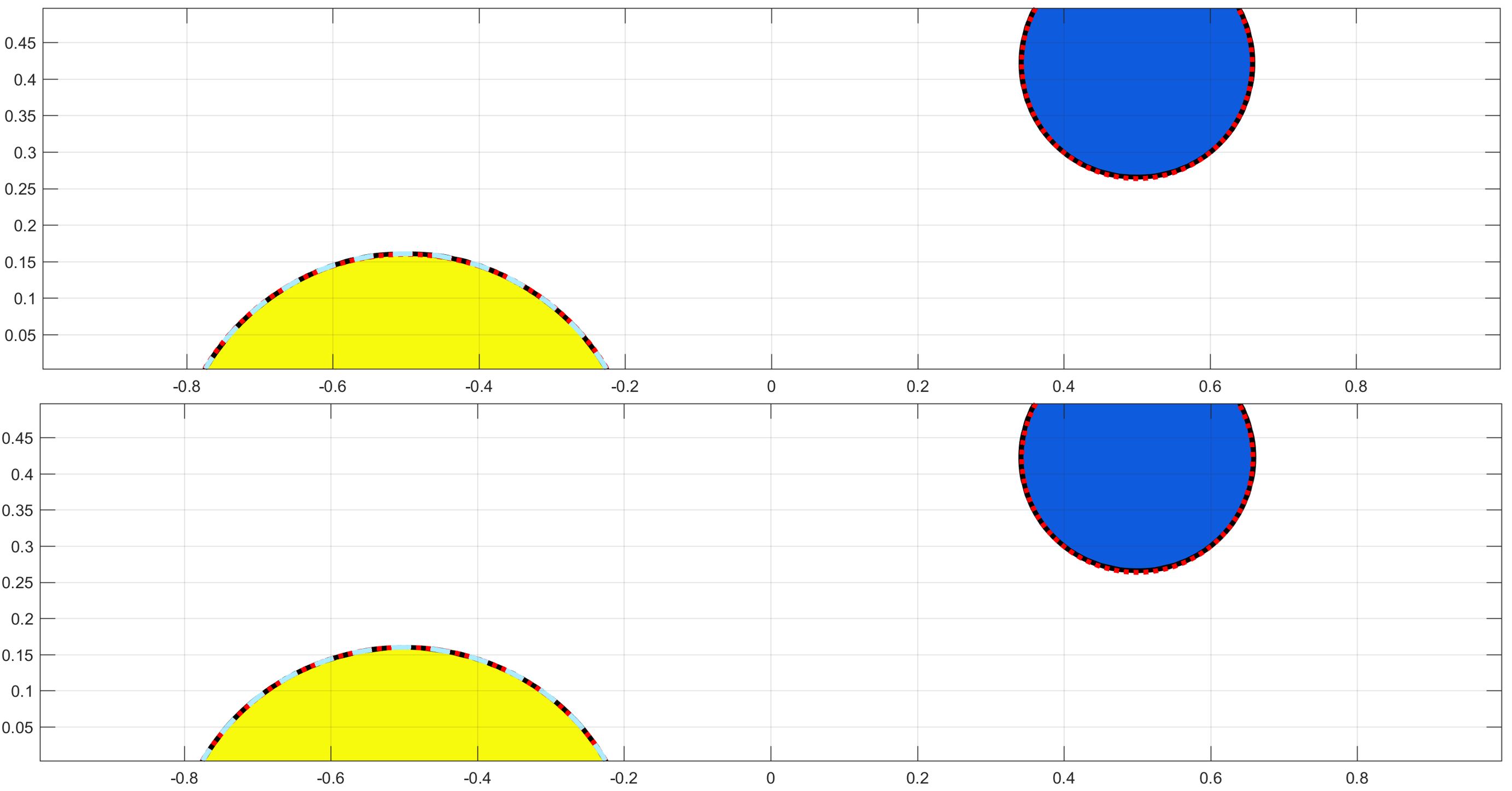}
	\caption{Evolution of the water and oil drops using Eq.(\ref{Eq CAC N-phase}) with $|\mathbf{g}|=0$, $\theta_{1,2}=60^0$, and $\theta_{3,2}=120^0$. Yellow: water (Phase~1); White: air (Phase~2); Blue: oil (Phase~3); Red dotted line: exact solution from Eq.(\ref{Eq EDrop}); Cyan dashed line: two-phase solution from Eq.(\ref{Eq CAC two-phase}) in the left column of Fig.\ref{Fig EDrop}. From top to bottom, $t=0.0$, $t=0.2$, $t=0.4$, $t=1.0$, $t=1.4$, and $t=2.0$.
    \label{Fig EDrop NPhase}}
\end{figure}

Next, we supplement results of the $N$-phase model Eq.(\ref{Eq CAC N-phase}). The oil phase (Phase~3) is introduced, whose material properties are listed in Table~\ref{Table ED} as well. The domain size becomes $[-1,1]\times[0,0.5]$ while the grid size is the same as the two-phase cases. The water drop right now is on the bottom wall with a contact angle $\theta_{1,2}=60^0$, while the oil drop is attached to the top wall with a contact angle $\theta_{3,2}=120^0$.
Evolution of the drops is shown in Fig.\ref{Fig EDrop NPhase}. Not only both the water and oil drops finally match the exact solution Eq.(\ref{Eq EDrop}) but also the shape of the water drop at different moments is indistinguishable from the two-phase solution in the left column of Fig.\ref{Fig EDrop}. The final heights and spreading lengths of the two drops are measured and plotted in Fig.\ref{Fig EDrop HL} a) as well, and good agreement is obtained with both the exact and two-phase solutions. 
Fig.\ref{Fig EDrop Convergence} also shows the convergence behavior of the $N$-phase model Eq.(\ref{Eq CAC N-phase}). The behavior is similar to the two-phase one Eq.(\ref{Eq CAC two-phase}) in Fig.\ref{Fig EDrop Convergence} a) in terms of the height and spreading length of the water drop. Convergence in between 1st and 2nd order is again observed in Fig.\ref{Fig EDrop Convergence} b) in terms of $t_c$ after which the kinetic energy ($E_K=\int_\Omega \frac{1}{2} \rho \mathbf{u}\cdot \mathbf{u} d\Omega$) is less than $10^{-5}$. The kinetic energy now includes the contribution from the oil drop.

\begin{figure}[!t]
	\centering
	\includegraphics[scale=.4]{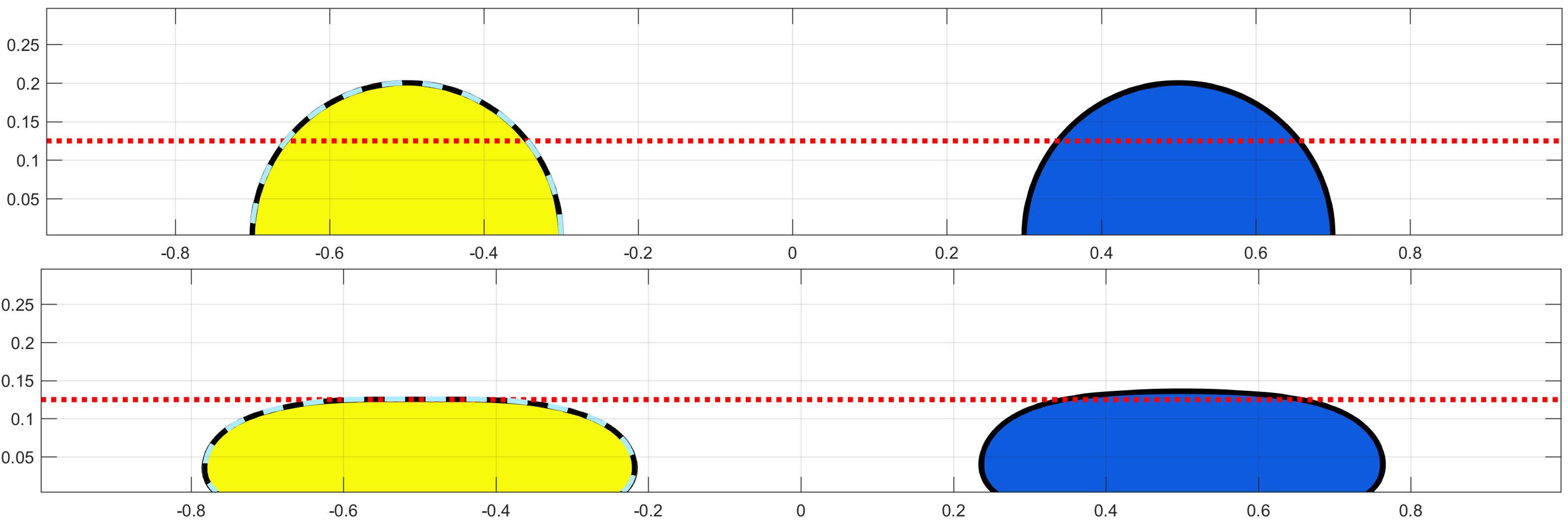}
	\includegraphics[scale=.4]{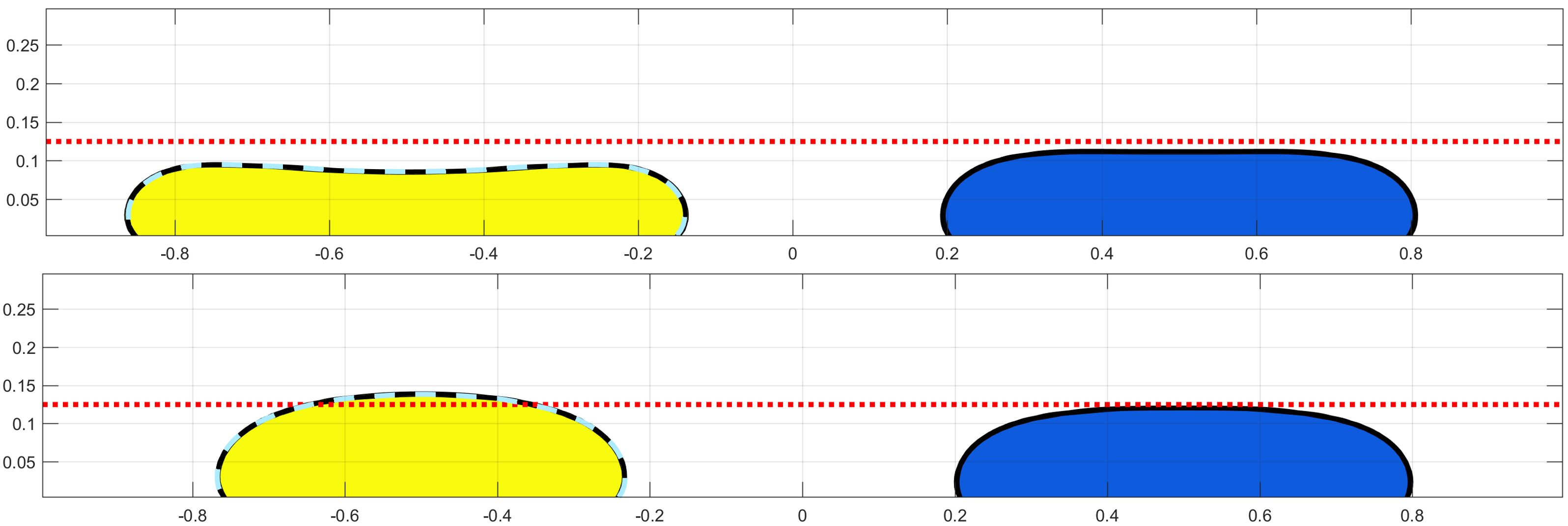}
	\includegraphics[scale=.4]{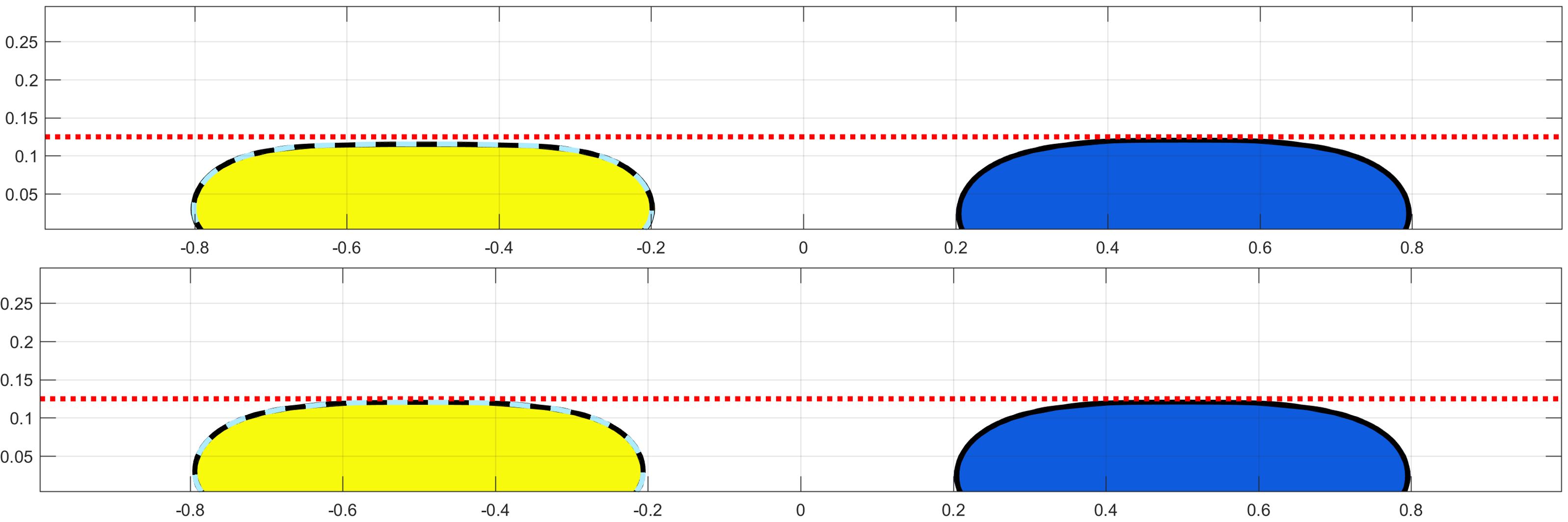}
	\caption{Evolution of the water and oil drops using Eq.(\ref{Eq CAC N-phase}) with $|\mathbf{g}|=10$, $\theta_{1,2}=135^0$, and $\theta_{3,2}=120^0$. Yellow: water (Phase~1); White: air (Phase~2); Blue: oil (Phase~3); Red dotted line: asymptotic solution from Eq.(\ref{Eq EDrop}); Cyan dashed line: two-phase solution from Eq.(\ref{Eq CAC two-phase}) in the left column of Fig.\ref{Fig EDrop g}. From top to bottom, $t=0.0$, $t=0.2$, $t=0.4$, $t=1.0$, $t=1.4$, and $t=3.0$.
    \label{Fig EDrop NPhase g}}
\end{figure}

Then, the gravity is added and the $N$-phase model Eq.(\ref{Eq CAC N-phase}) is again used. The surface tension between the oil and air is adjusted so that the final heights of both the water and oil drops, predicted from the asymptotic solution Eq.(\ref{Eq EDrop g}), are the same. The domain is $[-1,1]\times[0,0.3]$, and the magnitude of the gravity is $|\mathbf{g}|=10$ . The contact angle of the water drop on the bottom wall is $\theta_{1,2}=135^0$, while it is $\theta_{3,2}=120^0$ for the oil drop.
Evolution of the drops are shown in Fig.\ref{Fig EDrop NPhase g}. Both of the drops are compressed vertically and finally reach a similar height to the asymptotic prediction. Again, the water drop behaves identically to the two-phase solution in the left column of Fig.\ref{Fig EDrop g}. 
Fig.\ref{Fig EDrop HL} b) also includes the final heights of the two drops in this case, and they are in good agreement with both the asymptotic and two-phase solutions. We also investigate the mass conservation of the $N$-phase model, and the relative changes of $\Phi_p$, where $p$ is the index of the phases, are in the order of the round-off error, as shown in Fig.\ref{Fig EDrop Mass} b). In additional to that, the summation of the order parameters exactly satisfies Eq.(\ref{Eq Summation}), which is shown in Fig.\ref{Fig EDrop Mass} c).

\begin{figure}[!t]
	\centering
	\includegraphics[scale=.29]{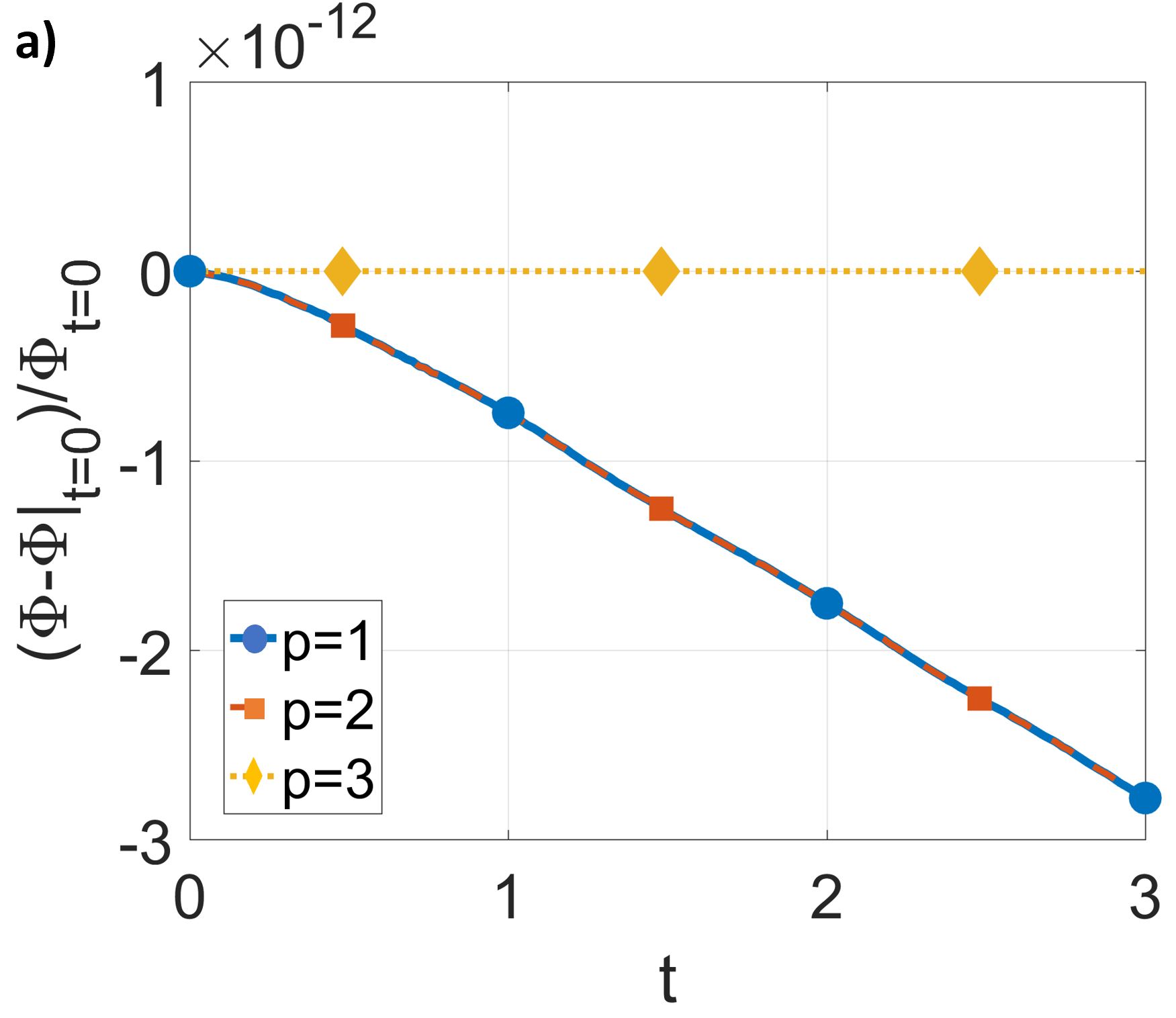}
	\includegraphics[scale=.29]{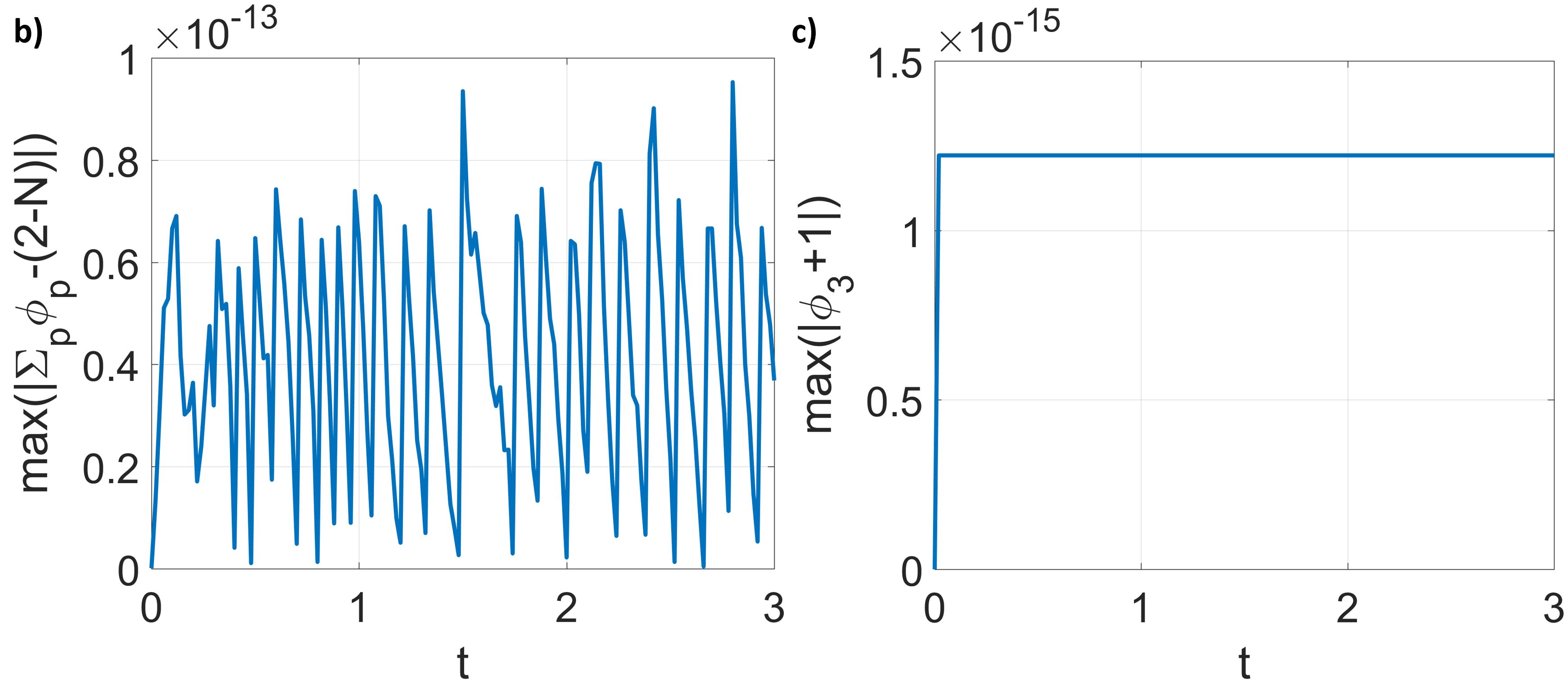}
	\caption{
	a) Relative changes of $\Phi$ ($=\int_{\Omega} \phi d\Omega$) from the $N$-phase solution versus time ($p$ is the phase index) with $\phi_3|_{t=0}=-1$.
	b) $\max|\sum_p \phi_p-(2-N)|$ from the $N$-phase solutions versus time with $\phi_3|_{t=0}=-1$.
	c) $\max|\phi_3+1|$ from the $N$-phase solutions versus time with $\phi_3|_{t=0}=-1$.
    \label{Fig EDrop NPhase RC}}
\end{figure}

The last property the $N$-phase model should satisfy is the \textit{consistency of reduction}. We repeat the $N$-phase case with $|\mathbf{g}|=10$ but only consider the left half of the domain, i.e., $-1 \leqslant x \leqslant 0$. Therefore, the oil drop disappears at the beginning, i.e., $\phi_3|_{t=0}=-1$. 
Evolution of the water drop from the $N$-phase model is shown in the left column of Fig.\ref{Fig EDrop g} as well using the cyan dashed line, and the difference from the two-phase solution is negligible. This also suggests that choosing $g_w(\phi)$ in Eq.(\ref{Eq F^w two-phase}) as a Sine or Hermite polynomial function has a negligible effect on the solution.
Fig.\ref{Fig EDrop NPhase RC} quantitatively validates that not only the mass conservation and the summation of the order parameters are exactly satisfied by the $N$-phase model Eq.(\ref{Eq CAC N-phase}) but also the \textit{consistency of reduction} since $\phi_3=-1$ is true at $\forall t>0$. 

\subsection{Couette flow}\label{Sec Couette}
\begin{figure}[!t]
	\centering
	\includegraphics[scale=.55]{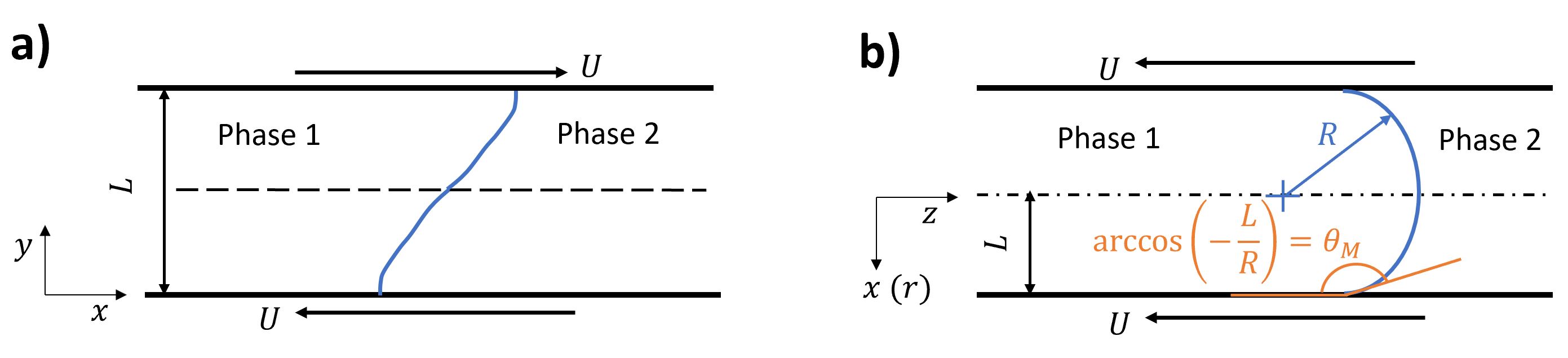}
	\caption{
	Schematics of
	a) the Couette flow, and 
	b) the Poiseuille flow.
    \label{Fig CP}}
\end{figure}

To demonstrate the proposed formulation in moving contact line problems, we consider the Couette flow in a reference frame moving with the contact line. The same problem was performed in \citep{Yueetal2010} to study the contact line dynamics of the Cahn-Hilliard model. 
Following the setup in \citep{Yueetal2010}, we consider a channel having a height $L$ and a length $4L$, see the schematic in Fig.\ref{Fig CP} a).
The top wall of the channel is moving horizontally with a velocity $U$, while the bottom wall is moving oppositely with the same speed. The steady state solution of the problem corresponds to a contact line moving at a constant speed $U$ with respect to a fixed bottom wall. 
Like those in \citep{Yueetal2010}, the capillary number is $Ca=\frac{\mu_1 U}{\sigma}=0.02$, the viscosity ratio is $\mu^*=\frac{\mu_2}{\mu_1}=1$, and the inertia is neglected. Another dimensionless number related to the mobility in the conservative Allen-Cahn model is $S_{CAC}=\mu_1 M=0.1$. 
The channel is discretized by $400 \times 100$ grid cells, and the time step size is $\frac{U \Delta t}{L}=1 \times 10^{-4}$. The homogeneous Neumann boundary condition is used at the left and right boundaries, while the no-slip boundary condition is used at the top and bottom. The contact angle at the top wall is either $\theta=90^0$ or $\theta=120^0$, and the same at the bottom. The interface is initially vertical and the initial velocity is identical to the steady state Couette flow without interfaces, i.e., $\mathbf{u}_C=\left\{ \frac{2U}{L}\left( y-\frac{L}{2} \right),  0 \right\}$.

\begin{figure}[!t]
	\centering
	\includegraphics[scale=.45]{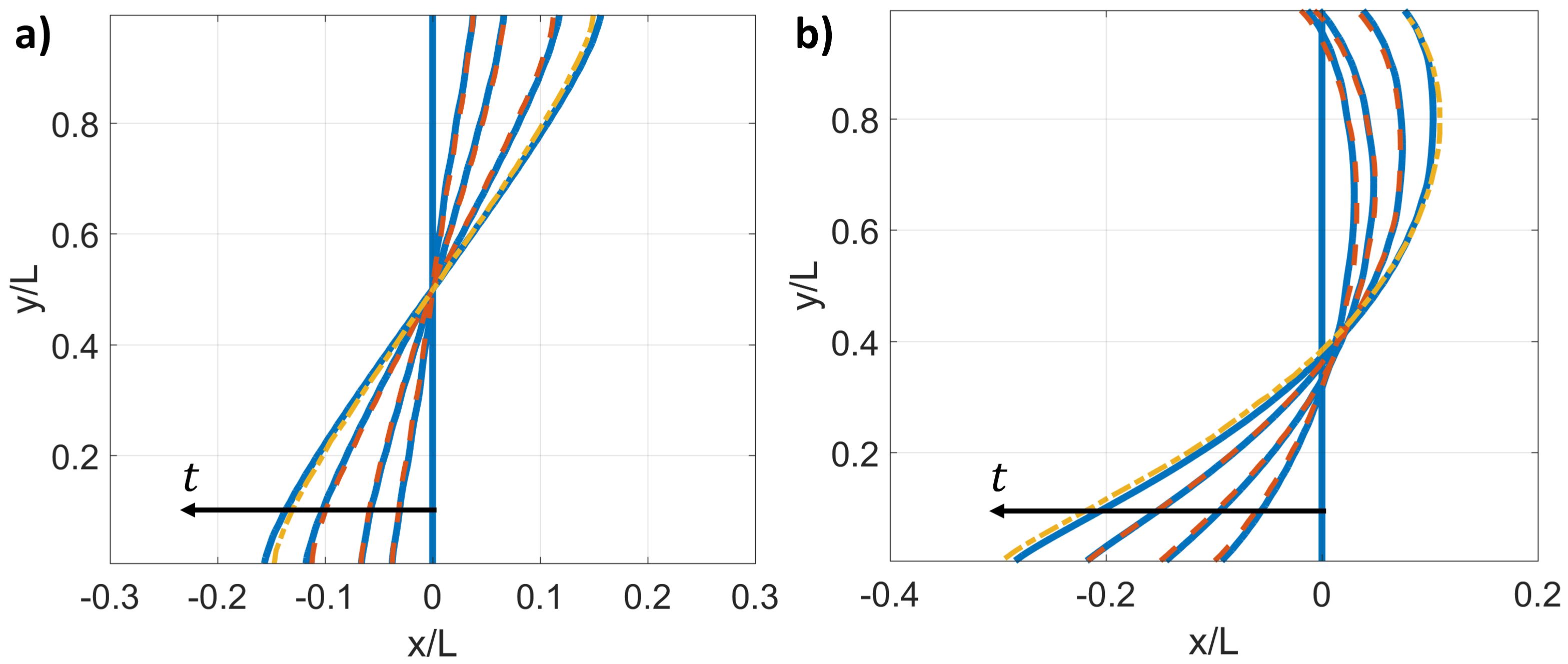}
	\caption{
	Results of the Couette flow.
	a) $\theta=90^0$.
	b) $\theta=120^0$.
	Blue solid lines: Interface at $\frac{Ut}{L}=0.00$, $0.05$, $0.10$, $0.25$, and steady state, from the proposed formulation with the conservative Allen-Cahn model.
	Yellow dash-dotted line: Interface at steady state from \citep{Yueetal2010} using the Cahn-Hilliard model.
	Red dashed lines: Interface at $\frac{U t}{L}=0.05$, $0.10$, and $0.25$, from the consistent and conservative Phase-Field method \citep{Huangetal2020} using the same Cahn-Hilliard model as the one in \citep{Yueetal2010}.
    \label{Fig Couette}}
\end{figure}

As shown in Fig.\ref{Fig Couette}, the steady state results from the proposed formulation agree very will with those reported in \citep{Yueetal2010} using the Cahn-Hilliard model, no matter the contact angle at the top and bottom walls is $90^0$ or $120^0$.
Since only the steady state results of the problem are provided in \citep{Yueetal2010}, we supplement the transitional results from the consistent and conservative Phase-Field method \citep{Huangetal2020} which uses the same Cahn-Hilliard model as the one in \citep{Yueetal2010}. The transitional results correspond to the acceleration of the moving contact line before it reaches the final speed $U$. Again in Fig.\ref{Fig Couette}, not only the steady state results but also the transitional ones from the proposed formulation match those from the Cahn-Hilliard model very well.

\subsection{Poiseuille flow}\label{Sec Poiseuille}
The Poiseuille flow in a reference frame moving with the contact line, reported in \citep{Yueetal2010}, is also considered, see the schematic in Fig.\ref{Fig CP} b). It corresponds to a fluid displacing another fluid in a capillary tube. 
The domain is an axisymmetric capillary tube with a radius $L$ and a length $6L$. The tube wall is moving backward with a velocity $U$, which is the average velocity of the fully-developed Poiseuille flow without interfaces, i.e., $\mathbf{u}_{P}=\{0,\frac{2U}{L^2}(L^2-x^2)\}$.
The dimensionless numbers $Ca$, $\mu^*$, and $S_{CAC}=0.1$ are defined identically to those in the Couette flow Section~\ref{Sec Couette}.
The capillary tube is discretized by $100 \times 600$ grid cells, and the time step size is $\frac{U \Delta t}{L}=1\times 10^{-4}$. The no-slip boundary condition is used at the tube wall, and the contact angle there is $\theta$. The Dirichlet boundary condition is used at the inlet and outlet with a velocity $\mathbf{u}_{io}=\{0,\frac{2U}{L^2}(L^2-x^2)-U\}$. The interface is initially vertical, and the initial velocity is identical to $\mathbf{u}_{io}$.

\begin{figure}[!t]
	\centering
	\includegraphics[scale=.45]{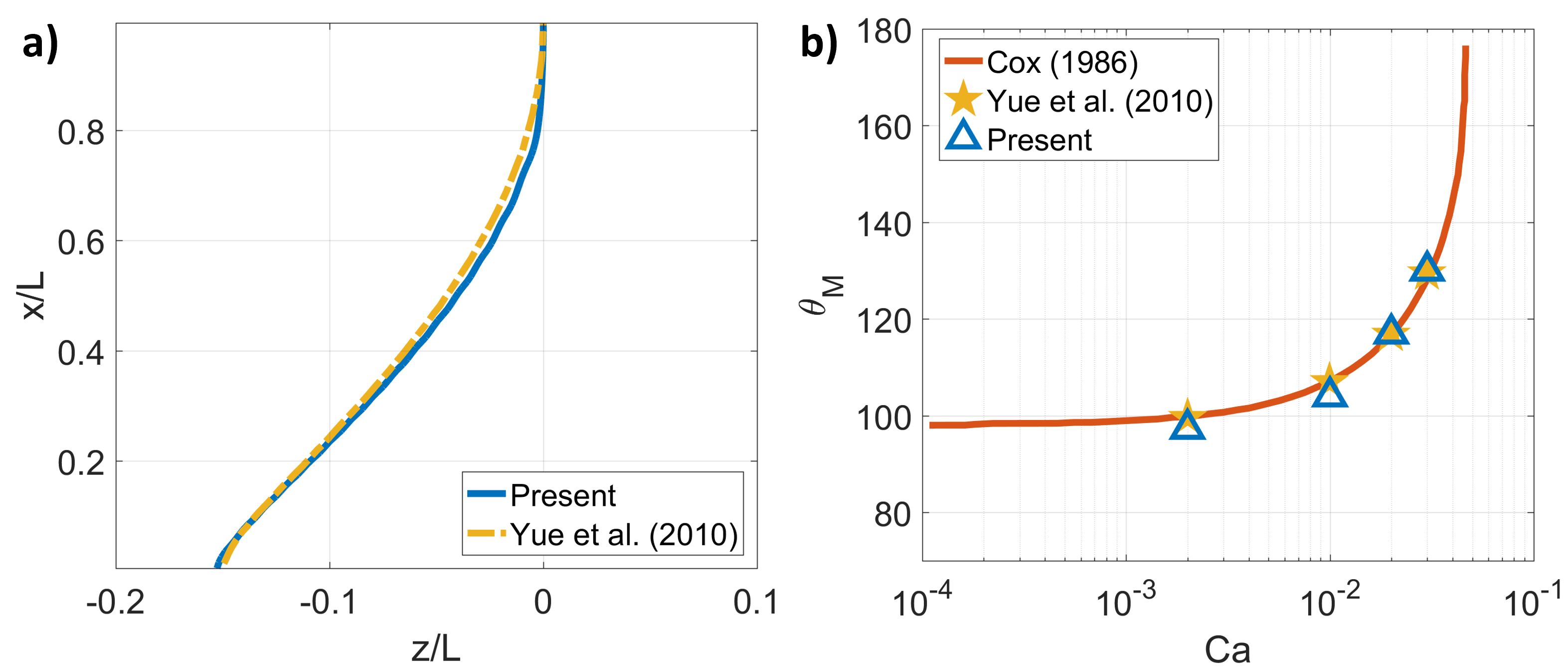}
	\caption{
	Results of the Poiseuille flow.
	a) Steady state interface with $Ca=0.02$, $\mu^*=1$, and $\theta=90^0$. The coordinate system has been adapted to the one in Yue et al. (2010) \citep{Yueetal2010} such that $x/L=0$ is at the tube wall and $x/L=1$ at the axis of symmetry.
	b) Apparent contact angle $\theta_M$ versus capillary number $Ca$, with $\mu^*=0.9$ and $\theta=98^0$, from the proposed formulation with the conservative Allen-Cahn model, Yue et al. (2010) \citep{Yueetal2010} using the Cahn-Hilliard model, and the theory of Cox (1986) \citep{Cox1986}.
    \label{Fig Poiseuille}}
\end{figure}

As shown in Fig.\ref{Fig Poiseuille} a), the steady state interface from the proposed formulation agrees well with the one in \citep{Yueetal2010} using the Cahn-Hilliard model, with $Ca=0.02$, $\mu^*=1$, and $\theta=90^0$. Notice that the coordinate system shown in Fig.\ref{Fig Poiseuille} a) has been adapted to the one in \citep{Yueetal2010} such that $x/L=0$ is at the tube wall and $x/L=1$ at the axis of symmetry. 
Further, we consider the apparent contact angle $\theta_M$ versus the capillary number $Ca$ with $\mu^*=0.9$ and $\theta=98^0$. Like those in \citep{Yueetal2010} and the references therein, the apparent contact angle is obtained from $\theta_M=\arccos(-L/R)$, under the assumption that the interface is a spherical cap with a radius $R$, as illustrated in Fig.\ref{Fig CP} b). In Fig.\ref{Fig Poiseuille} b), results from the proposed formulation agree with both from the Cahn-Hilliard model \citep{Yueetal2010} and the celebrated theory of Cox \citep{Cox1986}.
In \citep{Yueetal2010}, the adaptive mesh refinement (AMR) has been implemented. With a smaller $Ca$, the interface is less deformed. Then, AMR is able to locally increase the resolution near the interface, which is favorable in improving accuracy. This explains the minor discrepancy in Fig.\ref{Fig Poiseuille} b) when $Ca \leqslant 0.01$.

In the present section and Section~\ref{Sec Couette}, the Cahn-Hillard results from \citep{Yueetal2010} used the dimensionless diffusion length $S_{CH}=\sqrt{\mu_1 M_{CH}}/L=0.01$, where $M_{CH}$ is the mobility in the Cahn-Hilliard model. Moreover, the authors of \citep{Yueetal2010} related the dimensionless slip length in Cox's formula to $2.5S_{CH}$, which is used in Fig.\ref{Fig Poiseuille} b).
We discovery that quantitative matches between the Cahn-Hilliard model and the proposed formulation with the conservative Allen-Cahn model in contact-line dynamics are obtained if we correspond $S_{CH}=0.01$ of the Cahn-Hilliard model to $S_{CAC}=0.1$ of the conservative Allen-Cahn model. To further verify, explain, and generalize such a correspondence between the two Phase-Field models in contact-line dynamics will be interesting, but needs non-trivial additional works and is outside the scope of the present study. 

\subsection{Spreading drop}\label{Sec SD}
To further demonstrate the proposed formulation in dynamical problems with inertia, we consider spreading of a drop on a solid substrate, reported in \citep{Renardyetal2001} using the Volume-of-Fluid (VoF) method. 
The setup in \citep{Renardyetal2001} is followed: 
The computational domain is $[0,1]\times[0,1]$, whose left and right boundaries are periodic, and the top and bottom are no-slip. A circular drop of Phase~1 with a radius of $R_0=0.2$ is centered at $(0.5,0.85)$ and surrounded by Phase~2. The contact angle at the top wall is set to be $\theta=70.53^0$. The two phases have a matched density $\rho=0.1$ and viscosity $\mu=0.001$, and the surface tension between them is $\sigma=0.03$. The mobility follows $M \lambda=8 \times 10^{-3}$. The domain is discretized by $128 \times 128$ grid cells, and the time step size is $\Delta t=1 \times 10^{-4}$. 
Using the inertia-capillary time scale $T=\sqrt{ \frac{\rho R_0^3}{\sigma}}$ to determine the velocity scale, i.e., $U=\frac{R_0}{T}$, the Reynolds number and capillary number in this problem are $Re=\frac{\rho U R_0}{\mu}=24.5$ and $Ca=\frac{\mu U}{\sigma}=0.04$, respectively.

\begin{figure}[!t]
	\centering
	\includegraphics[scale=.4]{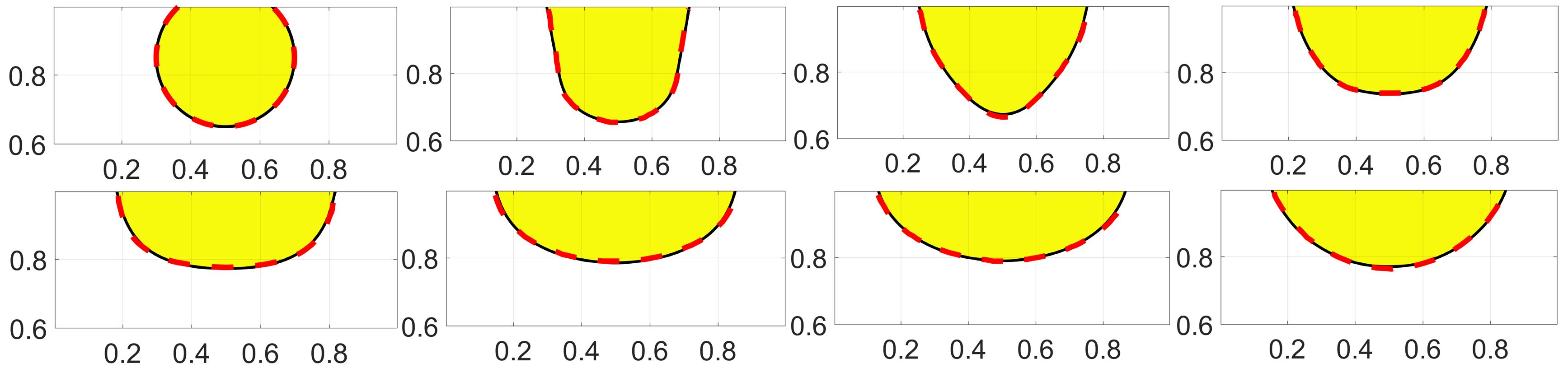}
	\caption{
	Results of the spreading drop. From left to right, top to bottom, $t=0.0$, $0.1$, $0.2$, $0.3$, $0.4$, $0.5$, $0.6$, and $1.0$. 
	Results from the proposed formulation with the conservative Allen-Cahn model label Phase~1 in yellow and Phase~2 in white. 
	Results from \citep{Renardyetal2001} using the Volume-of-Fluid (VoF) method are denoted by red dashed lines.
    \label{Fig SD}}
\end{figure}

The dynamical process of the problem is shown in Fig.\ref{Fig SD}. The initial configuration of the drop and the top wall intersect with an angle different from the assigned contact angle. Such a difference drives the drop to spread and deform, and finally reach the equilibrium shape. For comparison, the results in \citep{Renardyetal2001} using the Volume-of-Fluid method are also plotted in Fig.\ref{Fig SD}. The entire process predicted by the proposed formulation agrees very well with those in \citep{Renardyetal2001}, which demonstrates the effectiveness of the proposed formulation.

\subsection{Axisymmetric spreading drop}\label{Sec ASD}
Here, we present spreading of an axisymmetric drop on a solid substrate, mimicking the experiment in \citep{Eddietal2013}. 
The liquid drop is a mixture of Glycerine (79\%) and water (21\%), and its material properties are listed in Table \ref{Table ASD}, along with those of the surrounding air. The contact angle at the substrate is $\theta=85^0$. Initially, the spherical drop of a radius $R_0=0.5 \mathrm{mm}$ is released at $z=R_0$, in contact with the solid substrate at $z=0$.
\begin{table*}[!t]
\caption{Material properties in the axisymmetric spreading drop}
    \centering
    \includegraphics[scale=.3]{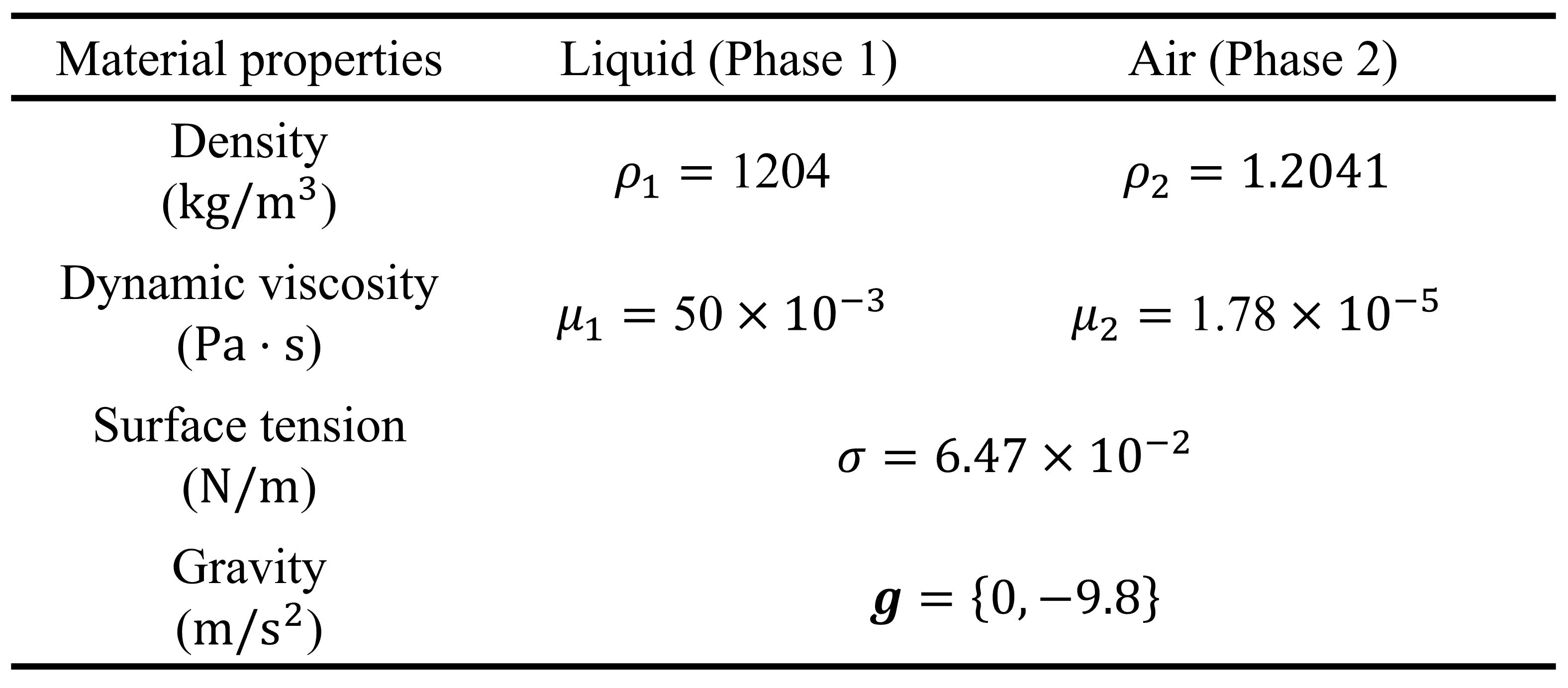}
    \label{Table ASD}
\end{table*}

The specific computational setup is as follows. We use a length scale $L_\mathrm{ref}=2R_0=1\mathrm{mm}$, density scale $\rho_\mathrm{ref}=1\mathrm{kg/m^3}$, and acceleration scale $a_\mathrm{ref}=1\mathrm{m/s^2}$ to non-dimensionalize the governing equations. As a result, the computational domain is $[0,1]\times[0,1.25]$. The top and right boundaries are the outflow boundary, the left is the axis of symmetric, and the bottom is the no-slip wall. Like in Section~\ref{Sec SD}, the mobility follows $M \lambda=8 \times 10^{-3}$. The domain is discretized by $128 \times 160$ grid cells, and the time step size is $\Delta t=5 \times 10^{-5}$.

\begin{figure}[!t]
	\centering
	\includegraphics[scale=.5]{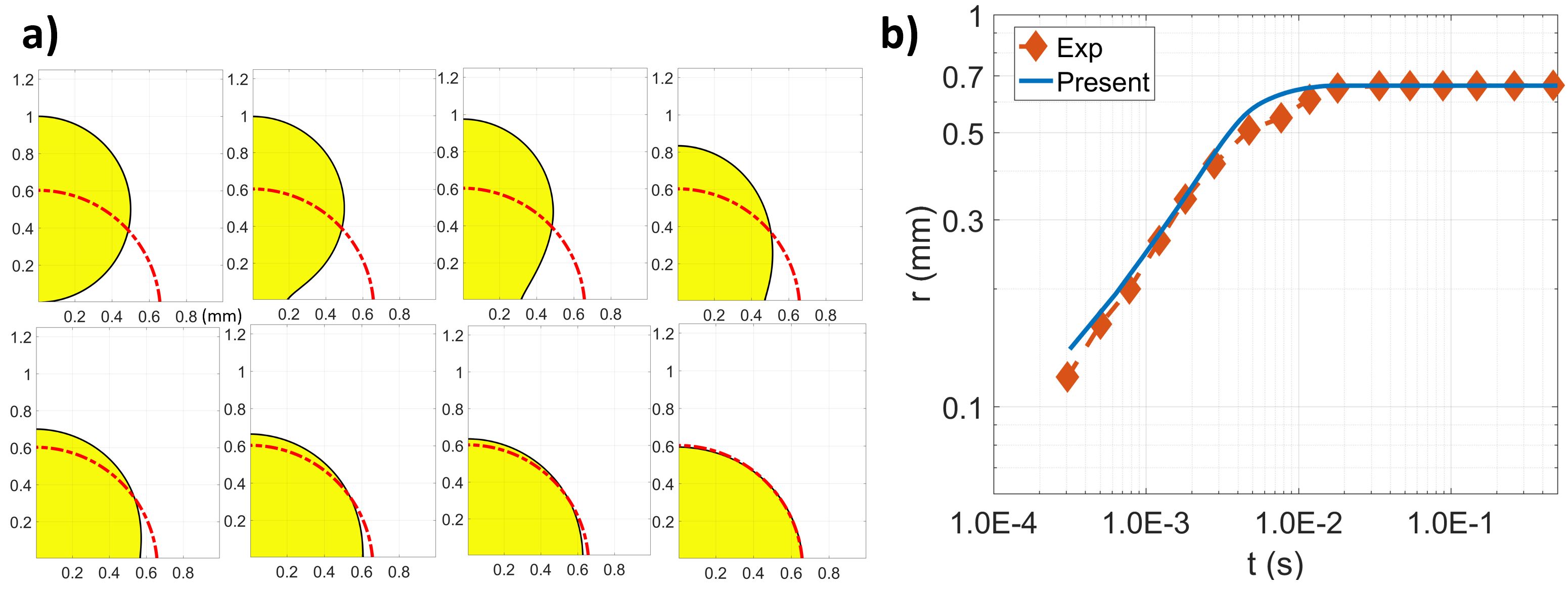}
	\caption{
	Results of the axisymmetric spreading drop. 
	a) Evolution of the drop. The $x$ ($r$) axis is horizontal, the $z$ axis is vertical, and the unit is millimeter (mm). From left to right, top to bottom, $t=0\mathrm{s}$, $6.32 \times 10^{-4}\mathrm{s}$, $1.58 \times 10^{-3}\mathrm{s}$, $3.16 \times 10^{-3}\mathrm{s}$, $4.74 \times 10^{-3}\mathrm{s}$, $6.32 \times 10^{-3}\mathrm{s}$, $7.91 \times 10^{-3} \mathrm{s}$, and steady state. Yellow: Liquid. White: Air. Red dashdotted line: exact steady state solution with a zero gravity.
	b) Radius $r$ (mm) of the wetted area versus time $t$ (s). The experimental data are from \citep{Eddietal2013}.
    \label{Fig ASD}}
\end{figure}

Fig.\ref{Fig ASD} shows the results in their dimensional forms. 
The evolution of the drop is shown in Fig.\ref{Fig ASD} a), along with the exact steady state solution with a zero gravity. The exact solution is obtained by matching the volume of a spherical cap to the volume of the liquid drop, i.e., $\frac{4}{3} \pi R_0^3 = \frac{1}{3} \pi h^2 (3R-h)$ where $R$ and $h=R(1-\cos(\theta))$ are the radius and height of the spherical cap, respectively. From Table \ref{Table ASD}, one can compute the Eötvös (or Bond) number $Eo=\frac{\rho_1 |\mathbf{g}| R_0^2}{\sigma}=0.0456$, which represents the ratio of the gravity force to the surface tension. With such a small $Eo$, the drop should finally be very close to the exact solution with a zero gravity, which is the case shown in Fig.\ref{Fig ASD} a).
Fig.\ref{Fig ASD} b) shows the radius of the wetted area versus time. The present results are compared to the experimental data in \citep{Eddietal2013}, and a good agreement is achieved.
The major dynamics are well captured but one may notice that the present results report a smoother transition than the experimental one in the approach of the drop to the stationary state. The difference is in an acceptable range, and can be caused by the experimental uncertainties, such as the roughness of the substrate.
The mobility in Sections~\ref{Sec SD} and \ref{Sec ASD} is obtained by trial and error, and the general trend we observed is that the contact line moves faster as the mobility increases. We expect future studies on theoretical analyses of the contact line dynamics of the conservative Allen-Cahn or the 2nd-order Phase-Field models, like \citep{Jacqmin2000,Qianetal2006,Yueetal2010,YueFeng2011,Xuetal2018} for the Cahn-Hilliard model, will provide more insights.

So far, we have demonstrated the proposed formulation with the conservative Allen-Cahn model in various equilibrium and dynamical problems quantitatively. The remaining cases will show some potential applications and most of those results are reported qualitatively.

\subsection{Bouncing drop}\label{Sec Bouncing drop}
Here, we consider a falling water drop bouncing back after it contacts the bottom wall, using the two-phase model Eq.(\ref{Eq CAC two-phase}). 
The circular drop, surrounded by the air, has a radius $R_0=1.25 \mathrm{mm}$, and is released above the bottom wall. The distance from the drop center to the bottom wall is $H_0=4R_0=5 \mathrm{mm}$. The material properties of the fluid phases considered are listed in Table~\ref{Table BD}.
\begin{table*}[!t]
\caption{Material properties in the bouncing drop}
    \centering
    \includegraphics[scale=.3]{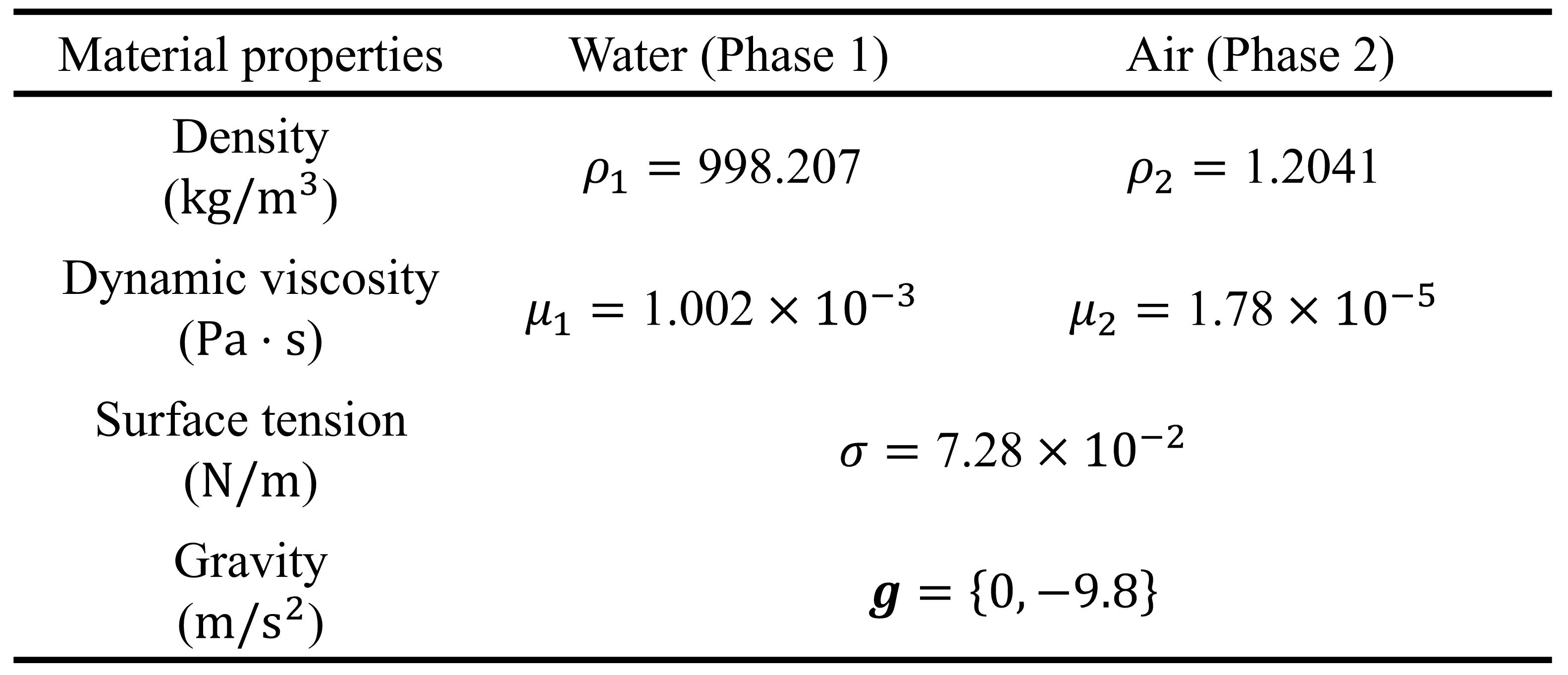}
    \label{Table BD}
\end{table*}
In the computation, non-dimensionalization is performed to the governing equations, based on $\rho_2$ (the air density), $H_0$ (the release height), and $a_\mathrm{ref}=1\mathrm{m/s^2}$ as the density, length, and acceleration scales, respectively. The acceleration scale is chosen for convenience so that the dimensionless value of the gravity is the same as the dimensional one. After the non-dimensionalization, the computational domain is $[-0.5,0.5]\times[0,1.5]$, and the circular drop is initially at $(0,1)$. The length of the domain is $4$ times the initial radius of the drop to prevent the drop from touching the lateral sides of the domain in the investigated cases of $\theta>90^0$. The boundaries are periodic at the lateral sides while no-slip at the top and bottom walls. The dimensionless grid size and time step are $h=0.01$ and $\Delta t=5\times 10^{-5}$, respectively. All the results reported in this section are in their dimensionless forms.

Fig.\ref{Fig BDrop} shows results with contact angle $\theta=165^0$ at the bottom wall. The drop remains circular as it is falling down. After the drop impacts on the bottom wall, it is strongly deformed to reduce the downward velocity and finally reaches a ``dumbbells-like'' shape. Then, the drop tries to restore the circular shape and jumps upward, leaving the bottom wall and finally arriving at a height lower than where it is initially released. This process repeats and the velocity is gradually reduced to zero. Finally, the drop settles down on the bottom wall and the equilibrium shape deviates slightly from the circular one because of the gravity. 

Different contact angles at the bottom wall are considered. We observe that the drop is unable to bounce back when the contact angle is less than or equal to $120^0$ and the water finally fills the bottom of the domain when the contact angle is less than or equal to $90^0$. The same behaviors are also reported in \citep{Dong2012}. Fig.\ref{Fig BDrop-CA} shows shapes of the drops from different contact angles at $t=0.46$, right after the first impact to the bottom wall, and at $t=4.00$. The ($y$-component) center of mass of the drop $y_c$ ($y_c=\int_{\Omega} y\frac{1+\phi}{2} d\Omega/\int_{\Omega} \frac{1+\phi}{2} d\Omega $) versus time is shown in Fig.\ref{Fig BDrop-yc} a). Until the second impact to the bottom wall, the centers of mass from $\theta=165^0$ and $\theta=150^0$ move very similarly, as shown in Fig.\ref{Fig BDrop-yc} a). However, with a smaller contact angle, length of the drop in contact with the bottom wall is larger, as shown in Fig.\ref{Fig BDrop-CA}. This can provide more dissipation, and as a result, the drop have a less chance to bounce back. On the other hand, each time when the drop impacts to the wall induces a large deformation of the drop, which also produces a strong dissipation due to the viscosity of the water. Therefore, from Fig.\ref{Fig BDrop-yc} a), peaks of the curves describing the motion of center of mass decay very fast for the drops that bounce back, e.g., those with $\theta=165^0$ and $150^0$.
For the drop that is unable to bounce back, e.g., the one with $\theta=120^0$, it oscillates on the bottom wall, and its center of mass curve has a higher frequency but there is less attenuation between the two neighboring peaks.
For the drop that will finally fill the bottom, e.g., those with $\theta=90^0$ and $\theta=60^0$, we observe a long-term but small-amplitude oscillation of the center of mass. This is caused by the capillary wave on the horizontal water-air interface, as shown in Fig.\ref{Fig BDrop-CA}.

Finally, we consider the effect of the mobility $M$. Fig.\ref{Fig BDrop-Mob} shows shapes of the drops with different mobilities (or $M\lambda$), and the mass centers ($y$ component) are shown in Fig.\ref{Fig BDrop-yc} b). With a larger mobility, the drop becomes more ``rigid'' and therefore less deforms, as shown in Fig.\ref{Fig BDrop-Mob}. On the other hand, a too ``soft'' drop, resulting from a small mobility, suffers from fictitious oscillation on the side close to the bottom wall. Even worse, the oscillation destroys the symmetry of the solution, and at the end produces a non-symmetry drop staying biased to left half of the domain. The drop with the smallest mobility finally is floating above the bottom wall because the interface is over-thickened. However, these unphysical behaviors are not observed in the cases with a larger mobility. As shown in Fig.\ref{Fig BDrop-yc} b), there is no significant difference due to the mobility before the first impact of the drop to the bottom wall. The one with the largest mobility can only bounce back once and settles down very fast. The one with the smallest mobility bounces back multiple times although the height it returns to after the first impact is lowest among the three cases. These behaviors suggest that a larger mobility produces more dissipation.
\begin{figure}[!t]
    \centering
    \includegraphics[scale=.4]{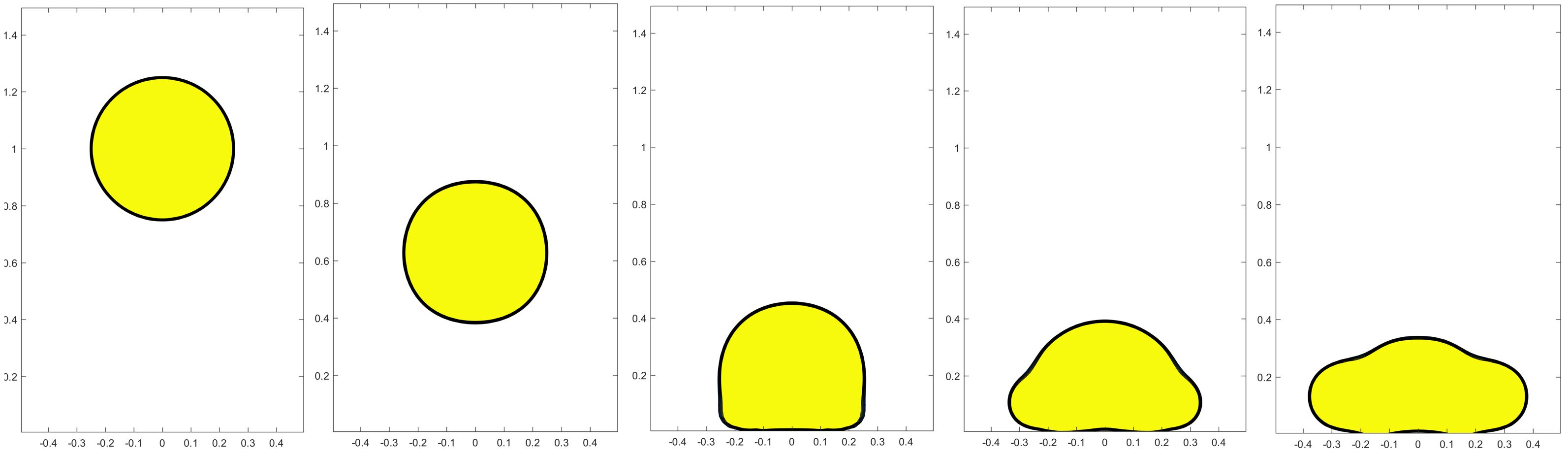}
	\includegraphics[scale=.4]{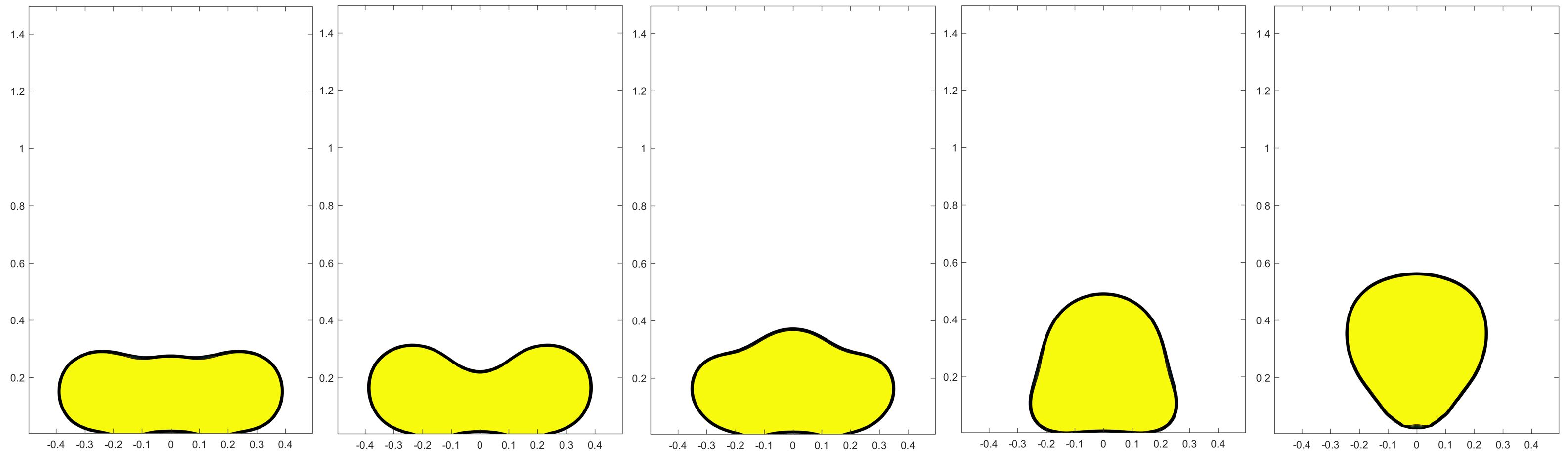}
	\includegraphics[scale=.4]{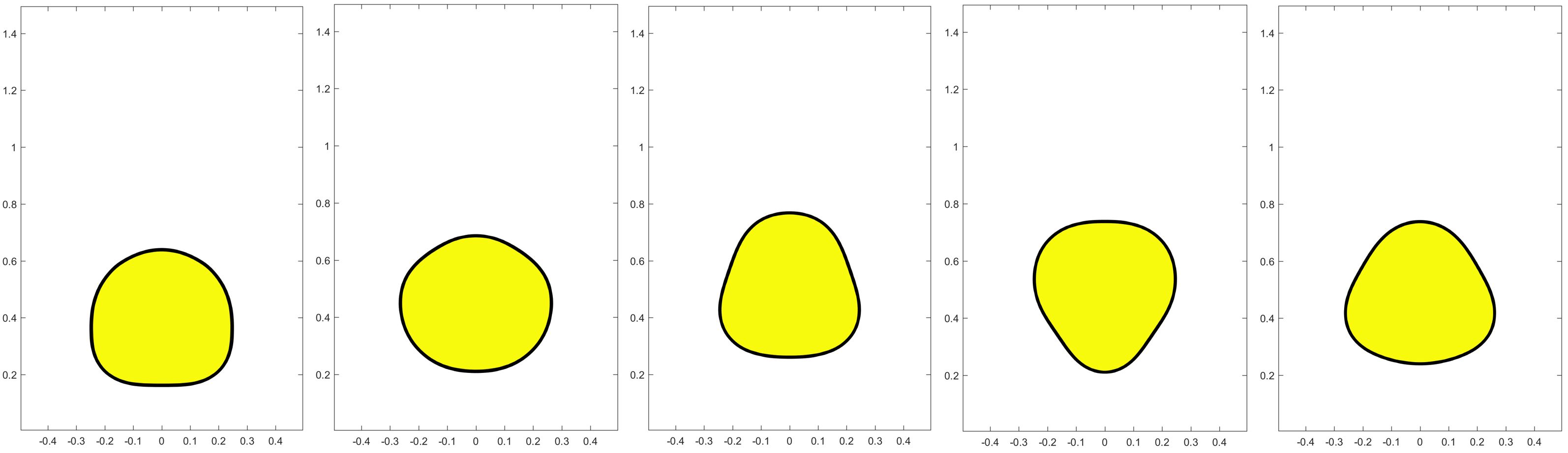}
	\includegraphics[scale=.4]{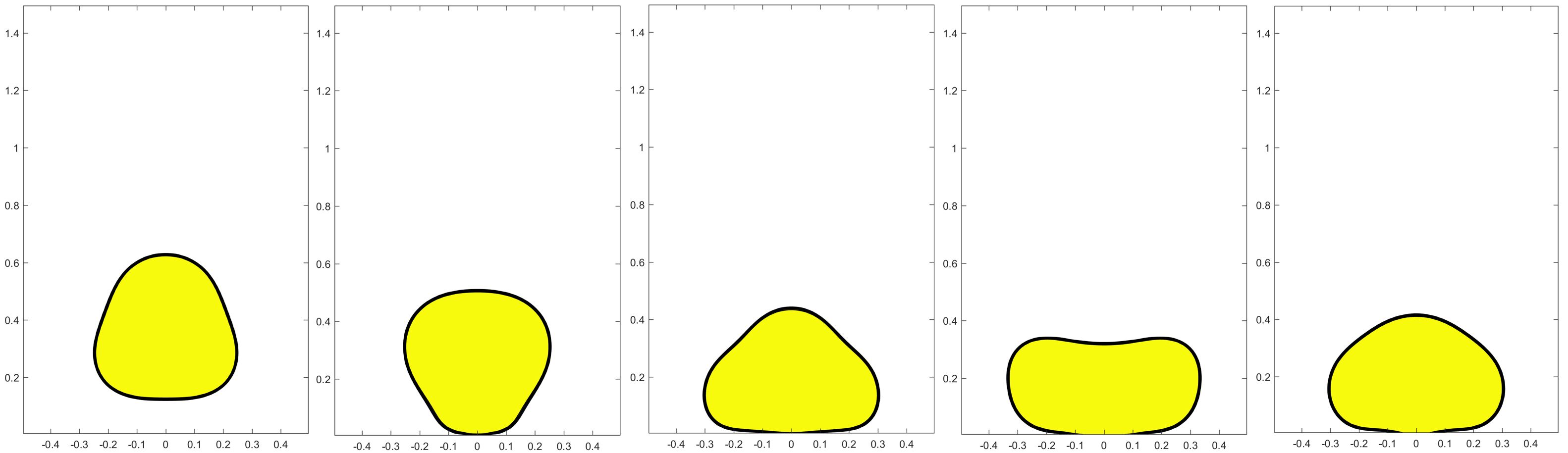}
    \caption{Results of the bouncing drop using the two-phase model Eq.(\ref{Eq CAC two-phase}) with $\theta=165^0$. Yellow: water (Phase 1); White: air (Phase 2); From left to right and top to bottom: $t=0.00$, $t=0.30$, $t=0.44$, $t=0.46$, $t=0.48$, $t=0.50$, $t=0.52$, $t=0.56$, $t=0.60$, $t=0.64$, $t=0.68$, $t=0.72$, $t=0.80$, $t=0.85$, $t=0.90$, $t=1.00$, $t=1.05$, $t=1.10$, $t=1.15$, and $t=1.20$.
    \label{Fig BDrop}}
\end{figure}
\begin{figure}[!t]
    \ContinuedFloat
    \captionsetup{list=off,format=cont}
    \centering
    \includegraphics[scale=.4]{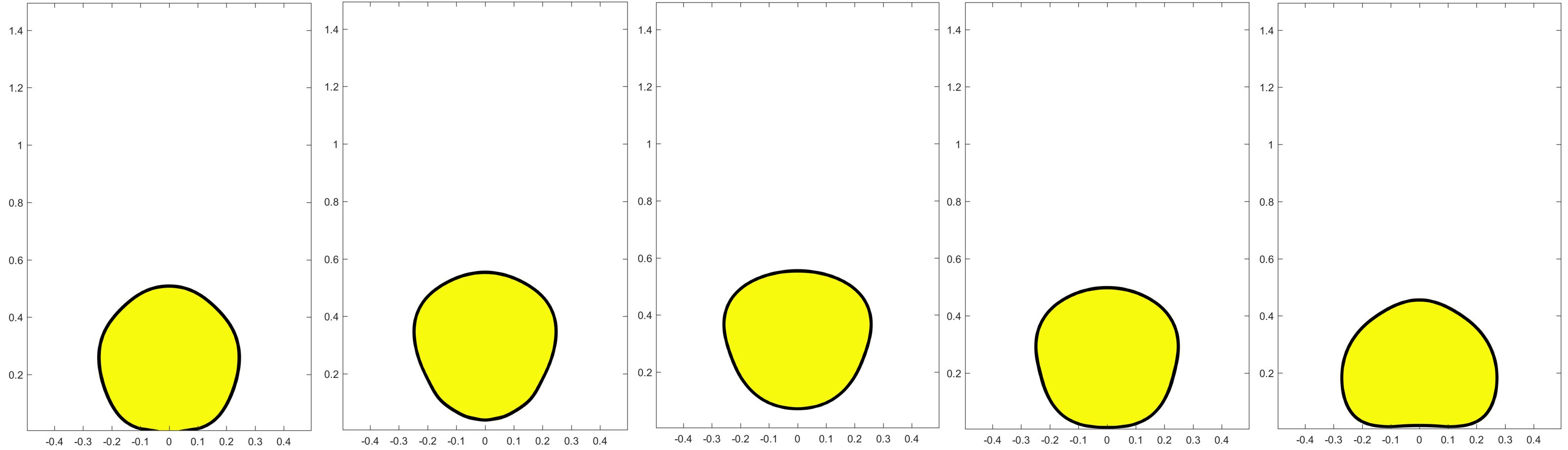}
	\includegraphics[scale=.4]{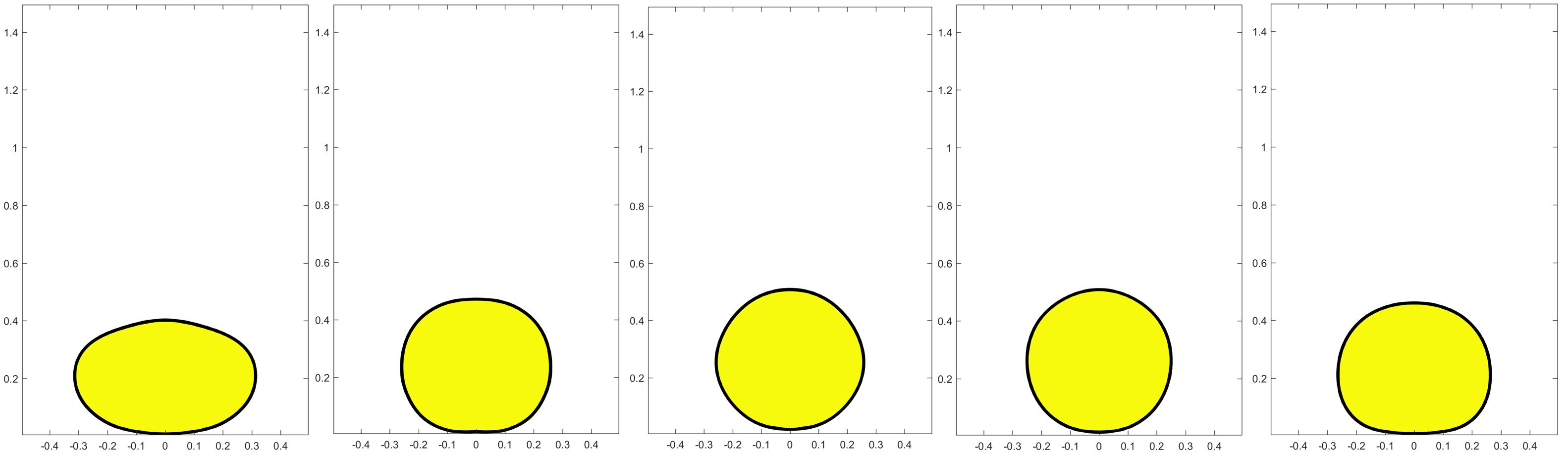}
	\includegraphics[scale=.4]{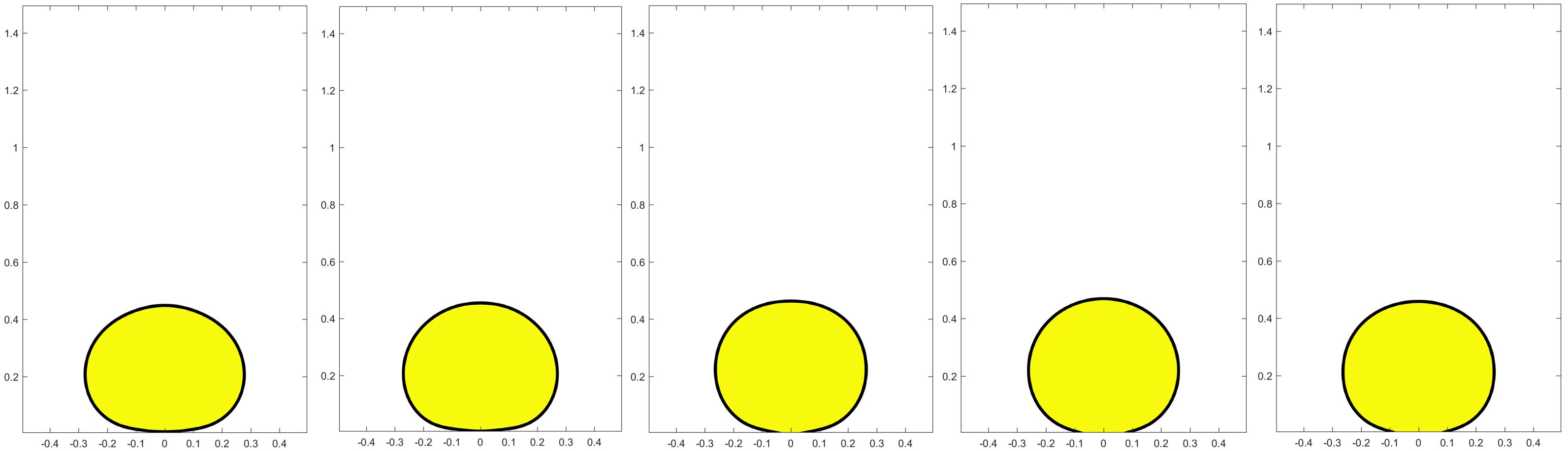}
	\includegraphics[scale=.4]{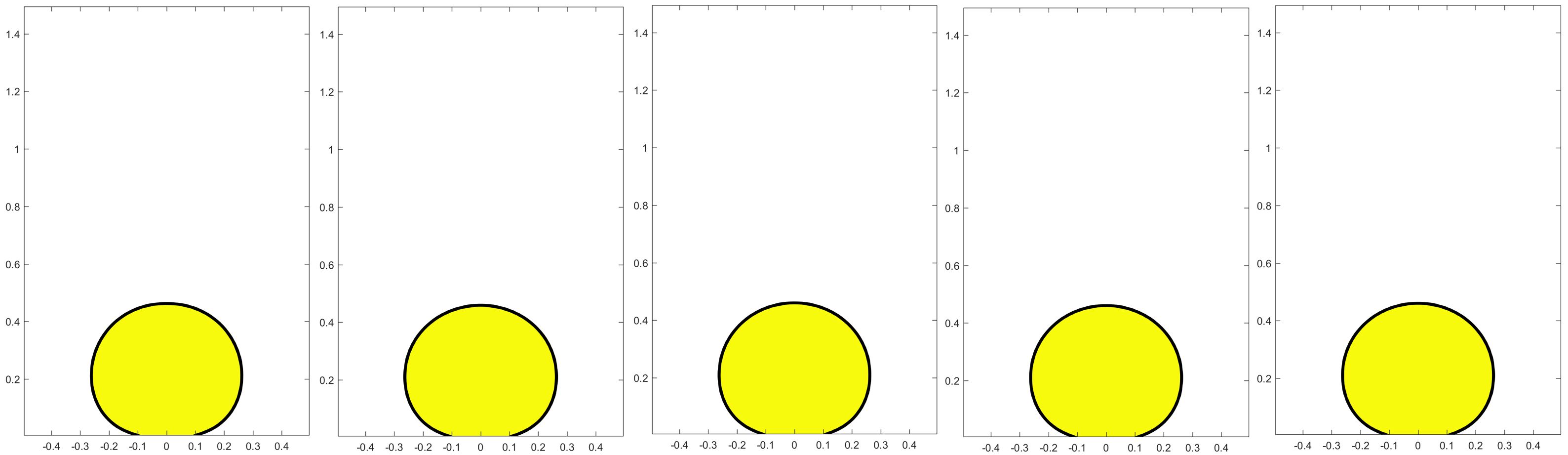}
    \caption{From left to right and top to bottom: $t=1.25$, $t=1.30$, $t=1.40$, $t=1.50$, $t=1.55$, $t=1.60$, $t=1.70$, $t=1.75$, $t=1.80$, $t=1.90$, $t=1.95$, $t=2.00$, $t=2.05$, $t=2.15$, $t=2.20$, $t=2.35$, $t=2.50$, $t=3.00$, $t=3.50$, and $t=4.00$.}
\end{figure}
\begin{figure}[!t]
    \centering
    \includegraphics[scale=.4]{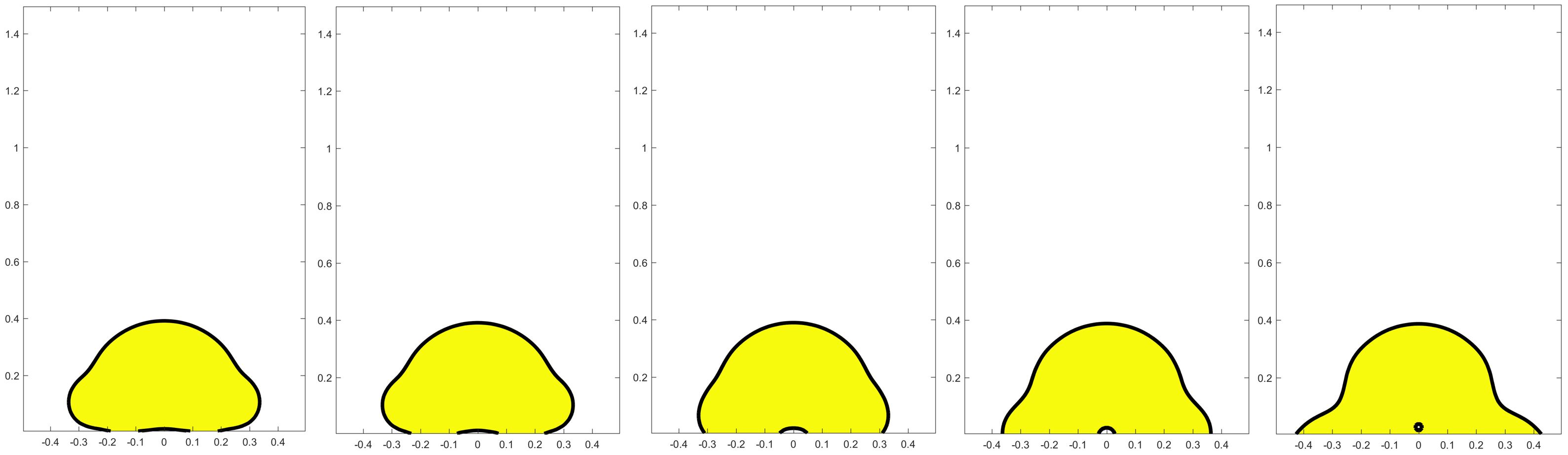}
	\includegraphics[scale=.4]{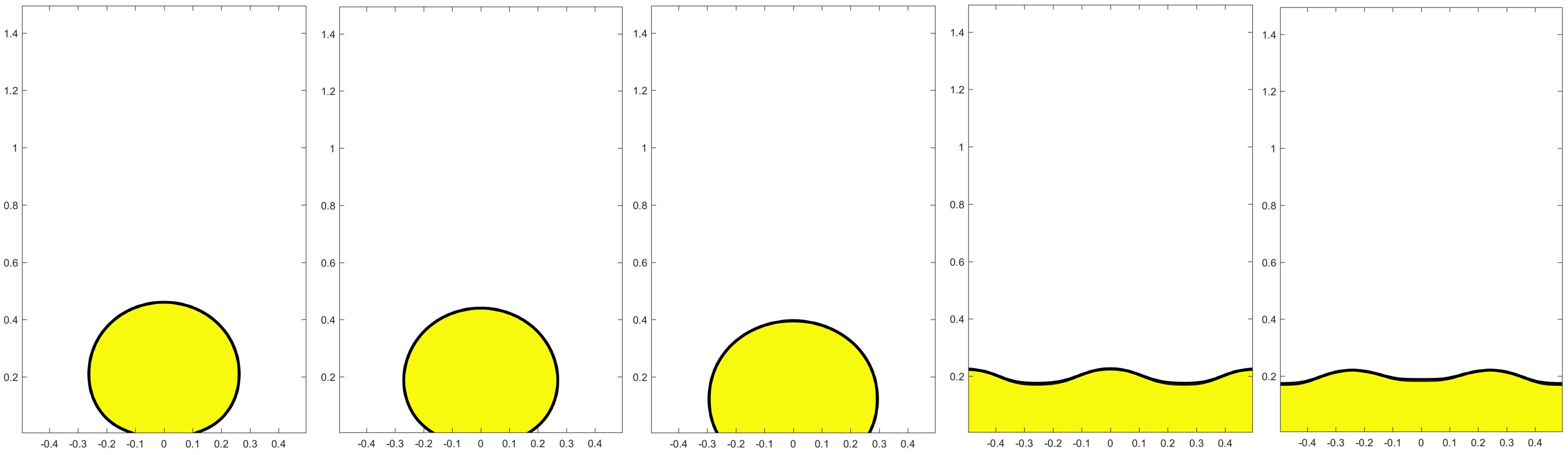}
    \caption{Shapes of the drops with different contact angles. Yellow: water (Phase 1); White: air (Phase 2); From left to right: $\theta=165^0$, $\theta=150^0$, $\theta=120^0$, $\theta=90^0$, and $\theta=60^0$. Top: $t=0.46$; Bottom: $t=4.00$.
    \label{Fig BDrop-CA}}
\end{figure}
\begin{figure}[!t]
    \centering
    \includegraphics[scale=.5]{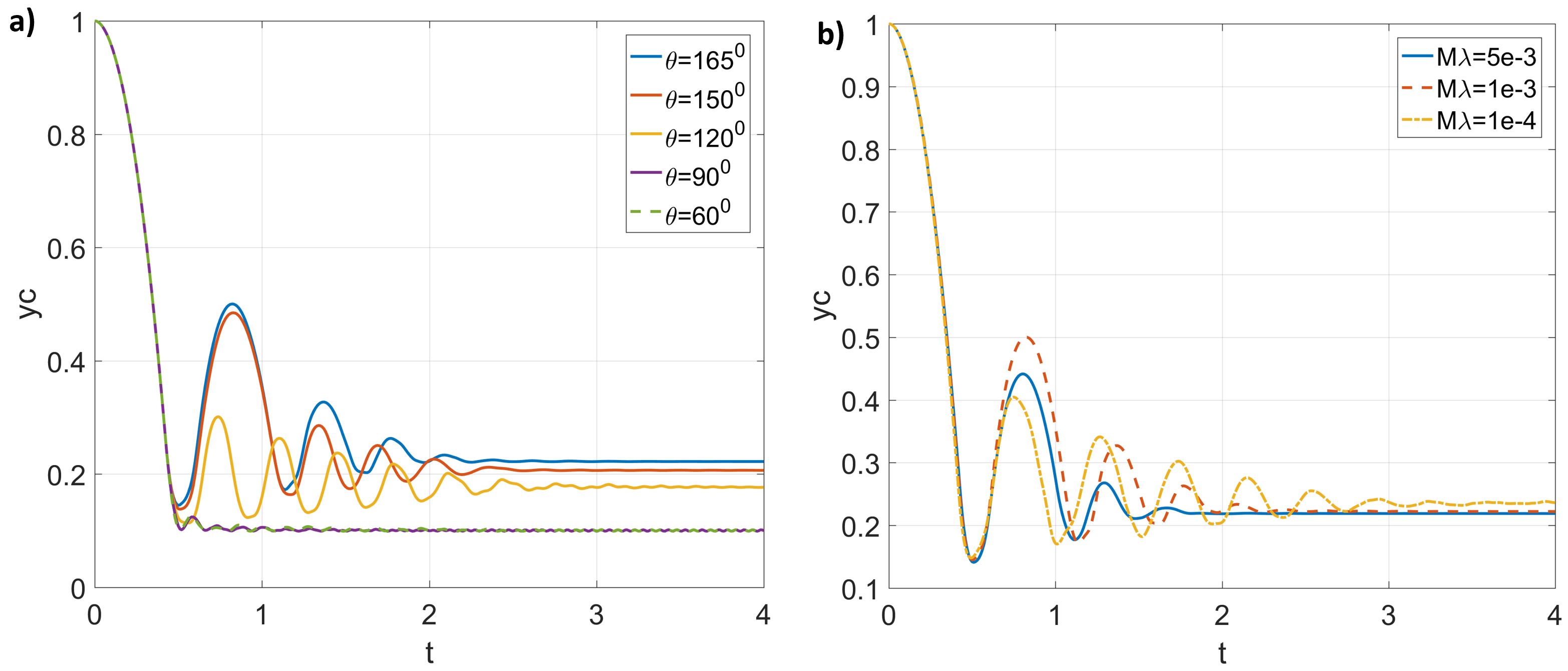}
    \caption{Mass center ($y$ component) of the drop versus time
    a) with different contact angles,
    b) with different mobilities.
    \label{Fig BDrop-yc}}
\end{figure}
\begin{figure}[!t]
    \centering
    \includegraphics[scale=.4]{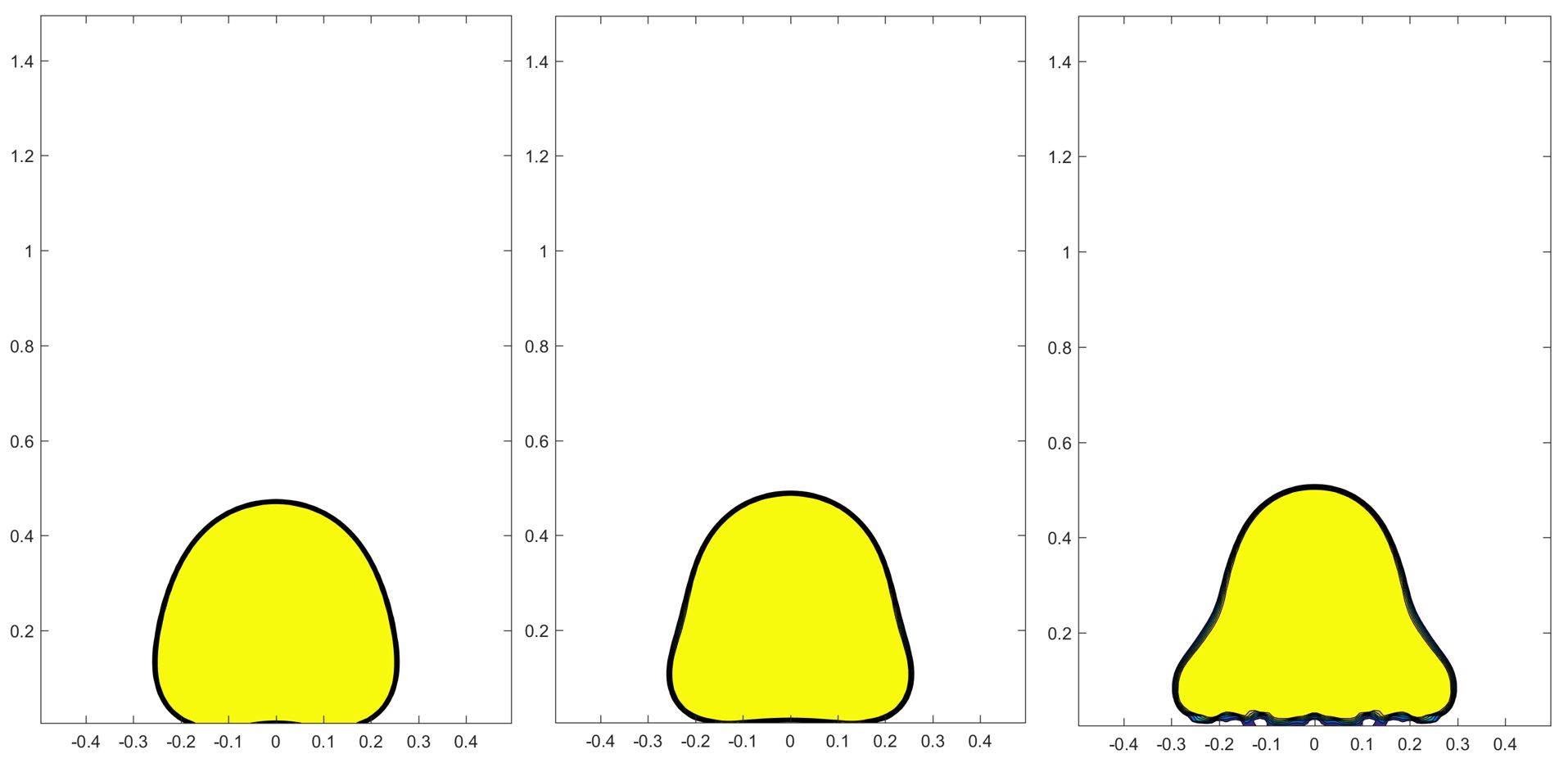}\\
    \includegraphics[scale=.4]{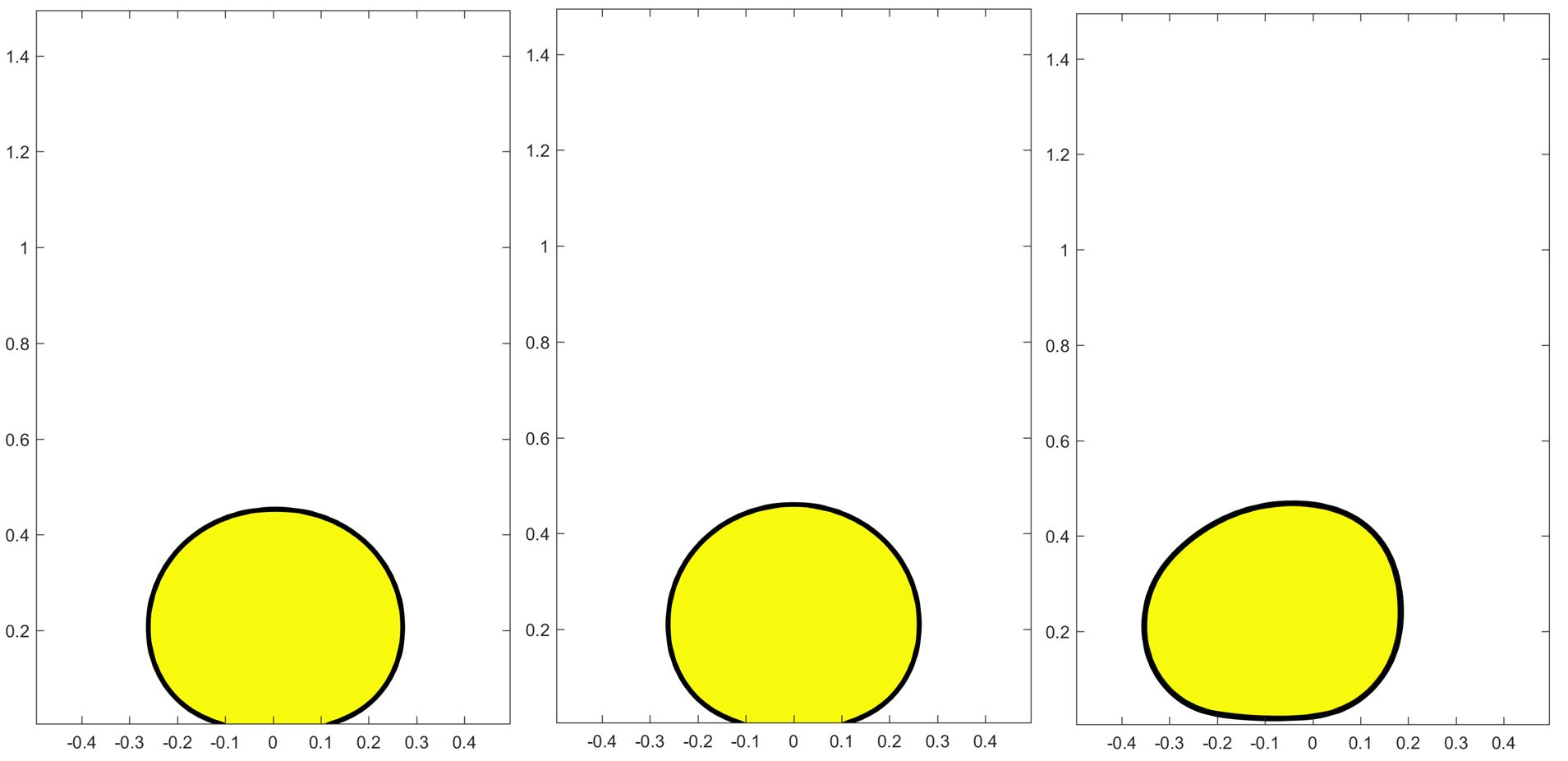}
    \caption{Shapes of the drops with different mobilities. Yellow: water (Phase 1); White: air (Phase 2); From left to right: $M\lambda=5 \times 10^{-3}$, $M\lambda=1\times10^{-3}$, and $M \lambda=1\times10^{-4}$. Top: $t=0.60$; Bottom: $t=4.00$.
    \label{Fig BDrop-Mob}}
\end{figure}

\subsection{Compound drop}\label{Sec Compound drop}
Here, we report a compound drop sliding on a horizontal solid wall using the $N$-phase model Eq.(\ref{Eq CAC N-phase}). 
Initially, the compound drop is semicircular with a radius $R_0$, composed of two quarter-circular drops. The left and right quarters are full of Phases 1 and 2, respectively, and they are surrounded by Phase 3. 
The Reynolds number and capillary number considered are $Re=\rho_1 U R_0/\mu_1=10$ and $Ca=\mu_1 U/\sigma_{1,2}=0.1$, respectively. Here, $U$ is determined from the inertia-capillary time scale $T=\sqrt{\rho_1 R_0^3/\sigma_{1,2}}$, i.e., $U=R_0/T$, which leads to the Weber number $We=\rho_1 U^2 R_0/\sigma_{1,2}=1$. 
The material properties of the other two phases are related to $\rho_1$, $\mu_1$, and $\sigma_{1,2}$, and are listed in Table~\ref{Table CD}.
\begin{table*}[!t]
\caption{Material properties in the compound drop}
    \centering
    \includegraphics[scale=.3]{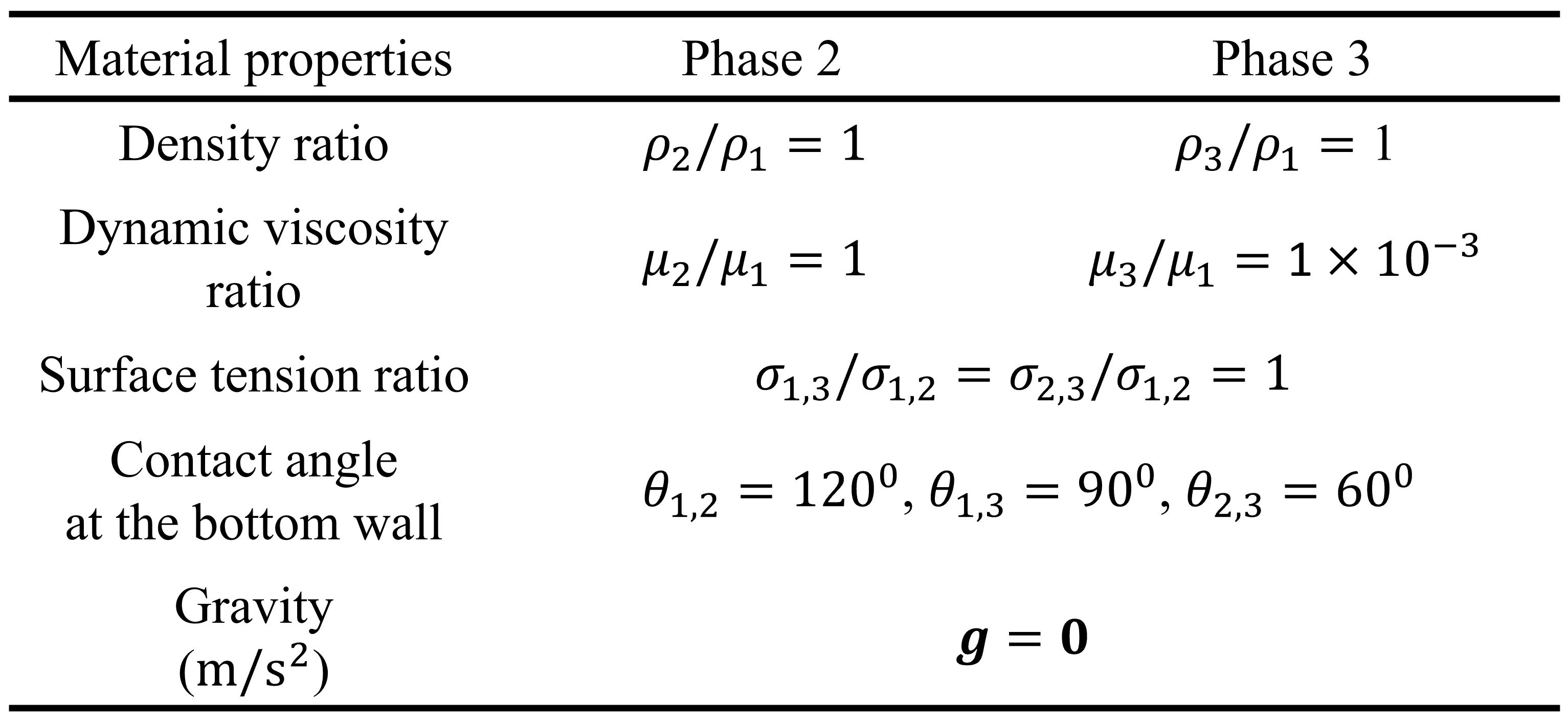}
    \label{Table CD}
\end{table*}

\begin{figure}[!t]
	\centering
	\includegraphics[scale=.4]{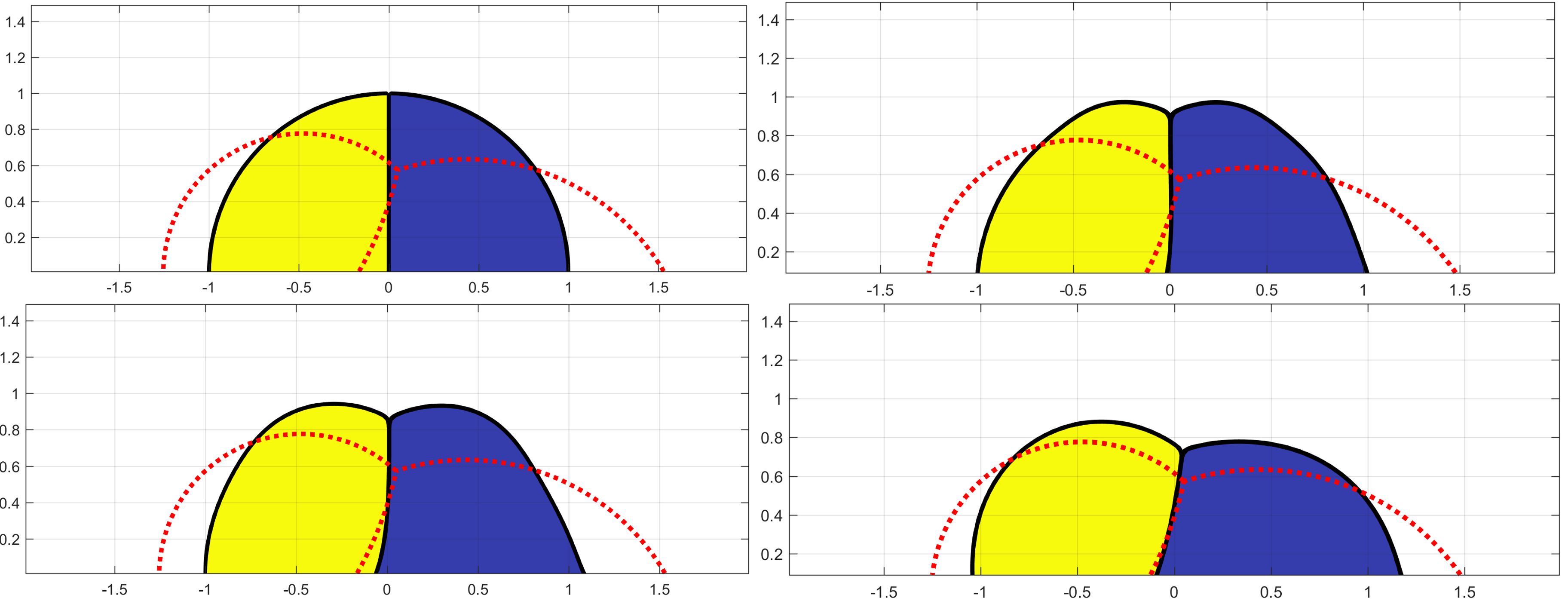}
	\includegraphics[scale=.4]{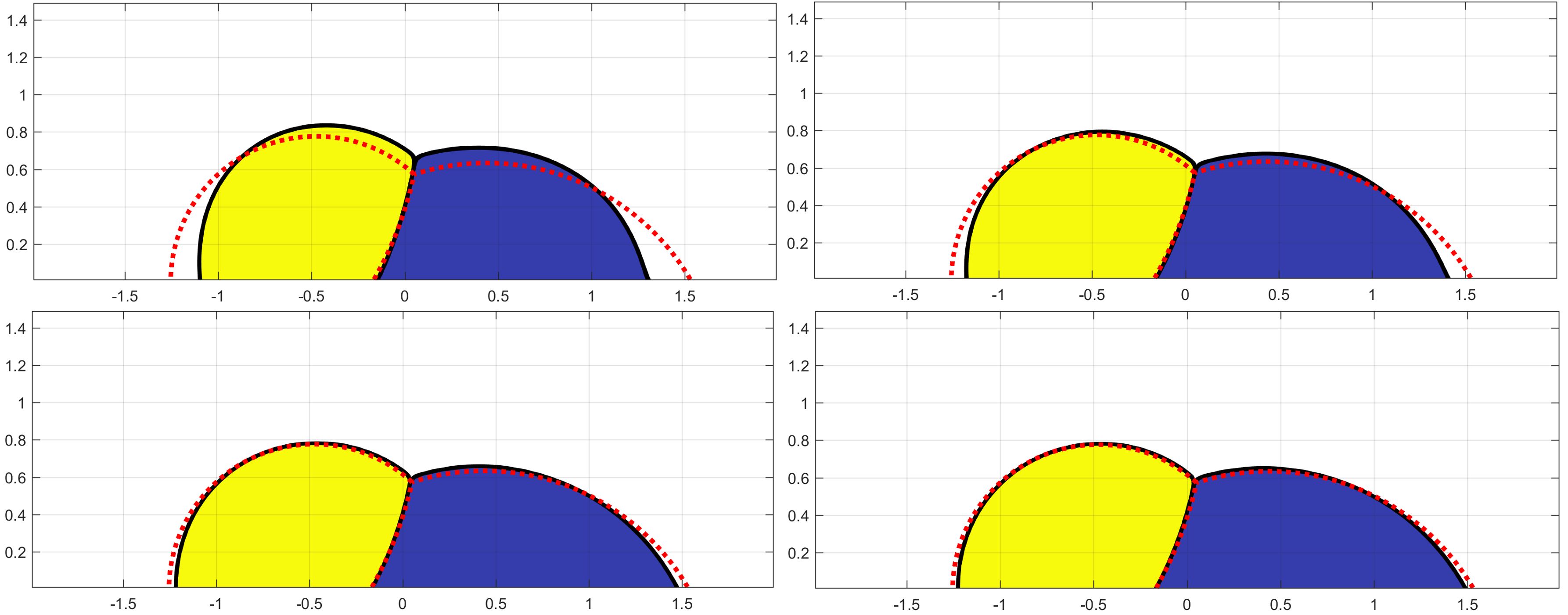}
	\includegraphics[scale=.4]{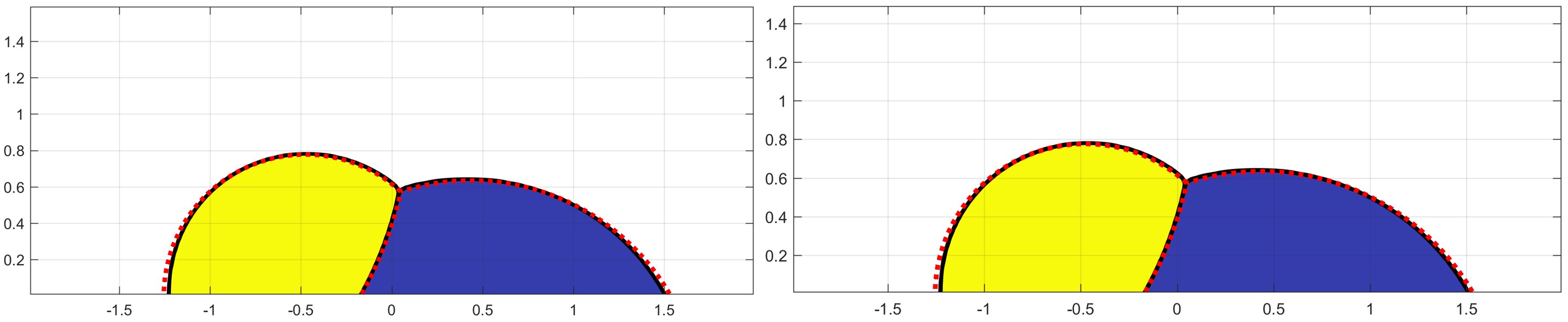}
	\caption{Evolution of the compound drop using Eq.(\ref{Eq CAC N-phase}) with a stationary bottom wall. The coordinate has been normalized by the initial radius of the compound drop $R_0$. Yellow: Phase 1; Blue: Phase 2; White: Phase 3; Red dotted line: exact solution from \citep{Zhangetal2016}. From top to bottom and left to right: $Ut/R_0=0.0$, $0.2$, $0.4$, $1.0$, $1.4$, $2.0$, $3.0$, $4.0$, $5.0$ and $6.0$, where $U=\sqrt{\sigma_{1,2}/(\rho_1 R_0)}$ is the inertia-capillary velocity scale.
    \label{Fig CDrops}}
\end{figure}

The computational domain is $[-2R_0,2R_0]\times[0R_0,1.2R_0]$, and the periodic and no-slip boundary conditions are assigned along the $x$ and $y$ axes, respectively. The compound drop is initially on the middle of the bottom wall. The space and time are discretized by $200\times60$ grid cells and $U \Delta t/R_0=1 \times 10^{-4}$.
Evolution of the drops are shown in Fig.\ref{Fig CDrops}, along with the exact solution from \citep{Zhangetal2016} for the equilibrium state.
The drops move towards the equilibrium shape, which agrees well with the exact solution. Quantitatively, the spreading lengths (normalized by $R_0$) of Phases 1 and 2 are $1.0547$ and $1.6871$, respectively, and the relative errors are 1.614\% and 1.166\% after comparing to the exact ones $1.0720$ and $1.7070$ from \citep{Zhangetal2016}.

Next, we investigate sliding motion of the compound drop when the bottom wall is moving backward with a speed $U$, and the setup is slightly changed as follows. Using the speed of the bottom wall, the corresponding Reynolds number, capillary number, and Weber number considered are $Re=\rho_1 U R_0/\mu_1=66.6$, $Ca=\mu_1 U/\sigma_{1,2}=0.075$, and $We=\rho_1 U^2 R_0/\sigma_{1,2}=5$, respectively. The dynamic viscosity ratios are changed to be $\mu_2/\mu_1=0.67$ and $\mu_3/\mu_1=0.33$. The domain height becomes $1.5R_0$, while the grid size remains the same. Results are shown in Fig.\ref{Fig CDrops W}, and the behaviors of the drops are significantly different from those on a stationary wall. We observe that the Phase 1 (yellow) drop climbs onto the Phase 2 (blue) drop, and thoroughly leave the bottom wall, sitting on the Phase 2 drop. Then, it crosses the Phase 2 drop and returns on the bottom wall. At the end, the Phase 1 drop is still in contact with the Phase 2 drop but moves in front of it. 
\begin{figure}[!t]
	\centering
	\includegraphics[scale=.4]{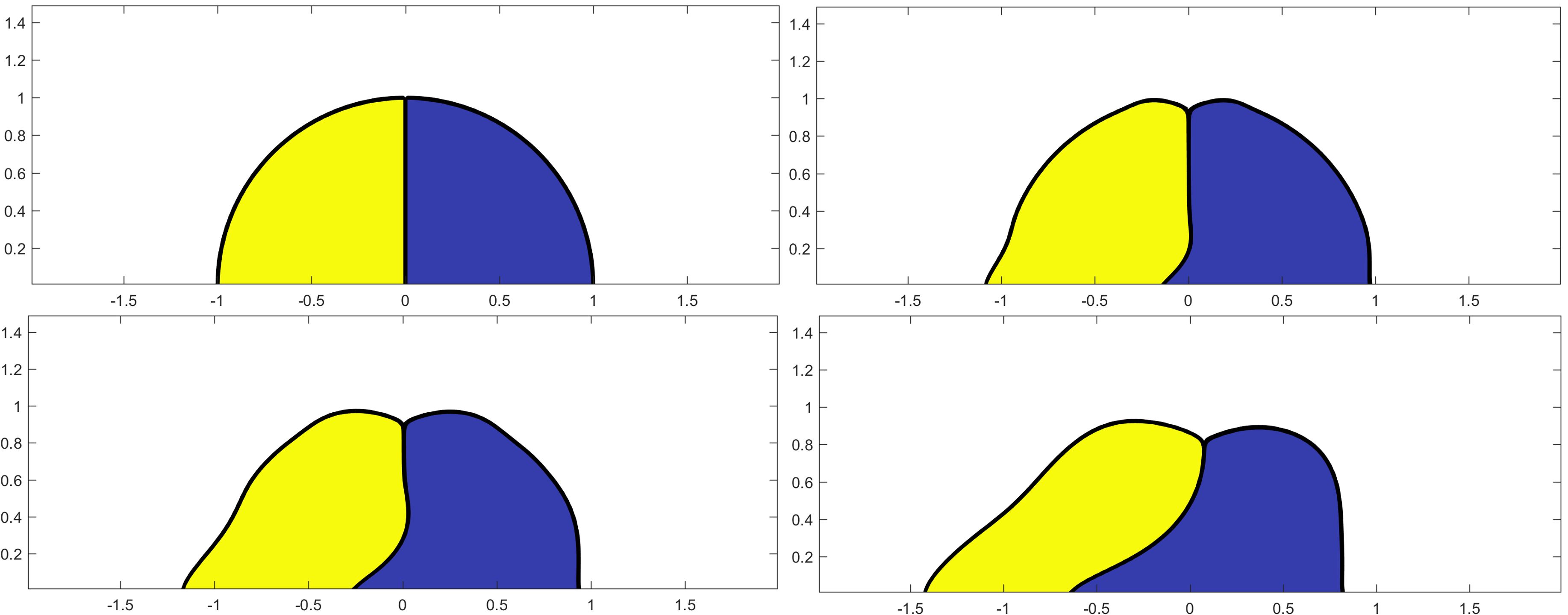}
	\includegraphics[scale=.4]{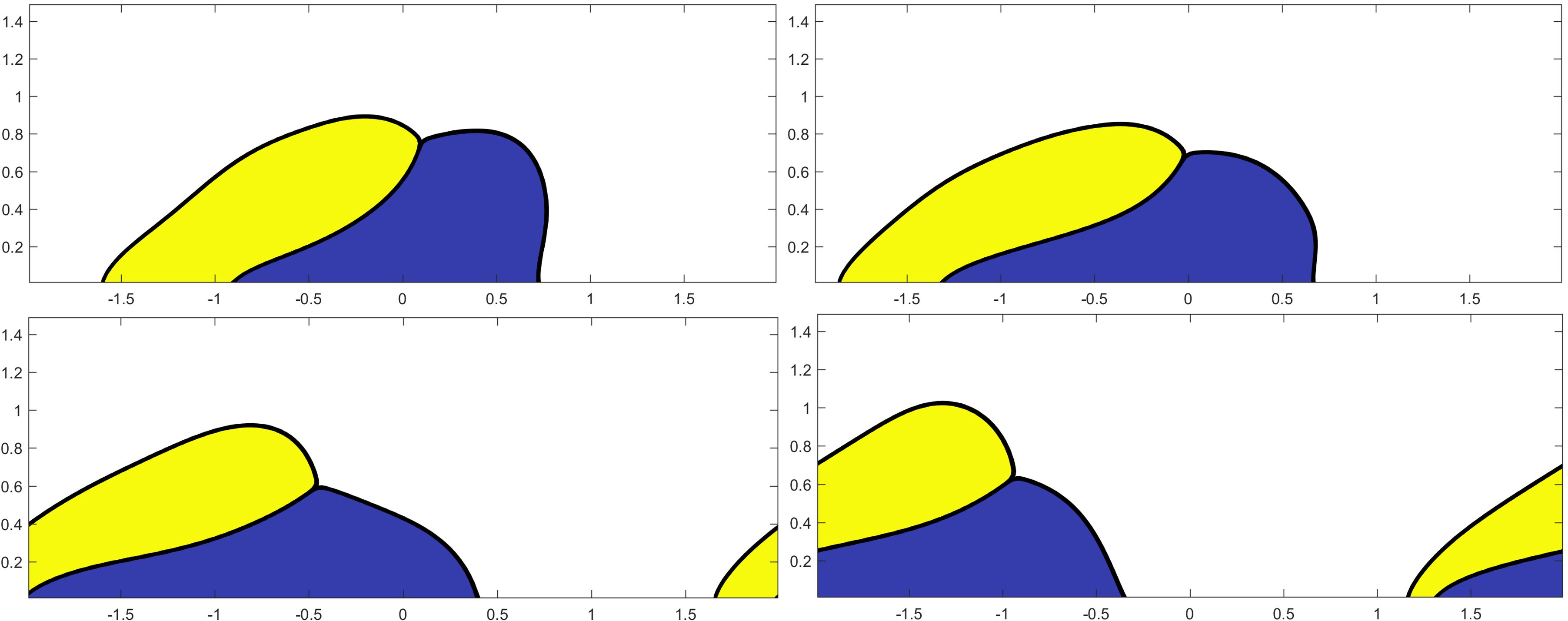}
	\includegraphics[scale=.4]{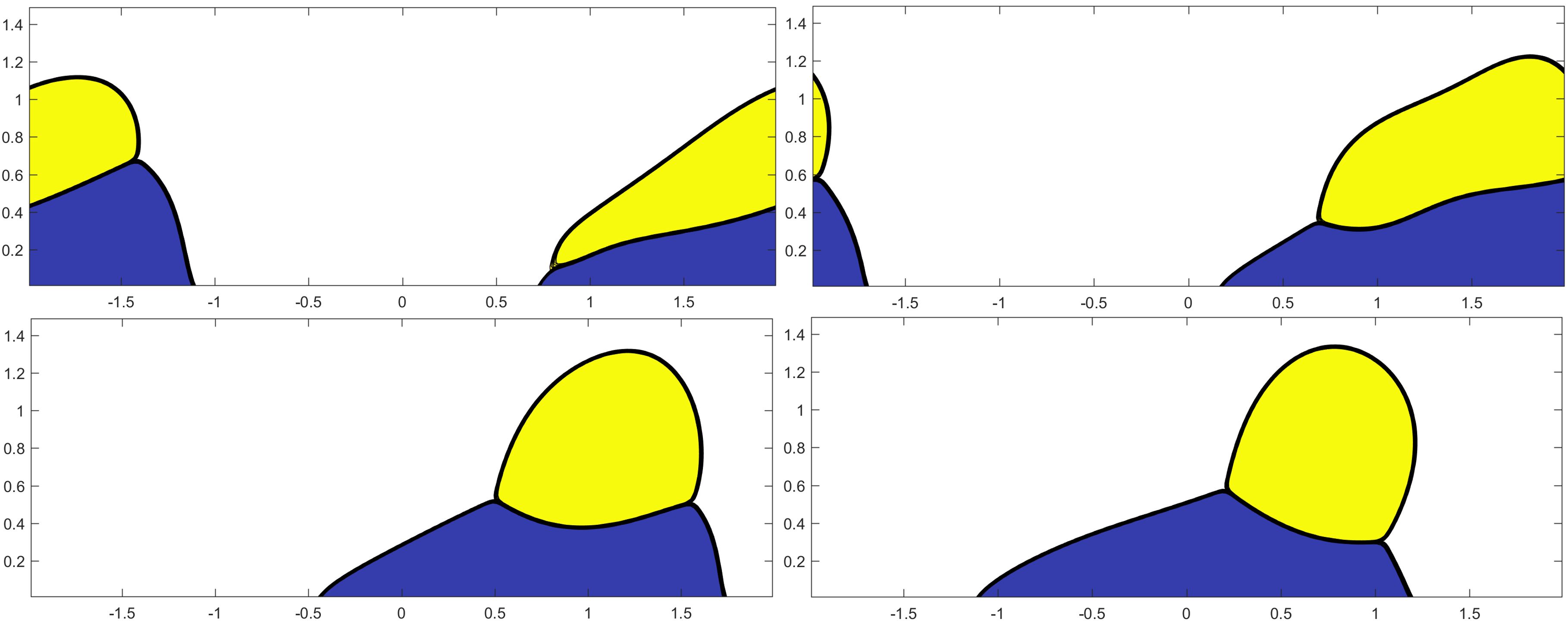}
	\caption{Evolution of the compound drop using Eq.(\ref{Eq CAC N-phase}) with a translating bottom wall. The coordinate has been normalized by the initial radius of the compound drop $R_0$. Yellow: Phase 1; Blue: Phase 2; White: Phase 3. From top to bottom and left to right: $Ut/R_0=0.0$, $0.2$, $0.4$, $1.0$, $1.4$, $2.0$, $3.0$, $4.0$, $5.0$, $6.0$, $7.0$, and $8.0$, where $U$ is the speed of the bottom wall.
    \label{Fig CDrops W}}
\end{figure}
\begin{figure}[!t]
    \ContinuedFloat
    \captionsetup{list=off,format=cont}
	\centering
	\includegraphics[scale=.4]{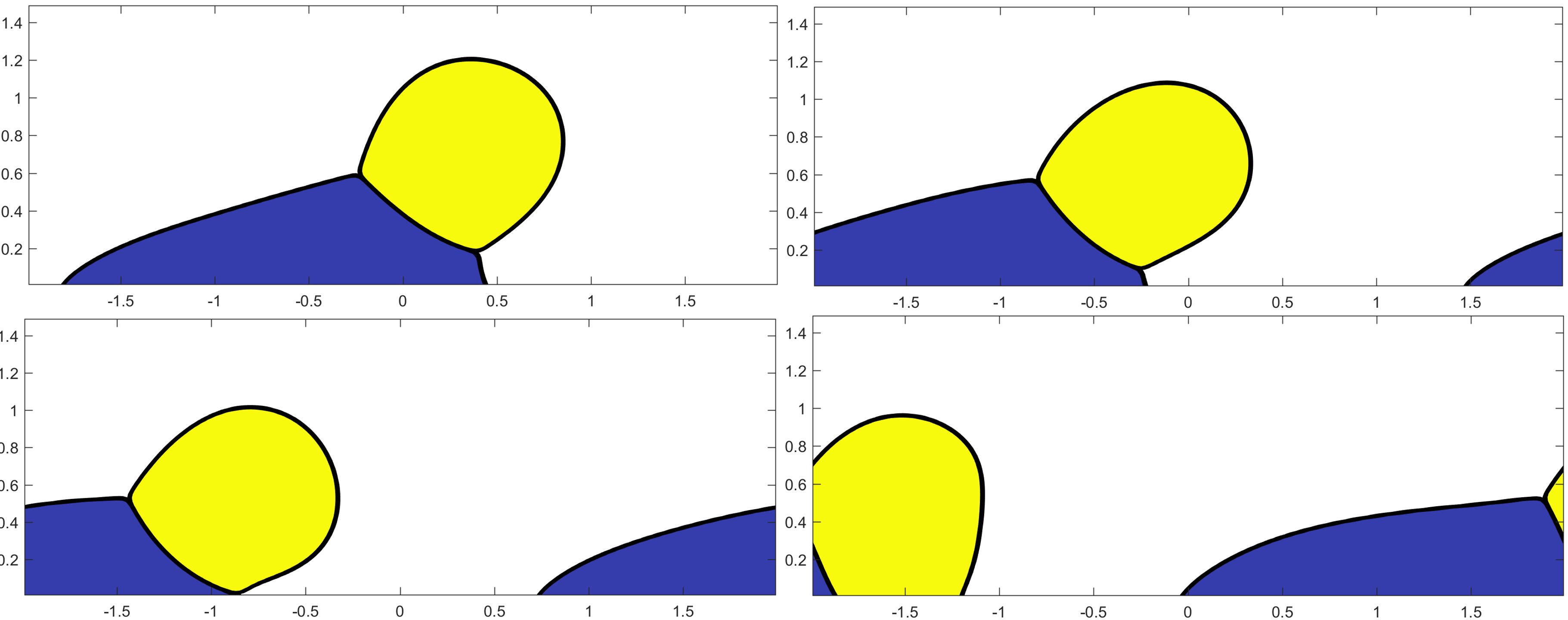}
	\includegraphics[scale=.4]{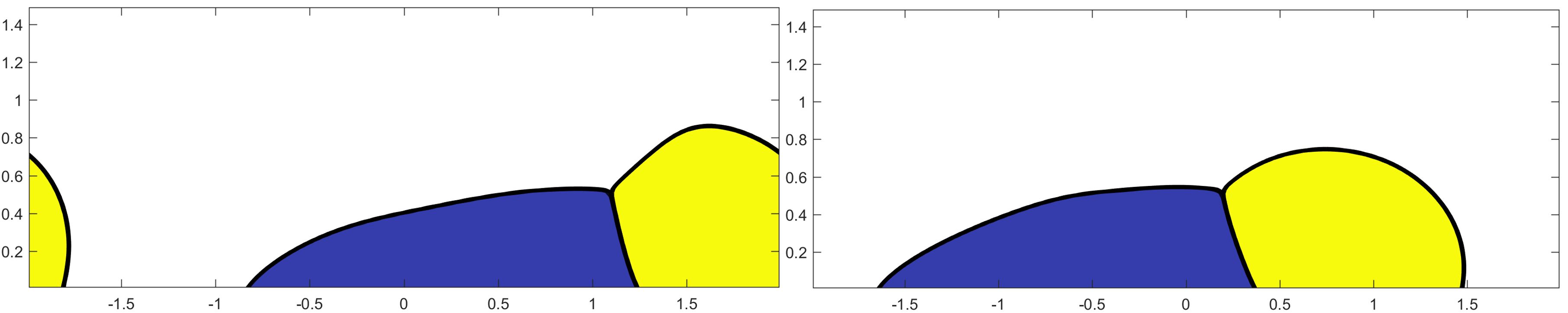}
	\caption{From top to bottom and left to right: $Ut/R_0=9.0$, $10.0$, $11.0$, $12.0$, $13.0$, and $14.0$, where $U$ is the speed of the bottom wall.}
\end{figure}

\section{Conclusions and future works}\label{Sec Conclusions}
In the present work, we proposed a general formulation to implement the contact angle boundary conditions for the second-order Phase-Field models. The original second-order Phase-Field models are modified by adding a Lagrange multiplier that enforces the mass conservation but does not change the summation of the order parameters and the \textit{consistency of reduction}. The newly introduced Lagrange multiplier is determined by the consistent and conservative volume distribution algorithm \citep{Huangetal2020B}. The proposed formulation is applicable to not only two-phase but also $N$-phase ($N \geqslant 2$) cases. Then, this novel formulation is physically coupled to the hydrodynamics using the consistent formulation \citep{Huangetal2020CAC} and can  be applied to large-density-ratio problems.
To demonstrate its effectiveness of moving contact line simulations, we apply the proposed formulation to the reduction-consistent multiphase conservative Allen-Cahn model \citep{Huangetal2020B}, whose two-phase version is equivalent to the one in \citep{BrasselBretin2011}.
The complete system is numerically solved by the consistent and conservative scheme \citep{Huangetal2020CAC,Huangetal2020B}, which preserves the mass conservation, the summation of the order parameters, and the \textit{consistency of reduction} exactly on the discrete level, as validated in the present study. 
Various numerical tests are performed in both 2D Cartesian and axisymmetric coordinates. The proposed formulation accurately reproduces the exact and/or asymptotic solutions for equilibrium problems, and captures important dynamical behaviors reported, e.g., in \citep{Yueetal2010,Zhangetal2016,Dong2012,Dong2017} using the Cahn-Hilliard models which are a 4th-order Phase-Field model, in \citep{Renardyetal2001} using the Volume-of-Fluid (VoF) method, and in \citep{Eddietal2013} performing experiments. 
Since the parallelization has not been implemented, only two-dimensional and axisymmetric results are reported. The extension of the proposed formulation to three-dimensional problems is straightforward without any modifications, and the physical properties demonstrated in the present study, such as the summation of the order parameters, mass conservation, and \textit{consistency of reduction}, will remain intact. However, an efficient parallel strategy with adaptive mesh refinement (AMR) is desired for three-dimensional problems, and this is a valuable future direction to proceed with the present study.

The present study leaves open the possibility of using the 2nd-order Phase-Field models for moving contact line problems, which has never been considered before. Therefore, it provides plenty of new opportunities to study in the future.
Generally speaking, the accuracy of the prediction heavily relies on the properties of the Phase-Field model and the contact angle boundary condition, i.e., the definitions of $\mathcal{L}$ in Eq.(\ref{Eq Phase-Field}) and $\mathcal{F}^w$ in Eq.(\ref{Eq Contact angle boundary}), and on the parameters therein. Since the pool of plausible Phase-Field models for moving contact line problems is greatly expanded, it is now not only possible but also desirable to investigate and clarify their performance. 
Unlike the Cahn-Hilliard models, there is little theoretical analysis of the 2nd-order Phase-Field models in moving contact line problems, e.g., the asymptotic analysis as the interface thickness tends to zero. Such an analysis is important to provide physical insights of determining the parameters in the models. We expect the present study will motivate consideration of using 2nd-order Phase-Field models in moving contact line problems as the effectiveness has been demonstrated.

\section*{Acknowledgments}
A.M. Ardekani would like to acknowledge the financial support from the National Science Foundation (CBET-1705371). This work used the Extreme Science and Engineering Discovery Environment (XSEDE) \cite{Townsetal2014}, which is supported by the National Science Foundation grant number ACI-1548562 through allocation TG-CTS180066  and TG-CTS190041.
G. Lin would like to acknowledge the support from National Science Foundation (DMS-1555072 and DMS-1736364, CMMI-1634832 and CMMI-1560834), and U.S. Department of Energy (DOE) Office of Science Advanced Scientific Computing Research program DE-SC0021142. 

\appendix
\section{Manufactured solution}\label{Appendix Manufactured}
\begin{figure}[!t]
	\centering
	\includegraphics[scale=.4]{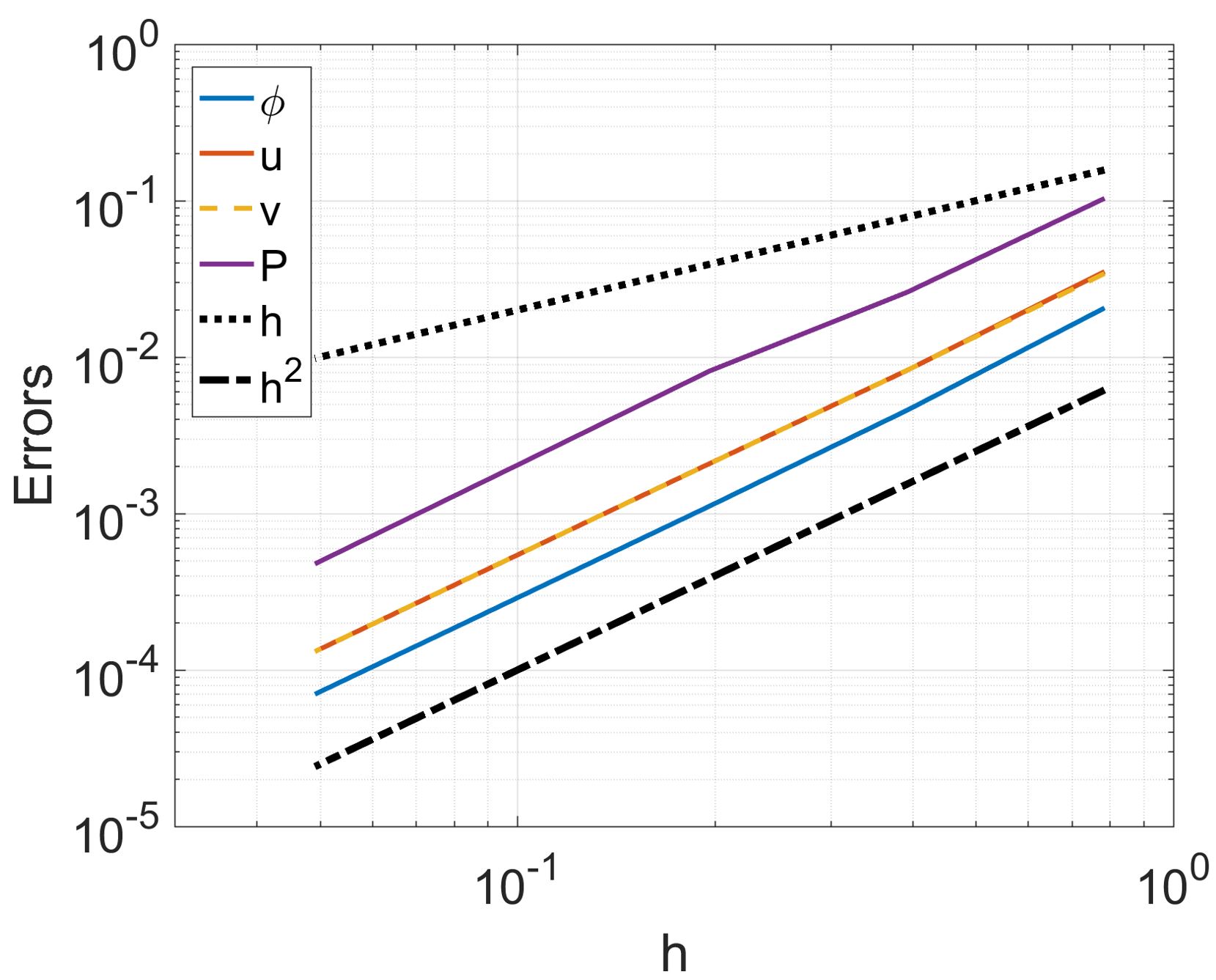}
	\caption{$L_1$ errors of $\phi$, $u$, $v$, and $P$ versus the cell size in the manufactured solution problem.
    \label{Fig Manufactured}}
\end{figure}

Here, we perform a manufactured solution test to the two-phase conservative Allen-Cahn model including the contact angle boundary condition to further demonstrate the convergence. In this problem, we assume that the exact solutions of the order parameter, velocity, and pressure are 
$\phi^E=\cos(x)\cos(y)\sin(t)$, 
$u^E=\sin(x)\cos(y)\cos(t)$, 
$v^E=-\cos(x)\sin(y)\cos(t)$, and 
$P^E=\cos(x)\cos(y)\sin(t)$, respectively. 
Then, a source term $S^{BC}$ is added to the contact angle boundary condition, i.e., 
$\mathbf{n} \cdot \nabla \phi=\mathcal{F}^w[\phi;\theta]+S^{BC}$,
where $S^{BC}$ is directly obtained with $\phi^E$, i.e.,
$S^{BC}=\mathbf{n} \cdot \nabla \phi^E-\mathcal{F}^w[\phi^E;\theta]$. 
In a similar manner, one can obtain the source terms added to the right-hand side of the Phase-Field model Eq.(\ref{Eq Phase-Field contact two-phase}) and the momentum equation Eq.(\ref{Eq Momentum}). We additionally assume $Q^E=\cos(x)\cos(y)\sin(t)$ to obtain the source term added to the right-hand side of the consistent formulation Eq.(\ref{Eq Q}).
The parameters used are $\rho_1=3$, $\rho_2=1$, $\mu_1=0.02$, $\mu_2=0.01$, $\sigma=0.0094$, $\mathbf{g}=\{1,-2\}$, $\eta=0.1$, and $M=0.001$.
The domain considered is $[-\pi,\pi]\times[-\pi,\pi]$ with the free-slip boundary condition. The contact angles at the boundaries are $90^0$ except the bottom one that is $75^0$. 
The initial conditions are $\phi^E$, $u^E$, $v^E$, and $P^E$ evaluated at $t=0$. The time step size is $\Delta t=1 \times 10^{-3}$, and the computations last till $t=1$.
We output $\phi$, $u$, $v$, and $P$, and depict the $L_1$ norms of $\phi-\phi^E$, $u-u^E$, $v-v^E$, and $P-P^E$, i.e., the averages of $|\phi-\phi^E|$ etc. over the domain, in Fig.\ref{Fig Manufactured}. All the variables are converging as the cell size is refined.

\bibliographystyle{plain}
\bibliography{refs.bib}

\end{document}